\documentclass[epj]{svjour}

\usepackage{graphics}
\usepackage{mathrsfs}
\usepackage{bm}
\usepackage{xr}
\usepackage{amsmath}
\usepackage{amssymb}
\usepackage{bbm}
\usepackage{amsfonts}
\usepackage{mathcomp}
\usepackage{graphicx}
\usepackage{lscape}
\usepackage{epsfig}
\usepackage{multirow}
\usepackage{array}
\usepackage[dvips]{color}

\newcommand{\be}{\begin{equation}}
\newcommand{\ee}{\end{equation}}
\newcommand{\ba}{\begin{eqnarray}}
\newcommand{\ea}{\end{eqnarray}}
\newcommand{\nn}{\nonumber}

\newcommand*{\mcal}[1]{\mathcal{#1}}                   
\newcommand*{\umat}[1]{\underline{\mathscr{#1}}}      
\newcommand*{\uvec}[1]{\underline{\bm{#1}}}           
\newcommand*{\vmc}[1]{\vec{\mathcal{#1}}}              
\newcommand{\LambShift}{L}											
\newcommand{\FineStructure}{\Delta}										
\newcommand{\HyperFineSplitting}{\mcal{A}}						
\newcommand{\sez}{\vec{e}_0}													
\newcommand{\sep}{\vec{e}_+}													
\newcommand{\sem}{\vec{e}_-}													
\newcommand{\mycell}[2]{\parbox{#1}{\vskip4pt #2 \vskip4pt}}	
\newcommand{\PV}{\mathrm{PV}}													
\newcommand{\PC}{\mathrm{PC}}													

\newcommand{\phm}{\phantom-}

\newcommand{\bokp}[3]{\left<\right.\hspace{-0.5ex}{#1}\left.\hspace{-0.5ex}\right|{#2}\left|\right.\hspace{-0.5ex}{#3}\left.\hspace{-0.5ex}\right)}

\providecommand*{\I}{\mathrm{i}}                           
\providecommand*{\ket}[1]{|#1\rangle}                      
\providecommand*{\rbra}[1]{(#1|}                           
\providecommand*{\rket}[1]{|#1)}                           
\providecommand*{\lrbra}[1]{\rbra{\widetilde{#1}}}    
\providecommand*{\rbracket}[2]{\rbra{#1}#2)}          
\providecommand*{\lrbracket}[2]{\lrbra{#1}#2)}       

\providecommand*{\klr}[1]{\left(#1\right)}					 

\renewcommand{\vec}[1]{\bm{#1}}                       
\renewcommand{\thesection}{\arabic{section}}
\renewcommand{\thesubsection}{\arabic{section}.\arabic{subsection}}
\renewcommand{\thetable}{\arabic{table}}

\begin{document}

\title{\Large Metastable states of hydrogen: their geometric phases and flux densities}

\author{Thomas~Gasenzer%
\thanks{T.Gasenzer@ThPhys.Uni-Heidelberg.DE}
\and
Otto~Nachtmann%
\thanks{O.Nachtmann@ThPhys.Uni-Heidelberg.DE}
\and
Martin-I.~Trappe%
\thanks{M.Trappe@ThPhys.Uni-Heidelberg.DE}
}

\institute{Institut f{\"u}r Theoretische Physik, Universit{\"a}t Heidelberg,\\ 
Philosophenweg 16, 69120 Heidelberg, Germany}

\date{Received: \today}

\abstract{We discuss the geometric phases and flux densities for the metastable states of hydrogen with principal quantum number $n=2$ being subjected to adiabatically varying external electric and magnetic fields. Convenient representations of the flux densities as complex integrals are derived. Both, parity conserving (PC) and parity violating (PV) flux densities and phases are identified. General expressions for the flux densities following from rotational invariance are derived. Specific cases of external fields are discussed. In a pure magnetic field the phases are given by the geometry of the path in magnetic field space. But for electric fields in presence of a constant magnetic field and for electric plus magnetic fields the geometric phases carry information on the atomic parameters, in particular, on the PV atomic interaction. We show that for our metastable states also the decay rates can be influenced by the geometric phases and we give a concrete example for this effect. Finally we emphasise that the general relations derived here for geometric phases and flux densities are also valid for other atomic systems having stable or metastable states, for instance, for He with $n=2$. Thus, a measurement of geometric phases may give important experimental information on the mass matrix and the electric and magnetic dipole matrices for such systems. This could be used as a check of corresponding theoretical calculations of wave functions and matrix elements.\\[2ex]
\hfill
{\small HD--THEP--11--05}
} 
\titlerunning{Metastable states of hydrogen and their geometric phases and flux densities}
\authorrunning{T. Gasenzer \textit{et al.}}
\maketitle

\section{Introduction}\label{s:IntroductionFD}

In this paper we study properties of geometric phases and geometric flux densities for metastable hydrogen atoms in external electric and magnetic fields. Geometric phases in quantum mechanics were introduced in \cite{Ber84} and have been studied extensively since then; see for instance \cite{Bar83,ShWi89,Nak90} and references therein. For a discussion of geometric phases for systems described by a non-hermitian Hamiltonian see \cite{Garrison1988177,Massar1996,KeKoMo2003,Berry2004,Heiss2004,NesCru2008} and references therein. In our group the adiabatic theorem and geometric phases for metastable states were studied in \cite{BeGaNa07_I,BeGaNa07_II}. Both, parity conserving (PC), and parity violating (PV) geometric phases were identified. One aim is to apply the theory developed in this way to the measurement of parity violation in light atoms like hydrogen with the longitudinal spin echo technique; see \cite{ABSE95,BeGaMaNaTr08_I,DeKGaNaTr11}. But, clearly, a measurement of geometric phases is very interesting by itself since these phases represent a deep quantum-mechanical phenomenon. For metastable states these phases are complex and, therefore, geometry also influences the decay rates of these states, as we shall demonstrate explicitly below.

In the present paper we are primarily interested in the structure of PC and PV geometric phases and flux densities, that is, what one can say on general grounds about their dependences on the external electric and magnetic fields. Since a measurement of PV geometric phases is one possibility to study atomic parity violation (APV) we briefly refer to recent work discussing the present status of this field. Standard reviews of APV can be found in \cite{Khrip91,Bou97}. A very recent survey of the past, present, and prospects of APV is given in \cite{Bou2011}. Experimental results for the heavy atoms Cs \cite{Bouchiat1982358,Wood1997,BeWi99}, Bi \cite{Macpherson1991}, Tl \cite{Edwards1995,Vet95}, Pb \cite{Mee93}, and Yb \cite{Budker2009} have been published. See also the review in \cite{PDG10}. A large effort is being undertaken to measure APV in Ra$^+$ \cite{Wansbeek2008,Versolato2011}, and plans for the future FAIR facility at GSI, Darmstadt, include a program of APV studies for highly charged ions \cite{PhysRevA.40.7362,PhysRevA.63.054105,Shabaev10,PhysRevA.81.062503,Surzhykov11}. The situation for APV in the lightest atoms, H and D, is nicely summarised in \cite{DuHo07,DuHo11}. In these latter papers it is also stressed that from the theory point of view H and D are the ideal candidates to study APV.

Our paper is organised as follows. In Section \ref{inexternalfield} we introduce the atomic systems which we want to study. In Section \ref{s:Geometricphasesand} we define the geometric phases and flux densities and derive useful representations for them in terms of complex integrals. Section \ref{s:4} is devoted to a study of the structure of these phases and flux densities following from rotational invariance. In Section \ref{s:5} we discuss specific cases. Section \ref{s:6} presents our conclusions. In Appendix \ref{s:AppendixA} we explain the notations used throughout our work and provide many useful formulae as well as essential numerical quantities. In addition, we give the non-zero parts of the mass matrix for the $n=2$ states of hydrogen. In the Appendices \ref{s:AppendixB}, \ref{s:AppendixC} and \ref{s:AppendixD} we present detailed proofs for the relations derived in Sections \ref{s:Geometricphasesand}, \ref{s:4}, and \ref{s:5}, respectively. In Appendix \ref{s:AppendixE} (online only) we give explicit formulae for various matrices used in our paper. If not stated otherwise we use natural units with $\hbar=c=1$.

\section{Metastable hydrogen states in external fields}\label{inexternalfield}

\begin{figure}[b!]
\centering
\includegraphics[scale=0.35]{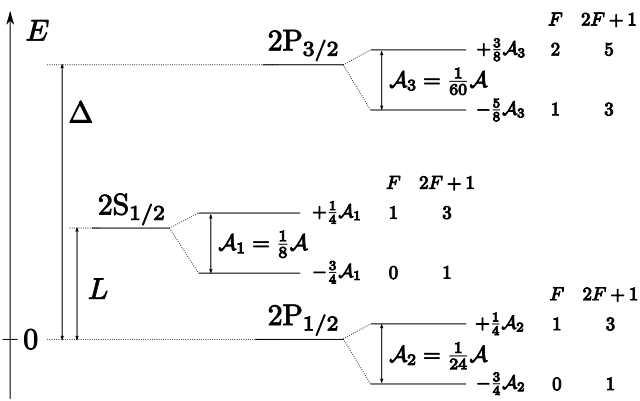}
\caption{Energy levels of the hydrogen states with principal quantum number $n=2$ in vacuum. The numerical values of the fine structure splitting $\Delta$, the Lamb shift $L$ and the ground state hyperfine splitting energy $\mcal A$ are given in Table \ref{t:valuesFD} of Appendix \ref{s:AppendixA}.}
\label{EnergyLevelsH}
\end{figure}

We are interested in the states of hydrogen with principal quantum number $n=2$. Their energy levels in vacuum are shown in Figure \ref{EnergyLevelsH}. The lifetimes $\tau$ of the 2S and 2P states of hydrogen in vacuum are $\tau_S=\Gamma_S^{-1}=0.1216\,$s and $\tau_P=\Gamma_P^{-1}=1.596\times 10^{-9}\,$s, respectively; see \cite{Sap04,LaShSo05}. Here $\Gamma_{S,P}$ are the decay rates. We have 16 states with $n=2$ for which we use a numbering scheme $\alpha=1,\dots,16$ as explained in Appendix \ref{s:AppendixA}, Table \ref{t:state.labelsFD}.

In this paper we shall consider $n=2$ hydrogen atoms at rest subjected to slowly varying electric and magnetic fields. In vacuum the 2S states of hydrogen are metastable and decay by two-photon emission to the ground state. The 2P states decay to the ground state by one-photon emission. Energetically allowed radiative decays from one $n=2$ state to another one are completely negligible. This remains true when we consider the $n=2$ states in an external, slowly varying, electromagnetic field in the adiabatic limit. There, by definition, the variation of the external fields has to be slow enough such that no transitions between the $n=2$ levels are induced. In this situation we can apply the standard Wigner-Weisskopf method \cite{WeWi30,WeWi30b}. The derivation of this method and its limitations are discussed in many textbooks and articles, see for instance \cite{MaWo95,BaRa97,Knight197299,Wang1974323,BeNa83,Nac90,BoBrNa95}. A derivation starting from quantum field theory can be found in \cite{PhysRev.121.350,RG1963239}. Let us note that for more complex situations than discussed in the present paper, for instance if radiative transitions are induced between the $n=2$ states by an external field, we would have to use other methods, master equations, the optical Bloch equation etc.; see \cite{BaRa97}.

Thus, for the situations we are considering the basic theoretical tool is the effective Schr\"odinger equation describing the evolution of the undecayed states with state vector $\rket{t}$ at time $t$, in the Wigner-Weisskopf approximation,
\begin{align}\label{2.1}
\I\frac{\partial}{\partial t}\rket{t}=\umat{M}(\vmc E(t),\vmc B(t))\rket{t}\ .
\end{align}
Here
\begin{align}\label{2.2}
\umat{M}(\vmc E(t),\vmc B(t))=\umat{\tilde M}_0-\uvec D\cdot\vmc E(t)-\uvec\mu\cdot\vmc B(t)
\end{align}
with $\umat{\tilde M}_0$ the mass matrix for the $n=2$ states in vacuum, $\uvec D$ and $\uvec\mu$ the electric and magnetic dipole operators, respectively, in the $n=2$ subspace, and $\vmc E$ and $\vmc B$ the electric and magnetic fields. We are interested in parity conserving (PC) and parity violating (PV) geometric phases. The mass matrix $\umat{\tilde M}_0$ will therefore be split into the PC part $\umat{M}_0$ and the PV part $\delta\,\umat{M}_{\PV}$,
\begin{align}\label{2.3}
\umat{\tilde M}_0=\umat{M}_0+\delta\,\umat{M}_{\PV}\ .
\end{align}
In the standard model of particle physics (SM) $\delta\,\umat{M}_{\PV}$ is determined by Z-boson exchange between the electrons of the hull and the quarks in the nucleus. Here, as in \cite{BeGaNa07_II}, we split off a (very small) numerical factor $\delta$ from the PV part of the mass matrix characterising the intrinsic strength of the PV terms. In Appendix \ref{s:AppendixA} we give explicitly $\delta$. The matrices $\umat{\tilde M}_0$, $\uvec D$ and $\uvec\mu$ are discussed further in Appendix \ref{s:AppendixA}, and their explicit forms used for numerical purposes are given in Appendix \ref{s:AppendixE}; see Tables \ref{t:H2.M0}, \ref{t:H2.D}, and \ref{t:H2.Mu}.

In the presence of electric fields the metastable 2S states will get a 2P admixture making them decay faster, see Figure 1 of \cite{BeGaNa07_II}. We are interested in the situation where the lifetime of the metastable states is still at least a factor of 5 larger than that of the other states. As shown in (27) of \cite{BeGaNa07_II} this limits us to electric fields
\begin{align}\label{2.5}
|\vmc E|\lesssim 250\,\mbox{V/cm}\ .
\end{align}

In \cite{BeGaNa07_I,BeGaNa07_II} the adiabatic theorem and geometric phases for metastable states were studied and in the present work we shall apply and extend the results obtained there. The mass matrix $\umat{M}$ in (\ref{2.1}) depends on the slowly varying parameters $\vmc E$ and $\vmc B$. Thus, we have a six-dimensional parameter space. Geometric phases are connected with the trajectories followed by the field strengths as function of time in this space.

In the following we shall, for general discussions, denote $\vmc E$ and $\vmc B$ collectively as parameters $K$,
\begin{align}\label{2.7}
\left(\begin{array}{c}
\vmc E\\
\vmc B
\end{array}\right)
=
\left(\begin{array}{c}
K_1\\
K_2\\
K_3\\
K_4\\
K_5\\
K_6
\end{array}\right)
\equiv K\ .
\end{align}
Indices $i,j\in\{1,2,3\}$ will be normal space indices, for instance $\mathcal E_i$, $\mathcal B_i$ etc., or refer to any three particular components of $K$. Indices $a$, $b$ shall refer to the components $K_a$, $a\in\{1,\dots,6\}$. The mass matrix (\ref{2.2}) shall be considered as function of the six parameters $K=K_a$
\begin{align}\label{2.8}
\umat{M}(\vmc E,\vmc B)\equiv\umat{M}(K)\ .
\end{align}
We shall assume that we work in a region of parameter space ($K$ space) where $\umat{M}(K)$ can be diagonalised. There are then 16 linearly independent right and left eigenvectors of $\umat{M}(K)$,
\begin{align}
\umat{M}(K)\rket{\alpha,K}&=E_\alpha(K)\rket{\alpha,K}\ ,\nn\\
\lrbra{\alpha,K}\umat{M}(K)&=\lrbra{\alpha,K}E_\alpha(K)\ ,\nn\\
(\alpha&=1,\dots,16)\ .\label{2.9}
\end{align}
These eigenvectors satisfy
\begin{align}\label{2.10}
\lrbracket{\alpha,K}{\beta,K}=\delta_{\alpha\beta}\ .
\end{align}
As normalisation condition we choose
\begin{align}\label{2.11}
&\rbracket{\alpha,K}{\alpha,K}=1\nn \\
&\mbox{(no summation over $\alpha$)}\ .
\end{align}
The complex energies are
\begin{align}\label{2.12}
E_\alpha(K)=E_{\alpha R}(K)-\frac{\I}{2}\Gamma_\alpha(K)
\end{align}
with $E_{\alpha R}$ the real part of the energy and $\Gamma_\alpha(K)=$\linebreak[4] $\big(\tau_\alpha(K)\big)^{-1}$ the decay rate, that is, the inverse lifetime of the state $\rket{\alpha,K}$. In the following we shall suppose
\begin{align}\label{2.13}
|E_\alpha(K)-E_\beta(K)|\ge c>0
\end{align}
for all $\alpha\not=\beta$ where $c$ is a constant. Exceptions where (\ref{2.13}) is not required to hold will be clearly indicated.

Below we shall make extensive use of the quasi projectors defined as
\begin{align}\label{2.14}
\mathbbm P_\alpha(K)=\rket{\alpha,K}\lrbra{\alpha,K}\ .
\end{align}
These satisfy
\begin{align}\label{2.15}
\mathbbm P_\alpha(K)\mathbbm P_\beta(K)&=\left\{
\begin{array}{cl}
\mathbbm P_\alpha(K) & \mbox{for } \alpha=\beta\ ,\\
0 & \mbox{for } \alpha\not=\beta\ ,
\end{array}\right.\\
\sum_\alpha\mathbbm P_\alpha(K)&=\mathbbm 1\ ,\label{2.16}
\end{align}
but, in general, the $\mathbbm P_\alpha(K)$ are non-hermitian matrices. Furthermore, we shall need the resolvent $\big(\zeta-\umat M(K)\big)^{-1}$ where $\zeta$ is arbitrary complex. With the help of the quasi projectors we get
\begin{align}\label{2.17}
\big(\zeta-\umat M(K)\big)^{-n}&=\sum_\alpha\big(\zeta-E_\alpha(K)\big)^{-n}\mathbbm P_\alpha(K)\ ,\nn\\
(n&=0,1,2,\dots)\ .
\end{align}

Now we come back to the effective Schr\"odinger equation (\ref{2.1}) which reads, replacing $(\vmc E,\vmc B)$ by $K$,
\begin{align}\label{2.18}
\I\frac{\partial}{\partial t}\rket{t}=\umat{M}(K(t))\rket{t}\ .
\end{align}
The state vector is expanded as
\begin{align}\label{2.19}
\rket{t}=\sum_{\alpha=1}^{16}\psi_\alpha(t)\rket{\alpha,K(t)}\ .
\end{align}
We always suppose slow enough variation of the parameter vector $K(t)$. As shown in Section 3 of \cite{BeGaNa07_II} we get then the solution of (\ref{2.18}) for the metastable states as follows.

We consider an initial metastable state at time $t=0$
\begin{align}\label{2.20}
\rket{t=0}=\sum_{\alpha\in I}\psi_\alpha(0)\rket{\alpha,K(0)}
\end{align}
where $\alpha$ only runs over the index set of the metastable states,
\begin{align}\label{2.21}
I=\{9,10,11,12\}\ ,
\end{align}
in our numbering scheme. Then we have, for $t\geq 0$,
\begin{align}\label{2.22}
\rket{t}=\sum_{\alpha\in I}\psi_\alpha(t)\rket{\alpha,K(t)}
\end{align}
where 
\begin{align}\label{2.23}
\psi_\alpha(t)&=\exp\big[-\I\varphi_\alpha(t)+\I\gamma_\alpha(t)\big]\psi_\alpha(0)\ ,\\
\label{2.24}
\varphi_\alpha(t)&=\int_0^t\mathrm d t'\,E_\alpha(K(t'))\ ,\\
\label{2.25}
\gamma_\alpha(t)&=\int_0^t\mathrm d t' \lrbra{\alpha,K(t')}\I\frac{\partial}{\partial t'}\rket{\alpha,K(t')}\ .
\end{align}
The quantities $\varphi_\alpha(t)$ and $\gamma_\alpha(t)$ are the familiar dynamic and geometric phases, respectively. For metastable states both will in general have real and imaginary parts.

Here and in the following the labels $\alpha=9,10,11$, and $12$ correspond to the states $\rket{\alpha,K}\equiv\rket{2\hat S_{1/2},F,F_3,\vmc E,\vmc B}$. These originate from the states $\rket{2 S_{1/2},F,F_3}$ with $(F,F_3)=(1,1)$, $(1,0)$, $(1,-1)$, and $(0,0)$, respectively, through the mixing with the 2P states according to the $\PV$, the $\vmc E$, and $\vmc B$ terms in the mass matrix (\ref{2.2}). This numbering has to be carefully defined, see Appendix \ref{s:AppendixA}, since we have to follow the states in their adiabatic motion along trajectories in parameter space. As explained in Appendix \ref{s:AppendixA} $(F,F_3)$ are then only labels of the states, no longer the total angular momentum quantum numbers. Thus, in order to avoid confusion, we shall stick to the labels $\alpha$ for our states in the following.

Below we shall study in detail the geometric phases for metastable states for the case that $K(t)$ makes a closed loop in parameter space. 

\section{Geometric phases and flux densities}\label{s:Geometricphasesand}

In this section we shall discuss general relations and properties for geometric phases and the corresponding flux densities defined below. These relations hold for any system with time evolution described by an effective Schr\"odinger equation
\begin{align}\label{3.1}
\I\frac{\partial}{\partial t}\rket{t}=\umat{M}(K(t))\rket{t}\ ,
\end{align}
with $N\times N$ matrices $\umat{M}(K)$, and having metastable states. The parameter vector $K$ can have any number of components and the dependence of $\umat M$ on $K$ need not be linear as for $\umat M(\vmc E,\vmc B)=\umat M(K)$ in Section \ref{inexternalfield}.

We consider now the system over a time interval\\ $\mbox{$0\leq t\leq T$}$ where the parameter vector $K(t)$ runs over a closed curve $\mathcal C$
\begin{align}\label{3.1a}
\mathcal C\,:\,t\rightarrow K(t),\; t\in [0,T],\; K(T)=K(0)\ .
\end{align}
The geometric phases (\ref{2.25}) acquired by the metastable states are then 
\begin{align}\label{3.2}
\gamma_\alpha(\mathcal C)\equiv\gamma_\alpha(T)&=\int_0^T\mathrm d t' \lrbra{\alpha,K(t')}\I\frac{\partial}{\partial t'}\rket{\alpha,K(t')}\nn\\
&=\int_{\mathcal C}\lrbra{\alpha,K}\I\,\mathrm d\rket{\alpha,K}\ ,
\end{align}
$\alpha\in I$, where $I$ is the index set of the metastable states. For our concrete hydrogen case $I$ is given in (\ref{2.21}). Here and in the following we use the exterior derivative calculus; see for instance \cite{Fla63}. Let $\mathcal F$ be a surface with boundary $\mathcal C$,
\begin{align}\label{3.3}
\partial\mathcal F=\mathcal C\ ,
\end{align}
and suppose that $\umat M(K)$ can be diagonalised  for all $K\in\mathcal F$, and (\ref{2.13}) holds for the eigenvalues. We get then 
\begin{align}\label{3.4}
\gamma_\alpha(\mathcal C)&=\I\int_{\partial\mathcal F}\lrbra{\alpha,K}\mathrm d\rket{\alpha,K}\nn\\
&=\I\int_{\mathcal F}\mathrm d\lrbra{\alpha,K}\mathrm d\rket{\alpha,K}\nn\\
&=\int_{\mathcal F}Y_{\alpha,ab}(K)\,\mathrm d K_a\wedge\mathrm d K_b\ .
\end{align}
Here we define the geometric flux densities $Y_{\alpha,ab}(K)$, the analogues for the metastable states of the quantities $\vec V$ of \cite{Ber84}, by
\begin{align}\label{3.5}
Y_{\alpha,ab}\,\mathrm d K_a\wedge\mathrm d K_b&=\I\,\mathrm d\lrbra{\alpha,K}\mathrm d\rket{\alpha,K}\ ,\nn\\
Y_{\alpha,ab}(K)+Y_{\alpha,ba}(K)&=0\ .
\end{align}
From (\ref{3.5}) we get easily
\begin{align}\label{3.6}
Y_{\alpha,ab}(K)\,\mathrm d K_a\wedge\mathrm d K_b&=+\I\big(\mathrm d\lrbra{\alpha,K}\big)\wedge\big(\mathrm d\rket{\alpha,K}\big)\nn\\
&\hspace{-2em}=-\I\sum_{\beta\not=\alpha}\lrbra{\alpha,K}\mathrm d\rket{\beta,K}\wedge\lrbra{\beta,K}\mathrm d\rket{\alpha,K}\ .
\end{align}
Here we use
\begin{align}\label{3.7}
\big(\mathrm d\lrbra{\alpha,K}\big)\rket{\beta,K}+\lrbra{\alpha,K}\mathrm d\rket{\beta,K}=0
\end{align}
which follows from (\ref{2.10}). Note that in (\ref{3.6}) $\alpha$ is the index of a metastable state, $\alpha\in I$, but in the sum over $\beta$ {\em all} states with $\beta\not=\alpha$ have to be included.

As a further relation following directly from (\ref{3.5}) we get the generalised divergence condition
\begin{align}\label{3.8}
\mathrm d\big(Y_{\alpha,ab}(K)\,\mathrm d K_a\wedge\mathrm d K_b\big)=\I\,\mathrm{dd}\lrbra{\alpha,K}\mathrm d\rket{\alpha,K}=0
\end{align}
which implies
\begin{align}\label{3.9}
\frac{\partial}{\partial K_a}Y_{\alpha,bc}(K)+\frac{\partial}{\partial K_b}Y_{\alpha,ca}(K)+\frac{\partial}{\partial K_c}Y_{\alpha,ab}(K)=0\ .
\end{align}
Let us now derive further representations and properties for the geometric flux densities. From (\ref{2.9}) and (\ref{2.10}) we get for $\beta\not=\gamma$
\begin{align}\label{3.10}
\lrbra{\beta,K}\umat M(K)\rket{\gamma,K}=0\ .
\end{align}
Taking the exterior derivative in (\ref{3.10}) gives
\begin{align}\label{3.11}
&[E_\beta(K)-E_\gamma(K)]\lrbra{\beta,K}\mathrm d\rket{\gamma,K}\nn\\
&\qquad\qquad+\lrbra{\beta,K}\big(\mathrm d\umat M(K)\big)\rket{\gamma,K}=0\ .
\end{align}
Since we suppose (\ref{2.13}) to hold for all $K\in\mathcal F$ we get for $\beta\not=\gamma$
\begin{align}\label{3.12}
\lrbra{\beta,K}\mathrm d\rket{\gamma,K}=-\frac{\lrbra{\beta,K}\big(\mathrm d\umat M(K)\big)\rket{\gamma,K}}{E_\beta(K)-E_\gamma(K)}\ .
\end{align}
Inserting this in (\ref{3.6}) gives
\begin{align}\label{3.13}
Y_{\alpha,ab}(K)&=\frac{\I}{2}\sum_{\beta\not=\alpha}[E_\alpha(K)-E_\beta(K)]^{-2}\nn\\
&\times\lrbra{\alpha,K}\frac{\partial\umat M(K)}{\partial K_a}\rket{\beta,K}\lrbra{\beta,K}\frac{\partial\umat M(K)}{\partial K_b}\rket{\alpha,K}\nn\\
&-(a\leftrightarrow b)\nn\\
&=\frac{\I}{2}\sum_{\beta\not=\alpha}[E_\alpha(K)-E_\beta(K)]^{-2}\nn\\
&\times\mathrm{Tr}\Big[\mathbbm P_\alpha(K)\frac{\partial\umat M(K)}{\partial K_a}\mathbbm P_\beta(K)\frac{\partial\umat M(K)}{\partial K_b}\nn\\
&\hspace{12em}-(a\leftrightarrow b)\Big]
\end{align}
where we use the quasi projectors (\ref{2.14}).

\begin{figure}[htb]
\centering
\includegraphics[scale=0.5]{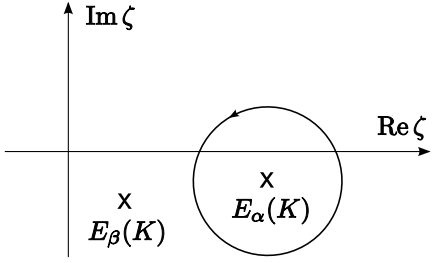}
\caption{The complex $\zeta$ plane with the (schematic) location of the energy eigenvalues $E_\alpha(K)$ and $E_\beta(K)$ with $\beta\not=\alpha$. The curve $S_\alpha$ encircles only $E_\alpha(K)$.}
\label{ComplexZetaPlane}
\end{figure}

We shall now derive an integral representation for\linebreak[4] $Y_{\alpha,ab}(K)$. Consider the complex $\zeta$ plane, see Figure \ref{ComplexZetaPlane}, where we mark schematically the position of the energy eigenvalues $E_\alpha(K)$ and $E_\beta(K)$, $\beta\not=\alpha$. Since we suppose (\ref{2.13}) to hold we can choose a closed curve $S_\alpha$ which encircles only $E_\alpha(K)$ but where all $E_\beta(K)$ with $\beta\not=\alpha$ are outside. The geometric flux densities (\ref{3.5}), (\ref{3.13}) are then given as a complex integral 
\begin{align}\label{3.14}
Y_{\alpha,ab}(K)=\frac{\I}{2}\frac{1}{2\pi\I}\oint_{S_\alpha}\mathrm d\zeta\,\mathrm{Tr}\Big[\frac{1}{(\zeta-\umat M(K))}\frac{\partial\umat M(K)}{\partial K_a}\nn\\
\hfill\times\frac{1}{(\zeta-\umat M(K))^2}\frac{\partial\umat M(K)}{\partial K_b}\Big]\ .
\end{align}
The proof of (\ref{3.14}) is given in Appendix \ref{s:AppendixB}. From (\ref{3.14}) we get convenient relations for the derivatives of $Y_{\alpha,ab}(K)$, see Appendix \ref{s:AppendixB},
\begin{align}\label{3.15}
\frac{\partial}{\partial K_a}&Y_{\alpha,bc}(K)=\frac{\I}{2}\frac{1}{2\pi\I}\oint_{S_\alpha}\mathrm d\zeta\Big\{\nn\\
&\mathrm{Tr}\Big[\frac{1}{(\zeta-\umat M(K))}\frac{\partial^2\umat M(K)}{\partial K_a\partial K_b}\frac{1}{(\zeta-\umat M(K))^2}\frac{\partial\umat M(K)}{\partial K_c}\Big]\nn\\
+&\mathrm{Tr}\Big[\frac{1}{(\zeta-\umat M(K))}\frac{\partial\umat M(K)}{\partial K_a}\frac{1}{(\zeta-\umat M(K))}\frac{\partial\umat M(K)}{\partial K_b}\nn\\
&\qquad\qquad\times\frac{1}{(\zeta-\umat M(K))^2}\frac{\partial\umat M(K)}{\partial K_c}\Big]\Big\}-(b\leftrightarrow c)\ ,
\end{align}
which can also be written as
\begin{align}\label{3.16}
&\frac{\partial}{\partial K_a}Y_{\alpha,bc}(K)=\frac{\I}{2}\Big\{\sum_{\beta\not=\alpha}\big[E_\alpha(K)-E_\beta(K)\big]^{-2}\nn\\
&\;\times\mathrm{Tr}\Big[\mathbbm P_\alpha(K)\frac{\partial^2\umat M(K)}{\partial K_a\partial K_b}\mathbbm P_\beta(K)\frac{\partial\umat M(K)}{\partial K_c}\nn\\
&\;\;-\mathbbm P_\alpha(K)\frac{\partial\umat M(K)}{\partial K_c}\mathbbm P_\beta(K)\frac{\partial^2\umat M(K)}{\partial K_a\partial K_b}
\Big]\nn\\
&+\sum_{\beta\not=\alpha}\big[E_\alpha(K)-E_\beta(K)\big]^{-3}\nn\\
&\;\times\mathrm{Tr}\Big[\hspace{-0.2em}-2\mathbbm P_\alpha(K)\frac{\partial\umat M(K)}{\partial K_a}\mathbbm P_\alpha(K)\frac{\partial\umat M(K)}{\partial K_b}\mathbbm P_\beta(K)\frac{\partial\umat M(K)}{\partial K_c}\nn\\
&\;\;+\mathbbm P_\alpha(K)\frac{\partial\umat M(K)}{\partial K_a}\mathbbm P_\beta(K)\frac{\partial\umat M(K)}{\partial K_b}\mathbbm P_\alpha(K)\frac{\partial\umat M(K)}{\partial K_c}\nn\\
&\;\;+\mathbbm P_\beta(K)\frac{\partial\umat M(K)}{\partial K_a}\mathbbm P_\alpha(K)\frac{\partial\umat M(K)}{\partial K_b}\mathbbm P_\alpha(K)\frac{\partial\umat M(K)}{\partial K_c}\Big]\nn\\
&+\sum_{\beta,\gamma\not=\alpha}\big[E_\alpha(K)-E_\beta(K)\big]^{-1}\big[E_\alpha(K)-E_\gamma(K)\big]^{-2}\nn\\
&\;\times\mathrm{Tr}\Big[\mathbbm P_\alpha(K)\frac{\partial\umat M(K)}{\partial K_a}\mathbbm P_\beta(K)\frac{\partial\umat M(K)}{\partial K_b}\mathbbm P_\gamma(K)\frac{\partial\umat M(K)}{\partial K_c}\nn\\
&\;\;+\mathbbm P_\beta(K)\frac{\partial\umat M(K)}{\partial K_a}\mathbbm P_\alpha(K)\frac{\partial\umat M(K)}{\partial K_b}\mathbbm P_\gamma(K)\frac{\partial\umat M(K)}{\partial K_c}\nn\\
&\;\;-\mathbbm P_\gamma(K)\frac{\partial\umat M(K)}{\partial K_a}\mathbbm P_\beta(K)\frac{\partial\umat M(K)}{\partial K_b}\mathbbm P_\alpha(K)\frac{\partial\umat M(K)}{\partial K_c}\nn\\
&\;\;-\mathbbm P_\beta(K)\frac{\partial\umat M(K)}{\partial K_a}\mathbbm P_\gamma(K)\frac{\partial\umat M(K)}{\partial K_b}\mathbbm P_\alpha(K)\frac{\partial\umat M(K)}{\partial K_c}\Big]\Big\}\nn\\
&-(b\leftrightarrow c)\ ,
\end{align}
see Appendix \ref{s:AppendixB}. From both, (\ref{3.15}) and (\ref{3.16}), we can easily check the divergence condition (\ref{3.9}).

In the following sections we shall use (\ref{3.13}) and (\ref{3.16}) to calculate numerically the geometric flux densities and their derivatives for metastable H atoms. We will be especially interested in the flux densities in three-dimensional subspaces of $K$ space. We will, for instance, consider the cases where the electric field $\vmc E$ is kept constant and only a magnetic field $\vmc B$ varies or vice versa. The geometric flux densities (\ref{3.5}), (\ref{3.13}) are then equivalent to three-di\-men\-sional complex vector fields. Indeed, let us consider the case that only three components of $K$, $K_{a_1}$, $K_{a_2}$ and $K_{a_3}$, are varied. The vectors 
\begin{align}\label{3.17}
\vec L\equiv\left(
\begin{array}{c}
L_{1}\\
L_{2}\\
L_{3}
\end{array}\right)
=\left(
\begin{array}{ccc}
\kappa_1^{-1}&0&0\\
0&\kappa_2^{-1}&0\\
0&0&\kappa_3^{-1}\\
\end{array}\right)\left(
\begin{array}{c}
K_{a_1}\\
K_{a_2}\\
K_{a_3}
\end{array}\right)
\end{align}
span the effective parameter space which is now three dimensional. In (\ref{3.17}) we multiply the $K_{a_i}$ with constants $1/\kappa_i$ which, in the following, will be chosen conveniently. We shall, for instance, always choose the $\kappa_i$ such that the $L_i$ have the same dimension for $i=1,2,3$. We define the geometric flux-density vectors in $\vec L$ space as 
\begin{align}\label{3.18}
J_{\alpha,i}^{(\vec L)}(\vec L)&=\sum_{j,k}\epsilon_{ijk}\,Y_{\alpha,a_ja_k}(K(\vec L))\,\kappa_j\kappa_k\ ,\nn\\
\vec J_\alpha^{(\vec L)}(\vec L)&=\left(
\begin{array}{c}
J_{\alpha,1}^{(\vec L)}(\vec L)\\[0.3em]
J_{\alpha,2}^{(\vec L)}(\vec L)\\[0.3em]
J_{\alpha,3}^{(\vec L)}(\vec L)
\end{array}\right)
\end{align}
where $i,j,k\in\{1,2,3\}$. The curve $\mathcal C$ (\ref{3.1a}) and the surface $\mathcal F$ (\ref{3.3}) live now in $\vec L$ space. With the ordinary surface element in $\vec L$ space
\begin{align}\label{3.19}
\mathrm df_i^{\vec L}=\frac12\epsilon_{ijk}\,\mathrm dL_j\wedge\mathrm dL_k
\end{align}
we get for the geometric phase (\ref{3.4})
\begin{align}\label{3.20}
\gamma_\alpha(\mathcal C)=\int_{\mathcal F}\vec J_\alpha^{(\vec L)}(\vec L)\,\mathrm d\vec f^{\vec L}\ .
\end{align}
From (\ref{3.9}) we find that
\begin{align}\label{3.21}
\mathrm{div}\vec J_\alpha^{(\vec L)}(\vec L)=0
\end{align}
wherever (\ref{2.13}) holds. That is, the vector fields $\vec J_\alpha^{(\vec L)}$ can have sources or sinks only at the points where the complex eigenvalues (\ref{2.12}) of $\umat M(K)$ become degenerate. More precisely, we see from (\ref{3.13}) that $\vec J_\alpha^{(\vec L)}$ can have such singularities only where $E_\alpha(K(\vec L))$ becomes degenerate with another eigenvalue $E_\beta(K(\vec L))$ $(\beta\not=\alpha)$. This is, of course, well known \cite{Ber84}. From (\ref{3.12}) we can also calculate the curl of $\vec J_\alpha^{(\vec L)}$:
\begin{align}\label{3.22}
\big(\mathrm{rot}\,\vec J_\alpha^{(\vec L)}(\vec L)\big)_i&=\epsilon_{ijk}\frac{\partial}{\partial L_j}J_{\alpha,k}^{(\vec L)}(\vec L)\nn\\
&=2\frac{\partial}{\partial K_{a_j}}Y_{\alpha,a_ia_j}(K(\vec L))\,\kappa_i\kappa_j^2\ .
\end{align}
Knowledge of both, $\mathrm{div}\vec J_\alpha^{(\vec L)}$ and $\mathrm{rot}\,\vec J_\alpha^{(\vec L)}$, will allow us an easy understanding of the behaviour of the geometric flux density vectors for concrete cases in Section \ref{s:5} below.

\section{Structure of phases and flux densities from rotational invariance}\label{s:4}

In this section we shall discuss what we can learn from rotational invariance about the geometric phases and flux densities for the metastable hydrogen states.

\subsection{Proper rotations}\label{ss:4.1}
We consider the mass matrix $\umat{M}(\vmc E(t),\vmc B(t))$ of (\ref{2.2}). Let $\mathrm R\in\mathrm{SO}(3)$ be a proper rotation
\begin{align}\label{4.0}
&\mathrm R\,:\,x_i\rightarrow R_{ij}\,x_j\ ,\nn\\
&\qquad R=(R_{ij})\, ,\quad \mathrm{det}\,R=1\ .
\end{align}
We denote by $\underline{\mathcal R}$ its representation in the $n=2$ subspace of the hydrogen atom. We have then
\begin{align}\label{4.1}
\underline{\mathcal R}\,\umat{M}(\vmc E,\vmc B)\,\underline{\mathcal R}^{-1}=\umat{M}( R\vmc E, R\vmc B)\ .
\end{align}
This shows that $\umat{M}(\vmc E,\vmc B)$ and $\umat{M}( R\vmc E, R\vmc B)$ have the same set of eigenvalues. Since we have assumed non-degeneracy of the eigenvalues of $\umat{M}(\vmc E,\vmc B)$, see (\ref{2.13}), the same holds for $\umat{M}( R\vmc E, R\vmc B)$. Moreover, $\mathrm{SO}(3)$ is a continuous and connected group, therefore the numbering of the eigenvalues, as explained in Appendix \ref{s:AppendixA}, cannot change with $ R$. Thus, we get
\begin{align}\label{4.2}
E_\alpha( R\vmc E, R\vmc B)=E_\alpha(\vmc E,\vmc B)\ .
\end{align}
For the resolvent, cf. (\ref{2.17}) with $n=1$, we find
\begin{align}\label{4.3}
&\underline{\mathcal R}\big(\zeta-\umat M(\vmc E,\vmc B)\big)^{-1}\underline{\mathcal R}^{-1}=\big(\zeta-\umat M( R\vmc E, R\vmc B)\big)^{-1}\, ,\\[0.5em]
&\sum_\alpha\big(\zeta-E_\alpha(\vmc E,\vmc B)\big)^{-1}\,\underline{\mathcal R}\,\mathbbm P_\alpha(\vmc E,\vmc B)\,\underline{\mathcal R}^{-1}\nn\\[-0.5em]
&\hspace{2em}=\sum_\alpha\big(\zeta-E_\alpha( R\vmc E, R\vmc B)\big)^{-1}\,\mathbbm P_\alpha( R\vmc E, R\vmc B)\ .\label{4.4}
\end{align}
With (\ref{4.2}) we get from (\ref{4.4})
\begin{align}\label{4.5}
\underline{\mathcal R}\mathbbm P_\alpha(\vmc E,\vmc B)\underline{\mathcal R}^{-1}=\mathbbm P_\alpha( R\vmc E, R\vmc B)\ .
\end{align}
Note that for the states themselves we can only conclude that $\rket{\alpha, R\vmc E, R\vmc B}$ and $\underline{\mathcal R}\rket{\alpha,\vmc E,\vmc B}$ must be equal up to a phase factor.

With the identification of $(\vmc E,\vmc B)$ and $K$ of (\ref{2.7}) we can now decompose the $6\times 6$ flux density matrices $Y_{\alpha,ab}$ (\ref{3.5}) into $3\times 3$ submatrices corresponding to the $\vmc E$ and $\vmc B$ and mixed $\vmc E,\vmc B$ differential forms (see Appendix C of \cite{BeGaNa07_II}). We have with $\alpha\in I$, the index of a metastable state,
\begin{align}\label{4.6}
Y_{\alpha,ab}(K)\,\mathrm d K_a\wedge\mathrm d K_b&=\mathcal I_{\alpha,jk}^{(\vmc E)}(\vmc E,\vmc B)\,\mathrm d \mathcal E_j\wedge\mathrm d \mathcal E_k\nn\\
&\;+\mathcal I_{\alpha,jk}^{(\vmc B)}(\vmc E,\vmc B)\,\mathrm d \mathcal B_j\wedge\mathrm d \mathcal B_k\nn\\
&\;+\mathcal I_{\alpha,jk}^{(\vmc E,\vmc B)}(\vmc E,\vmc B)\,\mathrm d \mathcal E_j\wedge\mathrm d \mathcal B_k\ ,
\end{align}
where
\begin{align}\label{4.7}
\mathcal I_{\alpha,jk}^{(\vmc E)}(\vmc E,\vmc B)+\mathcal I_{\alpha,kj}^{(\vmc E)}(\vmc E,\vmc B)&=0\ , \nn\\
\mathcal I_{\alpha,jk}^{(\vmc B)}(\vmc E,\vmc B)+\mathcal I_{\alpha,kj}^{(\vmc B)}(\vmc E,\vmc B)&=0\ .
\end{align}
From (\ref{2.2}), (\ref{3.5}), (\ref{3.13}), (\ref{3.14}), (\ref{4.6}) and (\ref{4.7}) we obtain
\begin{align}
\mathcal I_{\alpha}^{(\vmc E)}(\vmc E,\vmc B)&=\big(\mathcal I_{\alpha,jk}^{(\vmc E)}(\vmc E,\vmc B)\big)\ ,\nn\\
\mathcal I_{\alpha,jk}^{(\vmc E)}(\vmc E,\vmc B)&=\frac{\I}{2}\sum_{\beta\not=\alpha}\big[E_\alpha(\vmc E,\vmc B)-E_\beta(\vmc E,\vmc B)\big]^{-2}\nn\\
&\;\times \mathrm{Tr}\big[\mathbbm P_\alpha(\vmc E,\vmc B)\underline D_j\mathbbm P_\beta(\vmc E,\vmc B)\underline D_k-(j\leftrightarrow k)\big]\nn\\
&=\frac{\I}{2}\frac{1}{2\pi\I}\oint_{S_\alpha}\mathrm d\zeta\,\mathrm{Tr}\Bigg[\frac{1}{\zeta-\umat{M}(\vmc E,\vmc B)}\nn\\
&\hspace{5em}\times\underline D_j\frac{1}{(\zeta-\umat{M}(\vmc E,\vmc B))^2}\underline D_k\Bigg]\ ,\label{4.8}\\
\mathcal I_{\alpha}^{(\vmc B)}(\vmc E,\vmc B)&=\big(\mathcal I_{\alpha,jk}^{(\vmc B)}(\vmc E,\vmc B)\big)\ ,\nn\\
\mathcal I_{\alpha,jk}^{(\vmc B)}(\vmc E,\vmc B)&=\frac{\I}{2}\sum_{\beta\not=\alpha}\big[E_\alpha(\vmc E,\vmc B)-E_\beta(\vmc E,\vmc B)\big]^{-2}\nn\\
&\;\times \mathrm{Tr}\big[\mathbbm P_\alpha(\vmc E,\vmc B)\underline\mu_j\mathbbm P_\beta(\vmc E,\vmc B)\underline\mu_k-(j\leftrightarrow k)\big]\nn\\
&=\frac{\I}{2}\frac{1}{2\pi\I}\oint_{S_\alpha}\mathrm d\zeta\,\mathrm{Tr}\Bigg[\frac{1}{\zeta-\umat{M}(\vmc E,\vmc B)}\nn\\
&\hspace{5em}\times\underline\mu_j\frac{1}{(\zeta-\umat{M}(\vmc E,\vmc B))^2}\underline\mu_k\Bigg]\ ,\label{4.9}\\
\mathcal I_{\alpha}^{(\vmc E,\vmc B)}(\vmc E,\vmc B)&=\big(\mathcal I_{\alpha,jk}^{(\vmc E,\vmc B)}(\vmc E,\vmc B)\big)\ ,\nn\\
\mathcal I_{\alpha,jk}^{(\vmc E,\vmc B)}(\vmc E,\vmc B)&=\I\sum_{\beta\not=\alpha}\big[E_\alpha(\vmc E,\vmc B)-E_\beta(\vmc E,\vmc B)\big]^{-2}\nn\\
&\;\times \mathrm{Tr}\big[\mathbbm P_\alpha(\vmc E,\vmc B)\underline D_j\mathbbm P_\beta(\vmc E,\vmc B)\underline\mu_k\nn\\
&\qquad\quad-\mathbbm P_\alpha(\vmc E,\vmc B)\underline\mu_k\mathbbm P_\beta(\vmc E,\vmc B)\underline D_j\big]\nn\\
&=\I\frac{1}{2\pi\I}\oint_{S_\alpha}\mathrm d\zeta\,\mathrm{Tr}\Bigg[\frac{1}{\zeta-\umat{M}(\vmc E,\vmc B)}\nn\\
&\hspace{5em}\times\underline D_j\frac{1}{(\zeta-\umat{M}(\vmc E,\vmc B))^2}\underline\mu_k\Bigg]\ .\label{4.10}
\end{align}
As explained in general in (\ref{3.17}) ff. we introduce the geometric flux-density vectors $\vec J_{\alpha}^{(\vmc E)}(\vmc E,\vmc B)$ and $\vec J_{\alpha}^{(\vmc B)}(\vmc E,\vmc B)$ with components
\begin{align}\label{4.10b}
J_{\alpha,i}^{(\vmc E)}(\vmc E,\vmc B)=\epsilon_{ijk}\,\mathcal I_{\alpha,jk}^{(\vmc E)}(\vmc E,\vmc B)\ ,\nn\\
J_{\alpha,i}^{(\vmc B)}(\vmc E,\vmc B)=\epsilon_{ijk}\,\mathcal I_{\alpha,jk}^{(\vmc B)}(\vmc E,\vmc B)\ ;
\end{align}
see also Appendix C of \cite{BeGaNa07_II}.

From (\ref{4.6}) to (\ref{4.10}) we get the decomposition of the $6\times 6$ matrix $\big(Y_{\alpha,ab}\big)$ in terms of $3\times 3$ submatrices
\begin{align}\label{4.11}
\big(Y_{\alpha,ab}\big)=\left(
\begin{tabular}{c|c}
\parbox{2.3em}{\vskip5pt$\mathcal I_{\alpha}^{(\vmc E)}$\vskip4pt} & \parbox{4em}{\vskip5pt$\frac12 \mathcal I_{\alpha}^{(\vmc E,\vmc B)}$\vskip4pt}\\
\cline{1-2}
\parbox{5.3em}{\vskip5pt$-\frac12 \big(\mathcal I_{\alpha}^{(\vmc E,\vmc B)}\big)^{\mathrm{T}}$\vskip4pt} & \parbox{2.1em}{\vskip5pt$\mathcal I_{\alpha}^{(\vmc B)}$\vskip4pt}
\end{tabular}
\right)\ .
\end{align}

The rotational properties of $\mathcal I_{\alpha}^{(\vmc E)}$, $\mathcal I_{\alpha}^{(\vmc B)}$ and $\mathcal I_{\alpha}^{(\vmc E,\vmc B)}$ are now easily obtained from (\ref{4.1}) to (\ref{4.5}) and (\ref{4.8}) to (\ref{4.10}) using
\begin{align}\label{4.12}
\underline{\mathcal R}^{-1}\underline D_j\underline{\mathcal R}&= R_{jk}\underline D_k\ ,\nn\\
\underline{\mathcal R}^{-1}\underline\mu_j\underline{\mathcal R}&= R_{jk}\underline\mu_k\ .
\end{align}
We get
\begin{align}\label{4.13}
 R\, \mathcal I_{\alpha}^{(\vmc E)}(\vmc E,\vmc B)\, R^{\mathrm{T}}&=\mathcal I_{\alpha}^{(\vmc E)}( R\vmc E, R\vmc B)\ ,\nn\\
 R\, \mathcal I_{\alpha}^{(\vmc B)}(\vmc E,\vmc B)\, R^{\mathrm{T}}&=\mathcal I_{\alpha}^{(\vmc B)}( R\vmc E, R\vmc B)\ ,\nn\\
 R\, \mathcal I_{\alpha}^{(\vmc E,\vmc B)}(\vmc E,\vmc B)\, R^{\mathrm{T}}&=\mathcal I_{\alpha}^{(\vmc E,\vmc B)}( R\vmc E, R\vmc B)\ .
\end{align}

\subsection{Improper rotations}\label{ss:4.2}
For improper rotations, $\overline{R}\in\mathrm{O}(3)$ with $\mathrm{det}\,\overline{R}=-1$, we have no invariance due to the PV term $\delta\,\umat M_{\PV}$ in the mass matrix (\ref{2.3}). In fact, we are especially interested in PV effects coming from this term. In this subsection we shall decompose the geometric flux densities (\ref{4.6}) to (\ref{4.10b}) into PC and PV parts.

It is clearly sufficient to consider just one improper rotation, the parity transformation
\begin{align}\label{4.14}
\mathrm P\,:\,\vec x\rightarrow P\,\vec x=-\vec x\ .
\end{align}
In the $n=2$ subspace of the hydrogen atom this is represented by a matrix $\underline{\mathcal P}$ which transforms the electric and magnetic dipole operators as follows
\begin{align}\label{4.15}
\underline{\mathcal P}^{-1}\,\underline D_j\,\underline{\mathcal P}&=-\underline D_j\ ,\nn\\
\underline{\mathcal P}^{-1}\,\underline \mu_j\,\underline{\mathcal P}&=\underline \mu_j\ .
\end{align}
The mass matrix $\umat{\tilde M}_0$ (\ref{2.3}) is, of course, not invariant under $\mathrm P$ and we have
\begin{align}
\underline{\mathcal P}\,\umat M_0\,\underline{\mathcal P}^{-1}&=\umat M_0\ ,\label{4.16}\\
\underline{\mathcal P}\,\umat M_{\PV}\,\underline{\mathcal P}^{-1}&=-\umat M_{\PV}\ ,\label{4.17}\\
\underline{\mathcal P}\,\umat{\tilde M}_0\,\underline{\mathcal P}^{-1}&=\underline{\mathcal P}(\umat M_0+\delta\,\umat M_{\PV})\underline{\mathcal P}^{-1}\nn\\
&=\umat M_0-\delta\,\umat M_{\PV}\ .\label{4.18}
\end{align}
Clearly, since $\delta\approx 7.57\times 10^{-13}$ is very small (see Appendix \ref{s:AppendixA}), it is useful to consider the case where it is set to zero, that is, where parity is conserved. We denote the quantities corresponding to this case by $\umat M^{(0)}$, $E_\alpha^{(0)}$, $\mathbbm P_\alpha^{(0)}$ etc. We have then from (\ref{2.2}) and (\ref{2.3})
\begin{align}\label{4.19}
\umat M^{(0)}(\vmc E,\vmc B)&=\umat M_0-\uvec D\cdot\vmc E-\uvec\mu\cdot\vmc B\ ,\\
\umat M(\vmc E,\vmc B)&=\umat M^{(0)}(\vmc E,\vmc B)+\delta\,\umat M_{\PV}\ .\label{4.20}
\end{align}
For the PC quantities we find from (\ref{4.15}), (\ref{4.16}) and (\ref{4.19})
\begin{align}
\underline{\mathcal P}\,\umat M^{(0)}(\vmc E,\vmc B)\,\underline{\mathcal P}^{-1}&=\umat M^{(0)}(-\vmc E,\vmc B)\ ,\label{4.21}\\
E_\alpha^{(0)}(\vmc E,\vmc B)&=E_\alpha^{(0)}(-\vmc E,\vmc B)\ ,\label{4.22}\\
\underline{\mathcal P}\,\mathbbm P_\alpha^{(0)}(\vmc E,\vmc B)\,\underline{\mathcal P}^{-1}&=\mathbbm P_\alpha^{(0)}(-\vmc E,\vmc B)\ .\label{4.23}
\end{align}
Here (\ref{4.22}) and (\ref{4.23}) need some further discussion; see Appendix \ref{s:AppendixC}.

Due to time reversal (T) invariance\footnote{Here we disregard the T-violating complex phase in the Cabibbo-Kobayashi-Maskawa Matrix. This is justified since we are dealing with a flavour diagonal process.} and condition (\ref{2.13}) $E_\alpha(\vmc E,\vmc B)$ gets no contribution linear in $\delta$, that is, linear in the PV term $\delta\,\umat M_{\PV}$; see \cite{BrGaNa99,LoSa77}. Neglecting higher order terms in $\delta$ we have, therefore,
\begin{align}\label{4.24}
E_\alpha(\vmc E,\vmc B)=E_\alpha^{(0)}(\vmc E,\vmc B)=E_\alpha^{(0)}(-\vmc E,\vmc B)\ .
\end{align}
Rotational invariance (\ref{4.2}) implies then that $E_\alpha(\vmc E,\vmc B)$ must be a function of the P-even invariants one can form from $\vmc E$ and $\vmc B$:
\begin{align}\label{4.25}
E_\alpha(\vmc E,\vmc B)\equiv E_\alpha(\vmc E^2,\vmc B^2,(\vmc E\cdot\vmc B)^2)\ .
\end{align}

In Subsection \ref{ss:4.3} below we present the analogous analysis in terms of invariants for the geometric flux densities (\ref{4.8}) to (\ref{4.10b}).  For the case of no P-violation we easily find from (\ref{4.15}) and (\ref{4.21}) that $\mathcal I_{\alpha}^{(\vmc E)}(\vmc E,\vmc B)$ and $\mathcal I_{\alpha}^{(\vmc B)}(\vmc E,\vmc B)$ must be even, whereas $\mathcal I_{\alpha}^{(\vmc E,\vmc B)}(\vmc E,\vmc B)$ must be odd under $\vmc E\rightarrow -\vmc E$. This allows us to define generally, without any expansion in $\delta$, the PC and PV parts of the geometric flux densities as follows:
\begin{align}
\mathcal I_{\alpha}^{(\vmc E)\PC}(\vmc E,\vmc B)&=\frac12\left[\mathcal I_{\alpha}^{(\vmc E)}(\vmc E,\vmc B)+\mathcal I_{\alpha}^{(\vmc E)}(-\vmc E,\vmc B)\right]\ ,\label{4.26}\\
\mathcal I_{\alpha}^{(\vmc E)\PV}(\vmc E,\vmc B)&=\frac12\left[\mathcal I_{\alpha}^{(\vmc E)}(\vmc E,\vmc B)-\mathcal I_{\alpha}^{(\vmc E)}(-\vmc E,\vmc B)\right]\ ,\label{4.27}\\
\mathcal I_{\alpha}^{(\vmc B)\PC}(\vmc E,\vmc B)&=\frac12\left[\mathcal I_{\alpha}^{(\vmc B)}(\vmc E,\vmc B)+\mathcal I_{\alpha}^{(\vmc B)}(-\vmc E,\vmc B)\right]\ ,\label{4.28}\\
\mathcal I_{\alpha}^{(\vmc B)\PV}(\vmc E,\vmc B)&=\frac12\left[\mathcal I_{\alpha}^{(\vmc B)}(\vmc E,\vmc B)-\mathcal I_{\alpha}^{(\vmc B)}(-\vmc E,\vmc B)\right]\ ,\label{4.29}\\
\mathcal I_{\alpha}^{(\vmc E,\vmc B)\PC}(\vmc E,\vmc B)&=\frac12\left[\mathcal I_{\alpha}^{(\vmc E,\vmc B)}(\vmc E,\vmc B)-\mathcal I_{\alpha}^{(\vmc E,\vmc B)}(-\vmc E,\vmc B)\right]\ ,\label{4.30}\\
\mathcal I_{\alpha}^{(\vmc E,\vmc B)\PV}(\vmc E,\vmc B)&=\frac12\left[\mathcal I_{\alpha}^{(\vmc E,\vmc B)}(\vmc E,\vmc B)+\mathcal I_{\alpha}^{(\vmc E,\vmc B)}(-\vmc E,\vmc B)\right]\ .\label{4.31}
\end{align}
These combine to
\begin{align}\label{4.32}
\mathcal I_{\alpha}(\vmc E,\vmc B)=\mathcal I_{\alpha}^{\PC}(\vmc E,\vmc B)+\mathcal I_{\alpha}^{\PV}(\vmc E,\vmc B)\ ,
\end{align}
for $\mathcal I_\alpha=\mathcal I_\alpha^{(\vmc E)}$, $\mathcal I_\alpha^{(\vmc B)}$ and $\mathcal I_\alpha^{(\vmc E,\vmc B)}$.

Using now the expansion in the small PV parameter $\delta$ up to linear order we get for the PC fluxes exactly the expressions (\ref{4.8}) to (\ref{4.10}) but with $\umat M(\vmc E,\vmc B)$, $\mathbbm P_{\alpha,\beta}(\vmc E,\vmc B)$, $E_\alpha(\vmc E,\vmc B)$ replaced by the corresponding quantities for $\delta=0$, that is, $\umat M^{(0)}(\vmc E,\vmc B)$ etc. For the PV fluxes we have to expand the expressions (\ref{4.8}) to (\ref{4.10}) up to linear order in $\delta$. This is easily done and leads to
\begin{align}\label{4.33}
\mathcal I_{\alpha,jk}^{(\vmc E)\PV}(\vmc E,\vmc B)&=\delta\frac{\I}{2}\frac{1}{2\pi\I}\oint_{S_\alpha}\mathrm d\zeta\,\mathrm{Tr}\Bigg[\frac{1}{\zeta-\umat M^{(0)}(\vmc E,\vmc B)}\umat M_{\mathrm \PV}\nn\\
&\hspace{-6em}\times\frac{1}{\zeta-\umat M^{(0)}(\vmc E,\vmc B)}\underline D_j\frac{1}{\big(\zeta-\umat M^{(0)}(\vmc E,\vmc B)\big)^2}\underline D_k-(j\leftrightarrow k)\Bigg]
\end{align}
and analogous expressions for $\mathcal I_{\alpha}^{(\vmc B)\PV}$ and $\mathcal I_{\alpha}^{(\vmc E,\vmc B)\PV}$. These, the derivation of (\ref{4.33}), and further useful expressions for PV fluxes, are given in Appendix \ref{s:AppendixC}.

\subsection{Expansions for flux densities}\label{ss:4.3}
We can now write down the expansions for the geometric flux densities (\ref{4.8}) to (\ref{4.10}) following from rotational invariance and the parity transformation properties. In these expansions we encounter invariant functions
\begin{align}\label{4.34}
g_r^\alpha&\equiv g_r^\alpha(\vmc E^2,\vmc B^2,(\vmc E\cdot\vmc B)^2)\ ,\nn\\
h_r^\alpha&\equiv h_r^\alpha(\vmc E^2,\vmc B^2,(\vmc E\cdot\vmc B)^2)\ ,\nn\\
r&=1,\dots, 15\ ,
\end{align}
which are, in general, complex valued. Our notation is such that the terms with the $g_r^\alpha$ are the parity conserving (PC) ones, the terms with the $h_r^\alpha$ the parity violating (PV) ones. We find the following:
\begin{align}
\mathcal I_{\alpha,jk}^{(\vmc E)}(\vmc E,\vmc B)&=\frac12\epsilon_{jkl}J_{\alpha,l}^{(\vmc E)}(\vmc E,\vmc B)\ ,\nn\\
\vec J_{\alpha}^{(\vmc E)}(\vmc E,\vmc B)&=\vec J_{\alpha}^{(\vmc E)\PC}(\vmc E,\vmc B)+\vec J_{\alpha}^{(\vmc E)\PV}(\vmc E,\vmc B)\nn\\
&=\vmc E\,\big[(\vmc E\cdot\vmc B)\,g_1^\alpha+h_1^\alpha\big]+\vmc B\,\big[g_2^\alpha+(\vmc E\cdot\vmc B)\,h_2^\alpha\big]\nn\\
&\quad+(\vmc E\times\vmc B)\,\big[(\vmc E\cdot\vmc B)\,g_3^\alpha+h_3^\alpha\big]\ ,\label{4.35}\\
\mathcal I_{\alpha,jk}^{(\vmc E,\vmc B)}(\vmc E,\vmc B)&=\mathcal I_{\alpha,jk}^{(\vmc E,\vmc B)\PC}(\vmc E,\vmc B)+\mathcal I_{\alpha,jk}^{(\vmc E,\vmc B)\PV}(\vmc E,\vmc B)\nn\\
&\hspace{-5.7em}=\frac12\epsilon_{jkl}\Big\{\mathcal E_l\big[g_4^\alpha+(\vmc E\cdot\vmc B)\,h_4^\alpha\big]+\mathcal B_l\,\big[(\vmc E\cdot\vmc B)\,g_5^\alpha+h_5^\alpha\big]\nn\\
&\hspace{-5.7em}\qquad+(\vmc E\times\vmc B)_l\,\big[g_6^\alpha+(\vmc E\cdot\vmc B)\,h_6^\alpha\big]\Big\}\nn\\
&\hspace{-5.7em}\quad+\delta_{jk}\,\big[(\vmc E\cdot\vmc B)\,g_7^\alpha+h_7^\alpha\big]\nn\\
&\hspace{-5.7em}\quad+(\mathcal E_j\mathcal E_k-\frac13\delta_{jk}\,\vmc E^2)\,\big[(\vmc E\cdot\vmc B)\,g_8^\alpha+h_8^\alpha\big]\nn\\
&\hspace{-5.7em}\quad+(\mathcal B_j\mathcal B_k-\frac13\delta_{jk}\,\vmc B^2)\,\big[(\vmc E\cdot\vmc B)\,g_9^\alpha+h_9^\alpha\big]\nn\\
&\hspace{-5.7em}\quad+(\mathcal E_j\mathcal B_k+\mathcal E_k\mathcal B_j-\frac23\delta_{jk}\,(\vmc E\cdot\vmc B))\,\big[g_{10}^\alpha+(\vmc E\cdot\vmc B)\,h_{10}^\alpha\big]\nn\\
&\hspace{-5.7em}\quad+\big[\mathcal E_j\,(\vmc E\times\vmc B)_k+\mathcal E_k\,(\vmc E\times\vmc B)_j\big]\,\big[(\vmc E\cdot\vmc B)\,g_{11}^\alpha+h_{11}^\alpha\big]\nn\\
&\hspace{-5.7em}\quad+\big[\mathcal B_j\,(\vmc E\times\vmc B)_k+\mathcal B_k\,(\vmc E\times\vmc B)_j\big]\,\big[g_{12}^\alpha+(\vmc E\cdot\vmc B)\,h_{12}^\alpha\big]\ ,\label{4.36}\\
\mathcal I_{\alpha,jk}^{(\vmc B)}(\vmc E,\vmc B)&=\frac12\epsilon_{jkl}J_{\alpha,l}^{(\vmc B)}(\vmc E,\vmc B)\ ,\nn\\
\vec J_{\alpha}^{(\vmc B)}(\vmc E,\vmc B)&=\vec J_{\alpha}^{(\vmc B)\PC}(\vmc E,\vmc B)+\vec J_{\alpha}^{(\vmc B)\PV}(\vmc E,\vmc B)\nn\\
&=\vmc E\,\big[(\vmc E\cdot\vmc B)\,g_{13}^\alpha+h_{13}^\alpha\big]\nn\\
&\quad+\vmc B\,\big[g_{14}^\alpha+(\vmc E\cdot\vmc B)\,h_{14}^\alpha\big]\nn\\
&\quad+(\vmc E\times\vmc B)\,\big[(\vmc E\cdot\vmc B)\,g_{15}^\alpha+h_{15}^\alpha\big]\ .\label{4.37}
\end{align}

In the next section we shall discuss specific examples where the general expansions (\ref{4.35}) to (\ref{4.37}) will prove to be very useful.

\section{Specific Cases}\label{s:5}

In this section we shall illustrate the structures of the geometric flux densities for specific cases. First, we analytically derive the geometric flux densities $\vec J_{\alpha}^{(\vmc B)}(\vmc E=\vec 0,\vmc B)$ in magnetic field space for vanishing electric field and compare the results with numerical calculations. In that case the flux densities give rise only to P-conserving geometric phases. We then analyse the structure of $\vec J_{\alpha}^{(\vmc E)}(\vmc E,\vmc B)$ in electric field space together with a constant magnetic field $\vmc B=\mathcal B_3\,\vec e_3$. Here, we will obtain P-violating geometric phases. As a further example we investigate the geometric flux densities in the mixed parameter space of $\mathcal E_1,\mathcal E_3$ and $\mathcal B_3$ together with a constant magnetic field $\vmc B=\mathcal B_2\,\vec e_2$.

\subsection{A magnetic field $\vmc B$ and $\vmc E=0$}\label{ss:5.1}
Starting from the general expression (\ref{4.37}) for $\vec J_{\alpha}^{(\vmc B)}(\vmc E,\vmc B)$, we immediately find for $\vec J_{\alpha}^{(\vmc B)}(\vmc E=\vec 0,\vmc B)$
\begin{align}\label{5.1}
\vec J_{\alpha}^{(\vmc B)}(\vec 0,\vmc B)=\vmc B\,g^\alpha_{14}
\end{align}
where $g^\alpha_{14}=g_{14}^\alpha(\vmc B^2)$ only depends on the modulus squared of the magnetic field.

Because of (\ref{3.21}) we have
\begin{align}\label{5.1b}
\mathbf\nabla_{\vmc B}\cdot\vec J_{\alpha}^{(\vmc B)}(\vec 0,\vmc B)=0 
\end{align}
for $b\equiv\vmc B^2\not=0$. Inserting (\ref{5.1}) in (\ref{5.1b}) we get
\begin{align}\label{5.2}
0=\mathbf\nabla_{\vmc B}\cdot\big(\vmc B\,g^\alpha_{14}(b)\big)&=3\,g^\alpha_{14}(b)+2\,b\,\partial_b g^\alpha_{14}(b)
\end{align}
with the solution
\begin{align}
g^\alpha_{14}(b)=a^\alpha b^{-\frac32}=a^\alpha|\vmc B|^{-3}
\end{align}
where $a^\alpha$ is an integration constant. Therefore, from the rotational invariance arguments above we find already
\begin{align}\label{5.6}
\vec J_{\alpha}^{(\vmc B)}(\vec 0,\vmc B)=a^\alpha\frac{\vmc B}{|\vmc B|^3}\ .
\end{align}
This is the field of a Dirac monopole of strength $a^\alpha$ at $\vmc B =\vec 0$. A detailed calculation of $a^\alpha$ for the 2S states is presented in Appendix \ref{s:AppendixD}. The resulting flux-density vector field is real and P-conserving. It reads
\begin{align}\label{5.7}
\vec J_{\alpha}^{(\vmc B)}(\vec 0,\vmc B)=\left\{
\begin{array}{cl}
-\frac{\vmc B}{\phantom{\big(}\hspace{-0.2em}|\vmc B|^3} & ,\quad\mbox{for }\alpha=9\ ,\\[0.3em]
\frac{\vmc B}{\phantom{\big(}\hspace{-0.2em}|\vmc B|^3} & ,\quad\mbox{for }\alpha=11\ ,\\[0.3em]
\mathbf 0 & ,\quad\mbox{for }\alpha=10,12\ .
\end{array}
\right.
\end{align}
We recall that the states with labels $\alpha=9,10,11$, and $12$ are connected to the 2S states with $(F,F_3)=(1,1)$, $(1,0)$, $(1,-1)$, and $(0,0)$, respectively; see Appendix \ref{s:AppendixA}, Table \ref{t:state.labelsFD} ff.

Comparing this exact result with numerical calculations, we can extract an estimate for the error of $\vec J_\alpha$ in parameter spaces other than that of the magnetic field. For $10^3$ equidistant grid points in a cubic parameter space volume $[-1\,\mathrm{mT},1\,\mathrm{mT}]^{3}$, at which the vectors $\vec J_{11}^{(\vmc B)}(\vec 0,\vmc B)$ are evaluated, we obtain numerically the following deviations from the vector field structure given in (\ref{5.7}):
\begin{align}\label{5.8}
||\vec J_{11}^{(\vmc B)}(\vec 0,\vmc B)|\cdot|\vmc B|^2-1|&\lesssim 5\times 10^{-12}\nn\ ,\\
\left|\frac{\vec J_{11}^{(\vmc B)}(\vec 0,\vmc B)}{|\vec J_{11}^{(\vmc B)}(\vec 0,\vmc B)|}\times\frac{\vmc B}{|\vmc B|}\right|&\lesssim 8\times 10^{-13}\ .
\end{align}
Since the data type used for the numerical calculations has a precision of approximately 16 digits, we find the results (\ref{5.8}) to be in good agreement with the analytic expression (\ref{5.7}).

\begin{figure}
\centering
\includegraphics[scale=1.1]{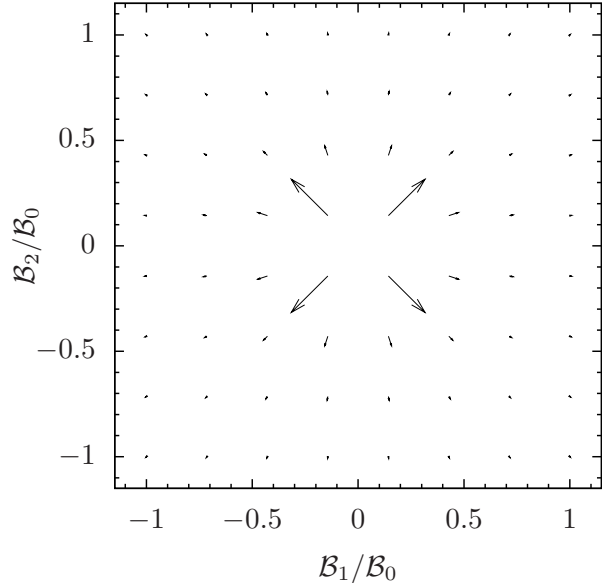}
\caption{Visualisation of the P-conserving flux-density vector field $\hat{\vec J}_{\alpha}^{(\vmc B)\PC}(\vec 0,\vmc B)$ (\ref{92a}) for $\alpha=11$ and $\eta=10^{-2}$ in magnetic field parameter space at $\mathcal B_3=0$.}
\label{B1B2B3Plot2}
\end{figure}

Figure \ref{B1B2B3Plot2} illustrates the numerical results for $\vec J_{\alpha}^{(\vmc B)}(\vec 0,\vmc B)$ with $\alpha=11$ in the $\mathcal B_3=0$ plane.

Here and in the following we find it convenient to plot  dimensionless quantities. Therefore, we choose reference values for the electric and magnetic field strengths
\begin{align}\label{92}
\mathcal E_0&=1\,\mathrm{V/cm}\ ,\nn\\
\mathcal B_0&=1\,\mathrm{mT}\ .
\end{align}
We label the axes in parameter space by $\mathcal B_i/\mathcal B_0$ and plot the vectors
\begin{align}\label{92a}
\hat{\vec J}_\alpha^{(\vmc B)}=\eta\,\vec J_\alpha^{(\vmc B)}\,\mathcal B_0^2\ .
\end{align}
Here $\eta$ is a rescaling parameter chosen such as to bring the vectors in the plots to a convenient length scale. The dimensionless geometric phases, see (\ref{3.2}), (\ref{3.4}), and (\ref{3.20}), are then given by the flux of this dimensionless vector field (\ref{92a}) through a surface in this space of $\mathcal B_i/\mathcal B_0$ and divided by $\eta$.

For any curve $\mathcal C=\partial \mathcal F$ in $\vmc B$ space we get for the geometric phases from (\ref{3.20}) and (\ref{5.7})
\begin{align}\label{92b}
\gamma_\alpha(\mathcal C)&=\int_{\mathcal F}\vec J_\alpha^{(\vmc B)}(\vmc B)\,\mathrm d\vec f^{(\vmc B)}\nn\\
&=\left\{
\begin{array}{cl}
-\Omega_{\mathcal C} & ,\quad\mbox{for }\alpha=9\ ,\\[0.3em]
+\Omega_{\mathcal C} & ,\quad\mbox{for }\alpha=11\ ,\\[0.3em]
0 & ,\quad\mbox{for }\alpha=10,12\ .
\end{array}
\right.
\end{align}
Here $\Omega_{\mathcal C}$ is the solid angle spanned by the curve $\mathcal C$. This is in accord with the expectation for a spin $1$ system; see \cite{Bar83}.

\subsection{An electric field $\vmc E$ together with a constant magnetic field}\label{ss:5.2}
We now consider the case of geometric flux densities in electric field space with a constant magnetic field $\vmc B=\mathcal B_3\,\vec e_3$ with $\mathcal B_3>0$. Here we find from (\ref{4.35})
\begin{align}
\vec J_{\alpha}^{(\vmc E)\PC}(\vmc E,\mathcal B_3\,\vec e_3)&=\vmc E\,\mathcal E_3\mathcal B_3\, g^\alpha_{1}+\mathcal B_3\left(
\begin{array}{c}
\mathcal E_2\mathcal E_3\mathcal B_3\,g_3^\alpha\\
-\mathcal E_1\mathcal E_3\mathcal B_3\,g_3^\alpha\\
g_2^\alpha
\end{array}
\right)\ ,\label{5.9}\\
\vec J_{\alpha}^{(\vmc E)\PV}(\vmc E,\mathcal B_3\,\vec e_3)&=\vmc E\,h_1^\alpha+\mathcal B_3\,\vec e_3\,\mathcal E_3\mathcal B_3\,h_2^\alpha\nn\\
&\quad+(\mathcal E_2\,\vec e_1-\mathcal E_1\,\vec e_3)\,\mathcal B_3\, h_3^\alpha\nn\\
&=\vmc E\,h_1^\alpha+\mathcal B_3\left(
\begin{array}{c}
\mathcal E_2\,h_3^\alpha\\
-\mathcal E_1\,h_3^\alpha\\
\mathcal E_3\mathcal B_3\,h_2^\alpha
\end{array}
\right)\ ,\label{5.9b}
\end{align}
where $g_{1,2,3}^\alpha$ and $h_{1,2,3}^\alpha$ may in general depend on $\vmc E^2$, $\mathcal B_3^2$ and $(\mathcal E_3\mathcal B_3)^2$ and $\alpha\in\{9,10,11,12\}$. In our case $\mathcal B_3$ is constant. We have, therefore,
\begin{align}\label{5.10}
g_i^\alpha&=g_i^\alpha(\vmc E_T^2,\mathcal E_3^2)\ ,\nn\\
h_i^\alpha&=h_i^\alpha(\vmc E_T^2,\mathcal E_3^2)\ ,\nn\\
(i&=1,2,3)
\end{align}
with
\begin{align}\label{5.11}
\vmc E_T^2=\mathcal E_1^2+\mathcal E_2^2\ .
\end{align}

In the following we give the results of numerical evaluations of the PC and PV flux-density vectors (\ref{5.9}) and (\ref{5.9b}), respectively. We split the PV vectors into the contributions from the nuclear-spin independent ($i=1$) and dependent ($i=2$) PV interactions
\begin{align}\label{96a}
\vec J^{(\vmc E)\PV}=\vec J^{(\vmc E)\PV_1}+\vec J^{(\vmc E)\PV_2}\ .
\end{align}
Here the $\vec J^{(\vmc E)\PV_i}$ are defined as in (\ref{4.33}) but with $\delta$ replaced by $\delta_i$ and $\umat M_\PV$ replaced by $\umat M_\PV^{(i)}$ ($i=1,2$); see (\ref{A0c}) and (\ref{A0d}). Again we shall plot dimensionless vectors
\begin{align}
\hat{\vec J}_{\alpha}^{(\vmc E)\PC}(\vmc E,\mathcal B_3\,\vec e_3)&=\eta_\alpha\,\vec J_{\alpha}^{(\vmc E)\PC}(\vmc E,\mathcal B_3\,\vec e_3)\,\mathcal E_0^2\ ,\label{96b}\\
\hat{\vec J}_{\alpha}^{(\vmc E)\PV_i}(\vmc E,\mathcal B_3\,\vec e_3)&=\eta'_{\alpha,i}\,\vec J_{\alpha}^{(\vmc E)\PV_i}(\vmc E,\mathcal B_3\,\vec e_3)\,\mathcal E_0^2\label{96c}
\end{align}
with $\mathcal E_0$ from (\ref{92}) and $\eta_{\alpha}$ and $\eta'_{\alpha,i}$ conveniently chosen.

\begin{figure}
\centering
\includegraphics[scale=1.1]{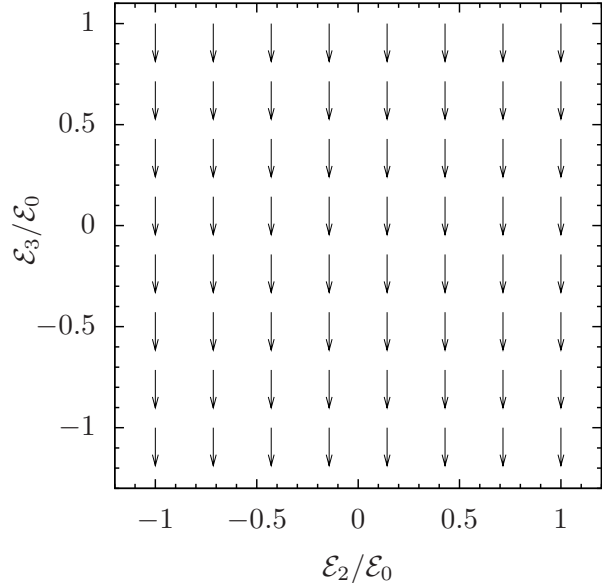}
\caption{Visualisation of the real part of the P-conserving flux-density vector field $\hat{\vec J}_{9}^{(\vmc E)\PC}(\vmc E,\mathcal B_3\,\vec e_3)$ (\ref{96b}) in electric field parameter space at $\mathcal E_1=1\,\mathrm{V/cm}$. A constant magnetic field with $\mathcal B_3=1\,\mathrm{mT}$ is applied. The scaling factor in (\ref{96b}) is chosen as $\eta_9=2.5\times 10^{4}$.}
\label{E1E2E3PlotPC2}
\end{figure}

As an example of a PC flux-density vector field we present in Figure \ref{E1E2E3PlotPC2} the results of a numerical calculation of $\vec J_{9}^{(\vmc E)\PC}(\vmc E,\mathcal B_3\,\vec e_3)$ for $\mathcal B_3=1\,\mathrm{mT}$. We recall that the state with $\alpha=9$ is connected to the 2S state with $(F,F_3)=(1,1)$; see Appendix \ref{s:AppendixA}. We plot the dimensionless vectors (\ref{96b}) with the scaling factor chosen as $\eta_9=2.5\times 10^{4}$. Comparing with (\ref{5.9}) we see that here the dominant term is the one proportional to $g_2^\alpha$:
\begin{align}\label{101e}
\vec J_{9}^{(\vmc E)\PC}(\vmc E,\mathcal B_3\,\vec e_3)\approx \mathcal B_3\,g_2^9(\vmc E_T^2,\mathcal E_3^2)\,\vec e_3
\end{align}
with $g_2^9(\vmc E_T^2,\mathcal E_3^2)$ being practically constant.

The numerical results shown in Figure \ref{E1E2E3PlotPC2} reveal a large sensitivity of the flux-density vector field $\vec J_{9}^{(\vmc E)\PC}$ to the normalisation of the dipole operator $\uvec D$. Indeed, suppose that we make in our calculations the replacement
\begin{align}
\uvec D\rightarrow\lambda\uvec D\label{101eb}
\end{align}
with $\lambda$ a positive real constant. From the mass matrix (\ref{2.2}) and from $\mathcal I_{9}^{(\vmc E)}$ in (\ref{4.8}) we find the following scaling behaviour for $\vec J_{9}^{(\vmc E)\PC}$
\begin{align}\label{101eaa}
\left.\vec J_{9}^{(\vmc E)\PC}(\vmc E,\mathcal B_3\,\vec e_3)\right|_{\lambda\uvec D}=\lambda^2\left.\vec J_{9}^{(\vmc E)\PC}(\lambda\vmc E,\mathcal B_3\,\vec e_3)\right|_{\uvec D}\ .
\end{align}
Since our calculations show that $\vec J_{9}^{(\vmc E)\PC}$ is practically constant for the range of fields explored here we get
\begin{align}\label{101eab}
\left.\vec J_{9}^{(\vmc E)\PC}(\vmc E,\mathcal B_3\,\vec e_3)\right|_{\lambda\uvec D}\approx\lambda^2\left.\vec J_{9}^{(\vmc E)\PC}(\vmc E,\mathcal B_3\,\vec e_3)\right|_{\uvec D}\ .
\end{align}
Therefore, measurements of $\vec J_{9}^{(\vmc E)\PC}(\vmc E,\mathcal B_3\,\vec e_3)$ for the setup considered here are highly sensitive to deviations of the normalisation of $\uvec D$ from the standard expression as given in Table \ref{t:H2.D}.

As an example we calculate the PC geometric phase for the following path in $\vmc E$ space
\begin{align}\label{101ed}
\mathcal C\hspace{-0.15em}:\; z\rightarrow \vmc E(z)&=\left(\begin{array}{c}
1.0+0.5\times\sin(200\,\pi\,z)\\
0.5\times\cos(200\,\pi\,z)\\
1.0
\end{array}
\right)\,\mathrm{V/cm}\ ,\nn\\
z&\in[0,1]\ .
\end{align}
That is, we consider a path circling $100$ times in the\linebreak[4] $\mathcal E_1-\mathcal E_2$ plane which is orthogonal to the flux direction for $\alpha=9$; see (\ref{101e}) and Figure \ref{E1E2E3PlotPC2}. For the constant magnetic field chosen there, $\mathcal B_3=1\,\mathrm{mT}$, we obtain for $\alpha =9$
\begin{align}\label{101phase1}
\gamma_9(\mathcal C)&= 5.89\times 10^{-4}-5.25\times 10^{-5}\,\I\ .
\end{align} 

\begin{figure}
\centering
\includegraphics[scale=1.1]{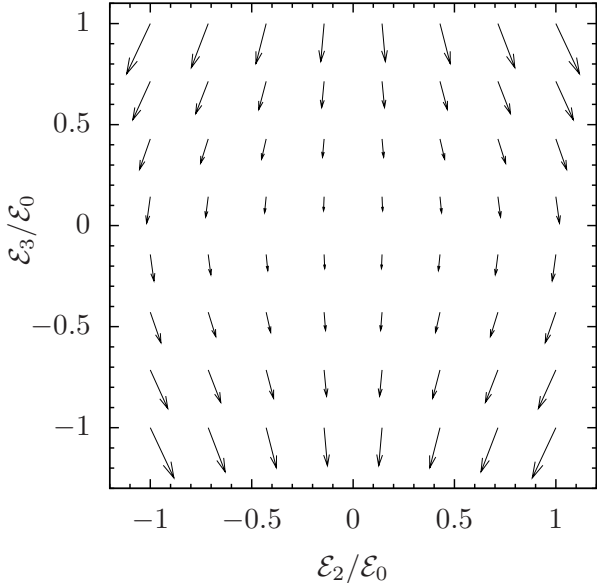}
\caption{Visualisation of the real part of the P-conserving flux-density vector field $\hat{\vec J}_{9}^{(\vmc E)\PC}(\vmc E,\mathcal B_3\,\vec e_3)$ (\ref{96b}) in electric field parameter space at $\mathcal E_1=1\,\mathrm{V/cm}$. A constant magnetic field with $\mathcal B_3=1\,\mu\mathrm{T}$ is applied. The scaling factor in (\ref{96b}) is chosen as $\eta_9=450$.}
\label{E1E2E3PlotPC21muT}
\end{figure}

Decreasing the constant magnetic field $\mathcal B_3$ we get larger geometric phases. Numerical results of the flux-density vector field $\vec J_{9}^{(\vmc E)\PC}(\vmc E,\mathcal B_3\,\vec e_3)$ for $\mathcal B_3=1\,\mu\mathrm T$ are presented in Figure \ref{E1E2E3PlotPC21muT}. There, the scaling factor is chosen as $\eta_9=450$. For the curve (\ref{101ed}), $\mathcal B_3=1\,\mu\mathrm{T}$, and $\alpha=9$ we obtain
\begin{align}\label{101phase2}
\gamma_9(\mathcal C)&= 0.0340-0.00596\,\I\ .
\end{align}

\begin{figure}
\centering
\includegraphics[scale=1.1]{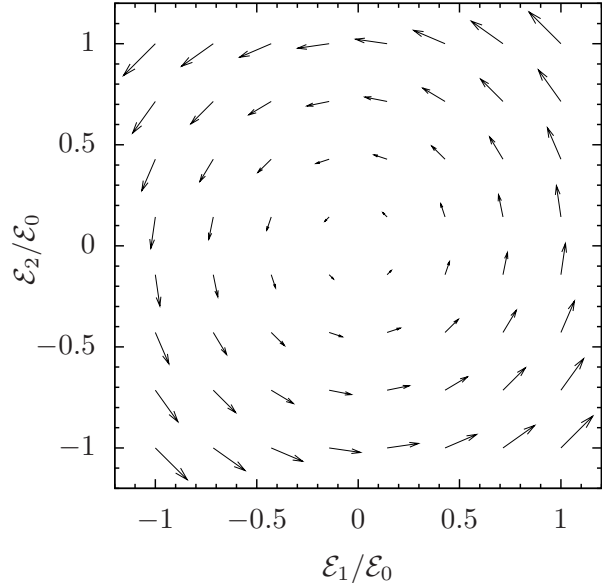}
\caption{Visualisation of the real part of the nuclear spin dependent P-violating flux-density vector field $\hat{\vec J}_{9}^{(\vmc E)\PV_2}(\vmc E,\vmc B)$ (\ref{96c}) in electric field parameter space at $\mathcal E_3=1\,\mathrm{V/cm}$ with an additional constant magnetic field $\vmc B=\mathcal B_3\,\vec e_3$, $\mathcal B_3=1\,\mu\mathrm{T}$. The scaling factor in (\ref{96c}) is chosen as $\eta'_{9,2}=400/\delta_2$.}
\label{E1E2E3Plot2}
\end{figure}

As an example of a PV flux-density vector field we present the numerical results for $\vec J_{9}^{(\vmc E)\PV_2}$ in Figure \ref{E1E2E3Plot2}.
We find both, the real and the imaginary part of \linebreak[4] $\vec J_{9}^{(\vmc E)\PV_2}(\vmc E,\mathcal B_3\,\vec e_3)$ to represent a flow, circulating around the $\vec e_3$-axis, with vanishing third component. That is, in (\ref{5.9b}) for $\alpha=9$ the terms involving $h_1^9$ and $h_2^9$, respectively, come out numerically to be negligible compared to the terms involving $h_3^9$. We may hence write
\begin{align}\label{5.16}
\vec J_{9}^{(\vmc E)\PV_2}(\vmc E,\mathcal B_3\,\vec e_3)&\approx\mathcal B_3\,h_3^9(\vmc E_T^2,\mathcal E_3^2)\left(
\begin{array}{c}
\mathcal E_2\\
-\mathcal E_1\\
0
\end{array}
\right)\ .
\end{align}
This is corroborated by numerical studies. For $|\mathcal E_j|\leq\mathcal E_0$ $(j=1,2,3)$ with $\mathcal E_0$ from (\ref{92}) we find
\begin{align}\label{5.12}
\frac{|\mathrm{Re}\,\vec e_3\,\vec J_{9}^{(\vmc E)\PV_2}|}{|\mathrm{Re}\,\vec J_{9}^{(\vmc E)\PV_2}|}\lesssim 6\times 10^{-11}
\end{align}
and
\begin{align}\label{5.15}
\frac{|\mathcal E_1\,\mathrm{Re}\,\vec e_1\,\vec J_{9}^{(\vmc E)\PV_2}+\mathcal E_2\,\mathrm{Re}\,\vec e_2\,\vec J_{9}^{(\vmc E)\PV_2}|}{|\mathrm{Re}\,\vec J_{9}^{(\vmc E)\PV_2}|\,\sqrt{\mathcal E_1^2+\mathcal E_2^2}}\lesssim 2.5\times 10^{-11}\ .
\end{align}
Results similar to (\ref{5.12}) and (\ref{5.15}) hold for the imaginary part $\mathrm{Im}\,\vec J_{9}^{(\vmc E)\PV_2}(\vmc E,\mathcal B_3\,\vec e_3)$ and for $\vec J_{9}^{(\vmc E)\PV_1}(\vmc E,\mathcal B_3\,\vec e_3)$. Thus, also $\vec J_{9}^{(\vmc E)\PV_1}$ has to a good approximation the structure (\ref{5.16}).

The antisymmetry of $\mathrm{Re}\,\vec J_{9}^{(\vmc E)\PV_2}(\vmc E,\mathcal B_3\,\vec e_3)$ with respect to $\vmc E\rightarrow -\vmc E$, see (\ref{5.9b}), is confirmed numerically at the same level of accuracy. We find
\begin{align}\label{5.13}
\left|\frac{|\mathrm{Re}\,\vec J_{9}^{(\vmc E)\PV_2}(\vmc E,\mathcal B_3\,\vec e_3)|}{|\mathrm{Re}\,\vec J_{9}^{(\vmc E)\PV_2}(-\vmc E,\mathcal B_3\,\vec e_3)|}-1\right|\lesssim 2.5\times 10^{-10}
\end{align}
and
\begin{align}\label{5.14}
&\qquad\frac{|\mathrm{Re}\,\vec J_{9}^{(\vmc E)\PV_2}(\vmc E,\mathcal B_3\,\vec e_3)+\mathrm{Re}\,\vec J_{9}^{(\vmc E)\PV_2}(-\vmc E,\mathcal B_3\,\vec e_3)|}{|\mathrm{Re}\,\vec J_{9}^{(\vmc E)\PV_2}(\vmc E,\mathcal B_3\,\vec e_3)|}\nn\\
&\lesssim 2.5\times 10^{-10}.
\end{align}
Similar numerical results are also obtained for\linebreak[4] $\mathrm{Im}\,\vec J_{9}^{(\vmc E)\PV_2}(\vmc E,\mathcal B_3\,\vec e_3)$ and for the real and imaginary parts of $\vec J_{9}^{(\vmc E)\PV_1}(\vmc E,\mathcal B_3\,\vec e_3)$.

\subsection{The mixed parameter space of $\mathcal E_1$, $\mathcal E_3$, $\mathcal B_3$ together with a constant magnetic field}\label{ss:5.3}
In our last example we discuss the parameter space spanned by the vectors
\begin{align}\label{5.17}
\vec L\equiv\left(
\begin{array}{c}
L_{1}\\
L_{2}\\
L_{3}
\end{array}\right)
=\left(
\begin{array}{ccc}
\mathcal E_0^{-1}&0&0\\
0&\mathcal E_0^{-1}&0\\
0&0&\mathcal B_0^{-1}\\
\end{array}\right)\left(\begin{array}{c}
\mathcal E_1\\
\mathcal E_3\\
\mathcal B_3
\end{array}
\right)\ ,
\end{align}
see (\ref{3.17}) ff. and (\ref{92}). In addition we assume the presence of a constant magnetic field $\vmc B=1\,\mu\mathrm{T}\,\vec e_2$. The values chosen for $\mathcal E_0$ and $\mathcal B_0$ in (\ref{92}) and (\ref{5.17}) should represent the typical range of electric and magnetic field variations, respectively, for a given experiment. Our choice here is motivated by the discussion of the longitudinal spin echo experiments in \cite{BeGaMaNaTr08_I}. For other experiments different choices of $\mathcal E_0$ and $\mathcal B_0$ will be appropriate.

From (\ref{3.17}), (\ref{3.18}), and (\ref{4.8}) to (\ref{4.11}) we find
\begin{align}\label{5.17b}
\vec J_\alpha^{(\vec L)}(\vmc E,\vmc B)=\left(
\begin{array}{ccc}
\mathcal E_0\mathcal B_0 & 0 & 0\\
0 & \mathcal E_0\mathcal B_0 & 0\\
0 & 0 & \mathcal E_0^2\\
\end{array}\right)\left(
\begin{array}{c}
\mathcal I_{\alpha,33}^{(\vmc E,\vmc B)}(\vmc E,\vmc B)\\
-\mathcal I_{\alpha,13}^{(\vmc E,\vmc B)}(\vmc E,\vmc B)\\
2\,\mathcal I_{\alpha,13}^{(\vmc E)}(\vmc E,\vmc B)
\end{array}\right)\ .
\end{align}
Here again $\alpha\in\{9,10,11,12\}$ labels the metastable state, see Table \ref{t:state.labelsFD} of Appendix \ref{s:AppendixA}, to which the geometric flux density corresponds. Inserting in (\ref{5.17b}) the expressions from (\ref{4.35}) and (\ref{4.36}) we get for the PC and PV parts of $\vec J_\alpha^{(\vec L)}(\vmc E,\vmc B)$
\begin{align}\label{5.18}
\vec J_{\alpha}^{(\vec L)\PC}(\vmc E,\vmc B)&=\left(
\begin{array}{c}
\mathcal B_3\mathcal E_3\,\tilde g_1^\alpha+\mathcal B_3\mathcal E_1\, \tilde g_2^\alpha\\
\mathcal B_3\mathcal E_3\,\tilde g_3^\alpha+\mathcal B_3\mathcal E_1\, \tilde g_4^\alpha\\
\tilde g_5^\alpha+\mathcal E_1\mathcal E_3\, \tilde g_6^\alpha
\end{array}
\right)\ ,\\
\vec J_{\alpha}^{(\vec L)\PV}(\vmc E,\vmc B)&=\left(
\begin{array}{c}
\tilde h_1^\alpha+\mathcal E_1\mathcal E_3\, \tilde h_2^\alpha\\
\tilde h_3^\alpha+\mathcal E_1\mathcal E_3\, \tilde h_4^\alpha\\
\mathcal B_3\mathcal E_3\, \tilde h_5^\alpha+\mathcal B_3\mathcal E_1\, \tilde h_6^\alpha
\end{array}
\right)\ ,\label{5.19}
\end{align}
where the functions $\tilde g_{i}^\alpha$ and $\tilde h_{i}^\alpha$, $i\in\{1,\dots,6\}$, may in general depend on $\mathcal E_1^2$, $\mathcal E_3^2$ and $\mathcal B_3^2$; see Appendix \ref{s:AppendixD}.

For ease of graphical presentation we shall multiply the $\vec J_{\alpha}^{(\vec L)}$ with scaling factors $\eta$. Thus we define
\begin{align}
\hat{\vec J}_\alpha^{(\vec L)\PC}(\vmc E,\vmc B)&=\eta_\alpha\,\vec J_\alpha^{(\vec L)\PC}(\vmc E,\vmc B)\ ,\label{5.19a}\\
\hat{\vec J}_\alpha^{(\vec L)\PV_i}(\vmc E,\vmc B)&=\eta_{\alpha,i}'\,\vec J_\alpha^{(\vec L)\PV_i}(\vmc E,\vmc B)\ ,\nn\\
(i&=1,2)\ .\label{5.19b}
\end{align}

\begin{figure}
\centering
\includegraphics[scale=1.1]{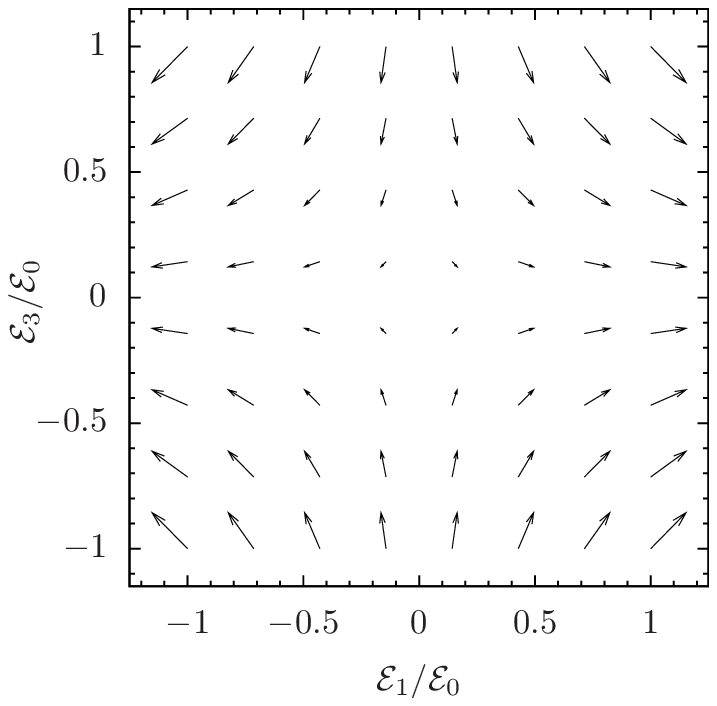}
\caption{Visualisation of the 1 and 2 components of the real part of the P-conserving flux-density vector field $\hat{\vec J}_{9}^{(\vec L)\PC}(\vmc E,\vmc B)$ (\ref{5.19a}) at $\mathcal B_3=143\,\mu\mathrm{T}$. The scaling factor is chosen as $\eta_9=2.5\times 10^{4}$.}
\label{E1E3B3PlotPC}
\end{figure}
\begin{figure}
\centering
\includegraphics[scale=1.1]{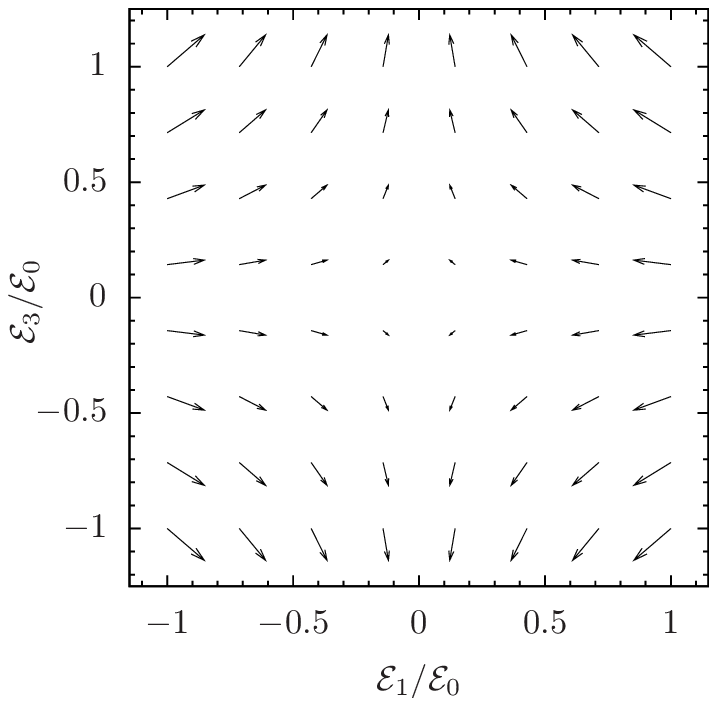}
\caption{Visualisation of the 1 and 2 components of the imaginary part of the P-conserving flux-density vector field $\hat{\vec J}_{9}^{(\vec L)\PC}(\vmc E,\vmc B)$ (\ref{5.19a}) at $\mathcal B_3=143\,\mu\mathrm{T}$. The scaling factor is chosen as $\eta_9=2.7\times 10^{5}$.}
\label{E1E3B3PlotPCImaginary}
\end{figure}
\begin{figure}
\centering
\includegraphics[scale=1.1]{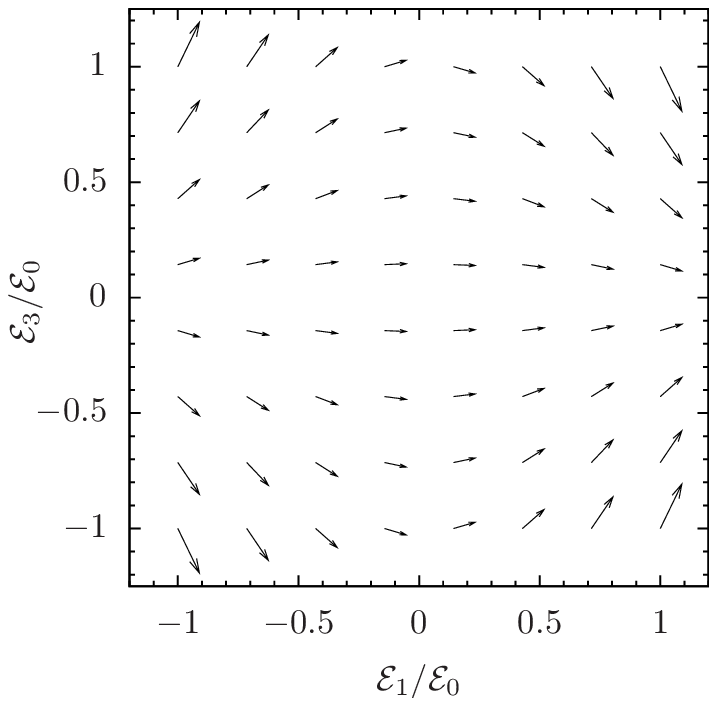}
\caption{Visualisation of the 1 and 2 components of the real part of the nuclear spin dependent P-violating flux-density vector field $\hat{\vec J}_{9}^{(\vec L)\PV_2}(\vmc E,\vmc B)$ (\ref{5.19b}) at $\mathcal B_3=143\,\mu\mathrm{T}$. The scaling factor is chosen as $\eta_{9,2}'=10^{4}/\delta_2$.}
\label{E1E3B3PlotPV2}
\end{figure}
\begin{figure}
\centering
\includegraphics[scale=1.1]{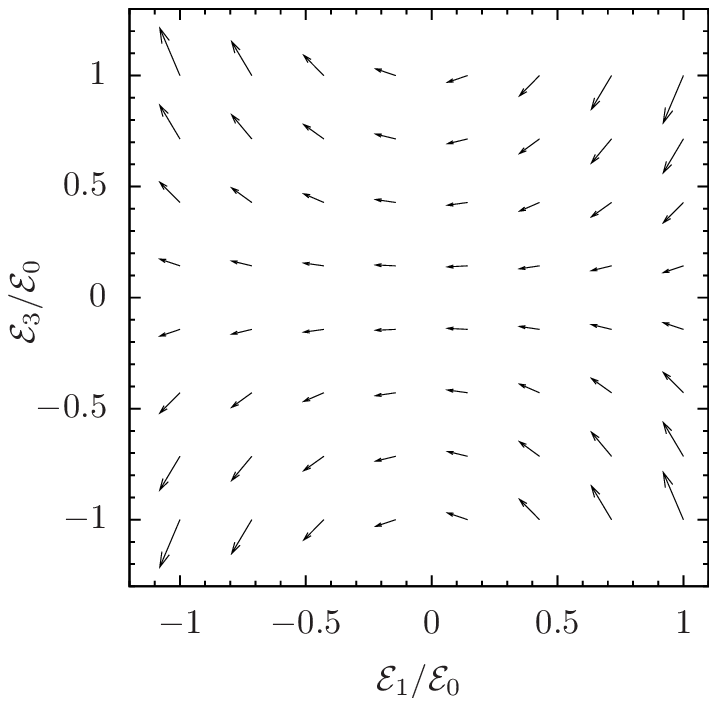}
\caption{Visualisation of the 1 and 2 components of the real part of the nuclear spin dependent P-violating flux-density vector field $\hat{\vec J}_{9}^{(\vec L)\PV_2}(\vmc E,\vmc B)$ (\ref{5.19b}) at $\mathcal B_3=86\,\mu\mathrm{T}$. The scaling factor is chosen as $\eta_{9,2}'=4\times10^{3}/\delta_2$.}
\label{E1E3B3PlotPV2at86muTgnuplot_script}
\end{figure}
\begin{figure}
\centering
\includegraphics[scale=1.1]{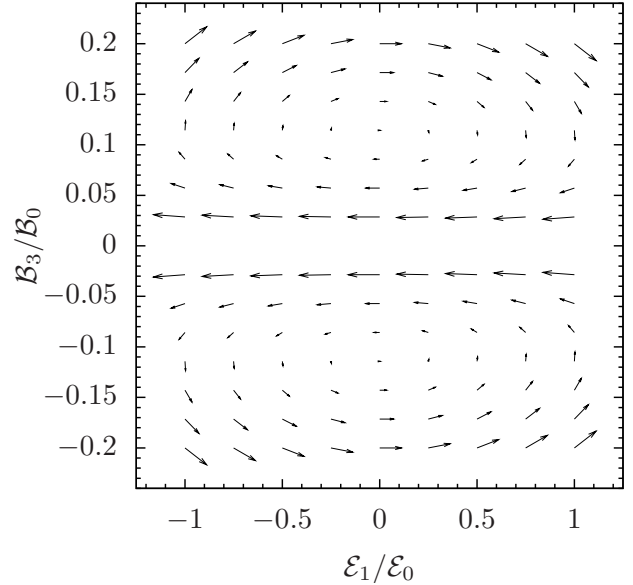}
\caption{Visualisation of the 1 and 3 components of the real part of the nuclear spin dependent P-violating flux-density vector field $\hat{\vec J}_{9}^{(\vec L)\PV_2}(\vmc E,\vmc B)$ (\ref{5.19b}) at $\mathcal E_3=0$. In order to resolve the structure more clearly the scaling factor is chosen as $\eta_{9,2}'=2\times10^{5}\,|\mathcal B_3/\mathcal B_0|^2/\delta_2$.}
\label{E1E3B3PlotPV2E30}
\end{figure}
In the Figures \ref{E1E3B3PlotPC} to \ref{E1E3B3PlotPV2E30} we illustrate the results of our numerical calculations of (\ref{5.19a}) and (\ref{5.19b}) for the geometric flux densities of the state with $\alpha=9$, that is, the state connected with the 2S, $(F,F_3)=(1,1)$ state; see Appendix \ref{s:AppendixA}, Table \ref{t:state.labelsFD}. For $30^3$ grid points in the parameter space volume $[-\mathcal E_0,\mathcal E_0]^2\times[-0.2\mathcal B_0,0.2\mathcal B_0]$ we find that numerically
\begin{align}\label{5.20}
|\mathrm{Re}\,\vec e_3\,\hat{\vec J}_{9}^{(\vec L)\PC}|&\lesssim 0.2\, \Big[(\mathrm{Re}\,\vec e_1\,\hat{\vec J}_{9}^{(\vec L)\PC})^2\nn\\
&\qquad\qquad+(\mathrm{Re}\,\vec e_2\,\hat{\vec J}_{9}^{(\vec L)\PC})^2\Big]^{\frac12}\ .
\end{align}
An analogous relation holds for $\mathrm{Im}\,\hat{\vec J}_{9}^{(\vec L)\PC}(\vmc E,\vmc B)$. This justifies the presentation of these flux-density vector fields in the $\mathcal E_1-\mathcal E_3$ plane as representative vector field structures. Investigating the dependencies of $\vec J_{9}^{(\vec L)}(\vmc E,\vmc B)$ on $\mathcal E_1$, $\mathcal E_3$ and $\mathcal B_3$ we find our numerical results to be consistent with the analytic structures in (\ref{5.18}) and (\ref{5.19}). We recall from (\ref{3.21}) that $\hat{\vec J}_9^{(\vec L)\PC}$ is divergence free and, thus, generated by a non-vanishing vortex distribution $\mathrm{rot}\,\vec J_{9}^{(\vec L)\PC}$. We have calculated this quantity numerically using (\ref{3.22}) and find $\mathrm{rot}\,\vec J_{9}^{(\vec L)\PC}$ to be in agreement with a direct evaluation using a fit function representing $\hat{\vec J}_9^{(\vec L)\PC}$.

Examples of PV flux-density vector fields are shown in Figures \ref{E1E3B3PlotPV2} to \ref{E1E3B3PlotPV2E30}. Again, the full three-dimensional vector fields are divergence free and thus represent flows without sources or sinks. Note that in Figure \ref{E1E3B3PlotPV2E30} we have chosen a scaling factor depending on $\mathcal B_3$ since the magnitude of the vector $\vec J_{9}^{(\vec L)\PV_2}(\vmc E,\vmc B)$ shows a strong increase towards $\mathcal B_3=0$.

The emergence of non-vanishing imaginary parts of the flux-density vector fields, for example, in the parameter space considered in the present section, see Figure \ref{E1E3B3PlotPCImaginary}, leads to an interesting phenomenon. Let us consider a closed curve $\mathcal C$, being the boundary of a surface $\mathcal F$, in $\vec L$ space (\ref{5.17}). We parametrise $\mathcal C$ as
\begin{align}\label{5.20a}
\mathcal C\hspace{-0.15em}:\; &z\rightarrow \vec L(z)\ ,\nn\\
&0\leq z\leq 1\ ,\nn\\
&\vec L(0)=\vec L(1)\ ,
\end{align}
cf. 
(\ref{3.1a}), (\ref{3.3}). We suppose that over a time interval $[0,T]$ this curve in parameter space is run through by the system in the following way:
\begin{align}\label{5.20b}
&t\rightarrow z(t)=\frac{t}{T}\ ,\nn\\
&0\leq t\leq T\ ,\nn\\
\mathcal C\hspace{-0.15em}:\; &t\rightarrow \vec L(z(t))\ .
\end{align}
We consider the adiabatic limit where $T$ becomes very large. We shall also consider that the system is run through the curve $\mathcal C$ in the reverse direction:
\begin{align}\label{5.20c}
&t\rightarrow \bar z(t)=\frac{T-t}{T}=1-z(t)\ ,\nn\\
&0\leq t\leq T\ ,\nn\\
\bar{\mathcal C}\hspace{-0.15em}:\; &t\rightarrow \vec L(\bar z(t))\ .
\end{align}

Suppose now that we have at $t=0$ the atom in the initial state $\psi_\alpha(0)$, see (\ref{2.23}). We change the parameters $\vec L$ along the curve $\mathcal C$ as in (\ref{5.20b}). From (\ref{2.23}) to (\ref{2.25}) we find the decrease of the norm of the state at time $T$ to be
\begin{align}\label{5.20d}
\frac{|\psi_\alpha(T)|^2}{|\psi_\alpha(0)|^2}=\exp\left[+2\,\mathrm{Im}\,\varphi_\alpha(T)-2\,\mathrm{Im}\,\gamma_\alpha(T)\right]\ .
\end{align}
Here $2\,\mathrm{Im}\,\varphi_\alpha(T)$ and $2\,\mathrm{Im}\,\gamma_\alpha(T)$ are the contributions due to the dynamic and geometric phases, respectively,
\begin{align}\label{5.20e}
2\,\mathrm{Im}\,\varphi_\alpha(T)&=T\,2\,\mathrm{Im}\,\int_0^1\mathrm dz\, E_\alpha(\vec L(z))\ ,\nn\\
2\,\mathrm{Im}\,\gamma_\alpha(T)&=2\,\mathrm{Im}\,\int_{\mathcal F}\vec J_\alpha^{(\vec L)}(\vec L)\,\mathrm d\vec f^{\vec L}\ ;
\end{align}
see (\ref{3.2}) and (\ref{3.20}). From (\ref{5.20d}) we can define an effective decay rate for the state $\alpha$ under the above conditions as
\begin{align}\label{5.20f}
\Gamma_{\alpha,\mathrm{eff}}(\mathcal C, T)&=\frac{1}{T}\left[-2\,\mathrm{Im}\,\varphi_\alpha(T)+2\,\mathrm{Im}\,\gamma_\alpha(T)\right]\nn\\
&=-2\,\mathrm{Im}\,\int_0^1\mathrm dz\, E_\alpha(\vec L(z))\nn\\
&\qquad+\frac{2}{T}\,\mathrm{Im}\,\int_{\mathcal F}\vec J_\alpha^{(\vec L)}(\vec L)\,\mathrm d\vec f^{\vec L}\ .
\end{align}
Note that this effective decay rate depends, of course, on the curve $\mathcal C$ and that the geometric contribution is suppressed by a factor $1/T$ relative to the dynamic contribution. From (\ref{5.20d}) and (\ref{5.20f}) we get for the decrease of the norm of the state
\begin{align}\label{5.20fa}
\frac{|\psi_\alpha(T)|^2}{|\psi_\alpha(0)|^2}=\exp\left[-\Gamma_{\alpha,\mathrm{eff}}(\mathcal C, T)\, T\right]\ .
\end{align}

Now we start again with the state $\psi_\alpha(0)$ at time $t=0$ but we change the parameters $\vec L$ along the reverse curve $\bar{\mathcal C}$ (\ref{5.20c}). It is easy to see that the dynamic term in (\ref{5.20f}) does not change whereas the geometric term changes sign,
\begin{align}\label{5.20g}
\Gamma_{\alpha,\mathrm{eff}}(\bar{\mathcal C}, T)&=-2\,\mathrm{Im}\int_0^1\mathrm dz\, E_\alpha(\vec L(z))\nn\\
&\qquad-\frac{2}{T}\,\mathrm{Im}\,\int_{\mathcal F}\vec J_\alpha^{(\vec L)}(\vec L)\,\mathrm d\vec f^{\vec L}\ .
\end{align}
Thus, the effective decay rate depends on the geometry and reversing the sense of the running through our closed curve in parameter space changes the sign of the geometric part.

\begin{figure}
\centering
\includegraphics[scale=0.6]{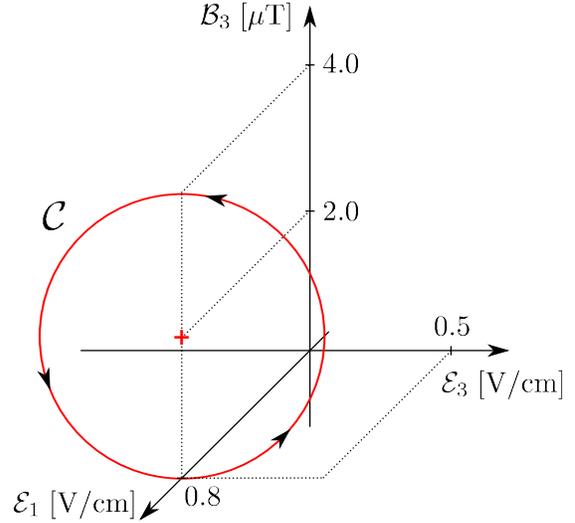}\vspace{1em}
\caption{The curve $\mathcal C$ (\ref{5.20h}) in the parameter space $\mathcal E_1,\,\mathcal E_3,\,\mathcal B_3$. The circle is run through twice.}
\label{Figure9}
\end{figure}

As a concrete example we choose a constant magnetic field $\mathcal B_2\,\vec e_2$ with $\mathcal B_2=1\,\mu\mathrm{T}$ and the following curve in $\vec L$ space
\begin{align}\label{5.20h}
\mathcal C\hspace{-0.15em}:\; z\rightarrow \vec L(z)&=\left(
\begin{array}{c}
\mathcal E_1(z)/\mathcal E_0\\
\mathcal E_3(z)/\mathcal E_0\\
\mathcal B_3(z)/\mathcal B_0
\end{array}
\right)\ ,\nn\\
\mathcal E_1(z)&=0.8\,\mathrm{V/cm}\ ,\nn\\
\mathcal E_3(z)&=0.5\times\cos(4\,\pi\,z)\,\mathrm{V/cm}\ ,\nn\\
\mathcal B_3(z)&=[2+2\times\sin(4\,\pi\,z)]\,\mu\mathrm T\ ,\nn\\
0&\leq z\leq 1\ ;
\end{align}
see Figure \ref{Figure9}.
We suppose as in (\ref{5.20b}) that $\mathcal C$ is run through in a time $T$ with $z(t)=t/T$. In this time the path in parameter space makes, according to (\ref{5.20h}), two loops. With $T=1\, \mathrm{ms}$ we can meet the adiabaticity requirements as spelt out in \cite{BeGaMaNaTr08_I} and in Equations (30), (31), and (37) of \cite{BeGaNa07_II}. The essential requirement here is that the frequency $\nu=2/T$ of the external field variation in (\ref{5.20h}) must be much less than the transition frequencies $\Delta E/h$ between the Zeeman levels for $\alpha=9,10,11$. For the external field of order $1\,\mu\mathrm{T}$ we get $\Delta E/h\simeq 10\,\mathrm{kHz}$ which gives the requirement
\begin{align}
\nu=\frac{2}{T}&\ll 10\,\mathrm{kHz}\ ,\nn\\
T&\gg 0.2\,\mathrm{ms}\ .
\end{align}
Calculating now the contributions to the effective decay rate (\ref{5.20f}) we find for the state $\alpha=9$ which is connected to the 2S state with $(F,F_3)=(1,1)$, see Appendix \ref{s:AppendixA}, the following
\begin{align}\label{5.20i}
-2\,\mathrm{Im}\int_0^1\mathrm dz\, E_9(\vec L(z))&= 1935.2\,\mathrm{s}^{-1}\ ,\\
\frac{2}{T}\,\mathrm{Im}\int_{\mathcal F}\vec J_9^{(\vec L)}(\vec L)\,\mathrm d\vec f^{\vec L}&= -1.8\,\mathrm{s}^{-1}\ .
\end{align}
This leads to
\begin{align}\label{5.20j}
\Gamma_{9,\mathrm{eff}}(\mathcal C, T)= 1933.4\,\mathrm{s}^{-1}\ .
\end{align}
For the reverse curve $\bar{\mathcal C}$ we find, instead,
\begin{align}\label{5.20k}
\Gamma_{9,\mathrm{eff}}(\bar{\mathcal C}, T)= 1937.0\,\mathrm{s}^{-1}\ .
\end{align}
Thus, under the above conditions the effective decay rates (\ref{5.20j}) and (\ref{5.20k}) differ by $1.9\,\tcperthousand$ and the corresponding decreases of the norms (\ref{5.20fa}) by $3.6\,\tcperthousand$. We emphasise that this difference has its origin in the geometric phase.

To give an example of a PV geometric phase we consider the following curve
\begin{align}\label{5.20l}
\mathcal C'\hspace{-0.15em}:\; z\rightarrow \vec L(z)&=\left(
\begin{array}{c}
\mathcal E_1(z)/\mathcal E_0\\
\mathcal E_3(z)/\mathcal E_0\\
\mathcal B_3(z)/\mathcal B_0
\end{array}
\right)\ ,\nn\\
\mathcal E_1(z)&=0\ ,\nn\\
\mathcal E_3(z)&=\mathcal E_0\,\sin(2\,\pi\,z)\ ,\nn\\
\mathcal B_3(z)&=0.1\,\mathcal B_0\,\cos(2\,\pi\,z)\ ,\nn\\
0&\leq z\leq 1\ .
\end{align}
For this curve the PC geometric phases vanish due to the antisymmetry of $\vec e_1\cdot\vec J_\alpha^{(\vmc E)\PC}(\vmc E,\mathcal B_3\,\vec e_3)$ under $(\mathcal E_1,\mathcal E_3)\rightarrow (-\mathcal E_1,-\mathcal E_3)$; see (\ref{5.18}). Thus, we get here
\begin{align}\label{5.20m}
\gamma_\alpha(\mathcal C')=\gamma_\alpha^{\PV}(\mathcal C')=\gamma_\alpha^{\PV_1}(\mathcal C')+\gamma_\alpha^{\PV_2}(\mathcal C')\ .
\end{align}
Numerically we find for $\alpha=9$ and $\mathcal B_2=1\,\mu\mathrm{T}$
\begin{align}\label{5.20n}
\gamma_9^{\PV_1}(\mathcal C')&=(0.00467-0.000457\,\I)\,\delta_1\ ,\\
\gamma_9^{\PV_2}(\mathcal C')&=(0.0942-0.00421\,\I)\,\delta_2\ .
\end{align}
Assuming now that we can circle the curve $\mathcal C'$ $N$ times we get as geometric phase $N\gamma_9(\mathcal C')$. Thus, the number of circlings acts as an enhancement factor for the small weak interaction effects in hydrogen. For $N=10^4$, for example, we obtain
\begin{align}\label{5.20o}
N\gamma_9^{\PV}(\mathcal C')&=(46.7-4.57\,\I)\,\delta_1+(942-42.1\,\I)\,\delta_2\ .
\end{align}
With $\delta_{1,2}$ from Table \ref{t:valuesFD} this gives a phase of the order of $10^{-9}$.

\section{Conclusions and outlook}\label{s:6}

In this article we have discussed the geometric phases and flux densities of the metastable states of hydrogen with principal quantum number $n=2$ in the presence of external electric and magnetic fields. We have provided expressions for the flux densities and their derivatives suited for an investigation of their general structure. This was achieved with the help of representations of the flux densities as complex integrals. For these integrals extensive use was made of resolvent methods, and the results turned out to be quite simple and easy to handle. Furthermore, employing proper and improper rotations we derived the general structure of the flux densities for the metastable states. We also obtained expansions of the flux densities in terms of P-conserving and P-violating contributions. The flux densities can be visualised in the case of three-dimensional parameter spaces as vector fields. We gave three examples of parameter spaces for which we compared analytical with numerical calculations. The results are consistent regarding the employed numerical precision. For vanishing electric field the flux densities in magnetic field space are real and P-conserving -- and so are the corresponding geometric phases. In this case the flux-density vector field can be computed analytically and turns out to be the field of a Dirac monopole sitting at $\vmc B=0$. In parameter spaces of both electric and magnetic field components the flux densities exhibit a rich structure of P-conserving as well as P-violating vector fields including both real and imaginary parts. In the general structures of the flux densities of those cases we encounter functions which are rotationally invariant; see (\ref{4.35}) to (\ref{4.37}). These functions contain all the information on the mass matrix and the electric and magnetic dipole matrices which one can obtain from the measurement of the geometric phases for the system considered. 

In Section \ref{s:5} we have calculated geometric phases for various situations and, surely, the question arises about their possible measurements. Present experiments can\linebreak[4] reach a precision in phase measurements of about $10^{-5}\,\mathrm{rad}$ \cite{DeKprivcom2}. Thus, the PC phases (\ref{92b}), (\ref{101phase1}), (\ref{101phase2}) and the change of decay rate (\ref{5.20j}) and (\ref{5.20k}) should be within reach of these experiments. The PV phases for hydrogen certainly need further theoretical and experimental efforts to bring them to a practically measurable level.

The general representations for the flux densities as complex integrals are, of course, easily transferred to other atomic systems having stable or metastable states, for instance, to the states of $n=2$ helium. Similarly, the analysis of the consequences of rotational invariance and of P violation in Section \ref{s:4} goes through unchanged for other atomic systems. The only requirement here is that the coupling of the system to the external electric and magnetic fields can be described as in (\ref{2.2}) with electric and magnetic dipole matrices $\uvec D$ and $\uvec\mu$, respectively. We note that measuring geometric phases for such systems can give valuable information on their atomic matrix elements. We have found for hydrogen that, on the one hand, there is the case of a pure magnetic field where the flux densities and geometric phases are simply given by the geometry of the path run through in parameter space; see Section \ref{ss:5.1}. On the other hand, a large sensitivity on the electric dipole matrix element was found for the flux densities in electric field space with the presence of a constant magnetic field in Section \ref{ss:5.2}. We also discussed the change of the effective decay rates for geometric reasons in Section \ref{ss:5.3}. All these phenomena should also occur for other atomic systems. Measurements of geometric phases could be an important testing ground for theoretical calculations of wave functions and matrix elements for atomic systems, for instance, for He.

\subsection*{Acknowledgements}

The authors would like to thank $\mbox{M. DeKieviet}$, $\mbox{G. Lach}$, $\mbox{P. Schmelcher}$, and $\mbox{A. Surzhykov}$ for useful discussions. Special thanks are due to $\mbox{M. Diehl}$ for providing us the information on the current status of the strange-quark contribution to the spin of the proton.\\
This project is partially funded by the Klaus Tschira Foundation gGmbH and supported by the Heidelberg Graduate School of Fundamental Physics and the Deutsche Forschungsgemeinschaft.


\appendix

\section*{Appendices}
\section{The $n=2$ states of hydrogen}\label{s:AppendixA}

\renewcommand{\theequation}{\thesection.\arabic{equation}}
 \setcounter{equation}{0}

In this appendix we collect the numerical values for the quantities entering our calculations for the hydrogen states with principal quantum number $n=2$. We specify our numbering scheme for these states. The expressions for the mass matrix at zero external fields and for the electric and the magnetic dipole operators are given in Appendix \ref{s:AppendixE}.

In Table \ref{t:valuesFD} we present for $^1_1$H, where the nuclear spin is $I=1/2$, the numerical values for the weak charges $Q_W^{(\varkappa)}$, $\varkappa=1,2$, the quantities $\Delta q+\Delta \bar q$, the Lamb shift $L = E(2\mathrm S_{1/2})-E(2\mathrm P_{1/2})$, the fine structure splitting $\Delta = E(2\mathrm P_{3/2}) - E(2\mathrm P_{1/2})$, and the ground state hyperfine splitting energy $\mcal A$. We have $\mcal A = E(1\mathrm S_{1/2},F=1)-E(1\mathrm S_{1/2},$ $F=0)$ for hydrogen. We define the weak charges as in Section~2 of \cite{BeGaNa07_II} which gives for the proton in the standard model (SM):
\begin{align}\label{A0a}
Q_W^{(1)}&=1-4\sin^2\vartheta_W\ ,\nn\\
Q_W^{(2)}&=-2\,(1-4\sin^2\vartheta_W)\nn\\
&\quad\times(\Delta u+\Delta \bar u-\Delta d-\Delta \bar d-\Delta s-\Delta \bar s)\ .
\end{align}
Here $\vartheta_W$ is the weak mixing angle and $\Delta q+\Delta \bar q$ denotes the total polarisation of the proton carried by the quarks and antiquarks of species $q$ $(q=u,d,s)$. Note that in \cite{BeGaNa07_II} and \cite{BoBrNa95} we adhered to the then usual notation of $\Delta q$ for what is now denoted as $\Delta q+\Delta \bar q$. The quantity $\Delta u+\Delta \bar u-\Delta d-\Delta \bar d$ is related to the ratio $g_A/g_V$ from neutron $\beta$ decay:
\begin{align}\label{A0a1}
\Delta u+\Delta \bar u-\Delta d-\Delta \bar d=-g_A/g_V\ .
\end{align}
The numerical value given in Table \ref{t:valuesFD} is from \cite{PDG10}. The total polarisation of the proton carried by strange quarks, $\Delta s+\Delta \bar s$, is still only poorly known experimentally. One finds values of $-0.12$ to very small and positive ones quoted in recent papers; see for instance \cite{PhysRevC.78.015207,PhysRevLett.101.072001,PhysRevD.80.034030,Alekseev2010227,Leader2011}. Therefore, we assume for our purposes
\begin{align}\label{A0a2}
-0.12\leq\Delta s+\Delta \bar s\leq 0\ .
\end{align}
Of course, the dependence of $Q_W^{(2)}$ on $\Delta s+\Delta \bar s$ is, in principle, very interesting, since this quantity can be determined in atomic P violation experiments with hydrogen.

We define (see (19) of \cite{BeGaNa07_II}) the dimensionless constants
\begin{align}\label{A0b}
\delta_\varkappa=-\frac{\sqrt{3}\,G}{64\pi\sqrt 2\, r_B^4 m_e}\frac{Q_W^{(\varkappa)}}{L}\qquad (\varkappa=1,2)
\end{align}
with Fermi's constant $G$, the Bohr radius $r_B$ and the electron mass $m_e$. We see from Table \ref{t:valuesFD} that varying $\Delta s+\Delta \bar s$ in the range (\ref{A0a2}) corresponds to a $10\%$ shift in $\delta_2$. Thus, a percent-level measurement of $\delta_2$ would be most welcome for a clarification of the role of strange quarks for the nucleon spin.
\bgroup
\tabcolsep=0.2em
\begin{table}[Hb!]
\begin{center}
\begin{tabular}{@{\hspace{0em}}c@{\hspace{0em}}c@{\hspace{1.0em}}c@{\hspace{1.0em}}c@{\hspace{0em}}}
\hline\noalign{\smallskip}
 		 & \multicolumn{2}{c}{${}_1^1$H} 					& Ref.\\   
\noalign{\smallskip}\hline\noalign{\smallskip}
$L/h$ 	 & \multicolumn{2}{r}{1057.8440(24) MHz\phantom{aaa}} 	&  \cite{NISTData}\\
$\Delta/h$  & \multicolumn{2}{r}{10969.0416(48) MHz\phantom{aaa}} 	& \cite{NISTData}\\
$\mcal A/h$ & \multicolumn{2}{r}{1420.405751768(1) MHz\phantom{aaa}} 	& \cite{Kar05}\\   
\noalign{\smallskip}\hline\noalign{\smallskip}
$Q_W^{(1)}$ & \multicolumn{2}{c}{0.04532(64)} 					& (11) of \cite{BeGaNa07_II}\\ 
$\delta_1$  & \multicolumn{2}{c}{$-2.78(4)\times 10^{-13}$} 			& (20) of \cite{BeGaNa07_II}\\   
\noalign{\smallskip}\hline\noalign{\smallskip}
\mycell{18mm}{$\Delta u+\Delta \bar u$\\[-0.0ex]$-\Delta d-\Delta \bar d$} & \multicolumn{2}{c}{$1.2694(28)$} 			&  \cite{PDG10}\\
\noalign{\smallskip}\hline\hline\noalign{\smallskip}
$\Delta s+\Delta \bar s$  & $-0.12$ 	& $0.00$ & (\ref{A0a2}) \\
\noalign{\smallskip}
$Q_W^{(2)}$ & $-0.1259(18)$		& $-0.1151(16)$ & \multirow{2}{*}{(\ref{A0a})-(\ref{A0b})}\\
\noalign{\smallskip}
$\delta_2$  & $7.74(11)\times 10^{-13}$	& $7.07(10)\times 10^{-13}$ & \\   
\noalign{\smallskip}\hline\noalign{\smallskip}
\end{tabular}
\caption{Values of parameters for $^1_1$H for numerical calculations. The weak mixing angle in the low energy limit, $\sin^2\vartheta_W = 0.23867(16)$, is taken from \cite{ErRa05}. The uncertainty in $\delta_1$ is dominated by the uncertainty of $\sin^2\vartheta_W$. The uncertainties quoted for $Q_W^{(2)}$ and $\delta_2$ for hydrogen are resulting from the error of the weak mixing angle. The variation of $Q_W^{(2)}$ and $\delta_2$ with $(\Delta s+\Delta \bar s)$ varying in the range (\ref{A0a2}) is given explicitly.}
\label{t:valuesFD}
\end{center}
\end{table}
\egroup

The mass matrix for zero external fields is given by
\begin{align}\label{A0c}
\umat{\tilde M}_0=\umat{M}_0+\delta_1\,\umat{M}_{\PV}^{(1)}+\delta_2\,\umat{M}_{\PV}^{(2)}\ ;
\end{align}
see (22) to (24) of \cite{BeGaNa07_II} and Table \ref{t:H2.M0} in Appendix \ref{s:AppendixE} below. The terms $\umat{M}_{\PV}^{(1)}$ and $\umat{M}_{\PV}^{(2)}$ correspond to the nuclear-spin independent and dependent PV interaction, respectively. As in (C.31) ff. of \cite{BeGaNa07_II} we define
\begin{align}\label{A0d}
\delta&=(\delta_1^2+\delta_2^2)^{\frac12}\ ,\nn\\
\umat{M}_{\PV}&=\sum_{i=1}^2\frac{\delta_i}{\delta}\umat{M}_{\PV}^{(i)}\ .
\end{align}

The $n=2$ states of hydrogen in the absence of P violation and for zero external fields are denoted by $|2L_J,$ $F,F_3)$, where $L$, $J$, $F$ and $F_3$ are the quantum numbers of the electron's orbital angular momentum, its total angular momentum, the total atomic angular momentum and its third component, respectively.

To give the matrices $\umat{M}_0$, $\umat{M}_{\PV}^{(1)}$, $\umat{M}_{\PV}^{(2)}$, $\uvec D$, and $\uvec\mu$ explicitly we use the following procedure. The hermitian part of $\umat{M}_0$, that is, $(\umat{M}_0+\umat{M}_0^\dagger)/2$ is given by the known energy levels of the $n=2$ hydrogen states; see Table \ref{t:valuesFD}. The decay matrix $\underline{\Gamma}=\I(\umat{M}_0-\umat{M}_0^\dagger)$ needs a little discussion. We have (for atoms at rest), see for instance (3.13) of \cite{BoBrNa95},
\begin{align}\label{A0e}
&\rbra{2L'_{J'},F',F'_3}\underline{\Gamma}\rket{2L_{J},F,F_3}=2\pi\sum_X\bokp{X}{\mathcal T}{2L'_{J'},F',F'_3}^*\nn\\
&\qquad\qquad\qquad\quad\times\delta(E_X-E_2)\bokp{X}{\mathcal T}{2L_{J},F,F_3}\ .
\end{align}
Here, $\ket{X}$ denotes the decay states, the atom in a $n=1$ state plus photons, and $\mathcal T$ is the transition matrix. Rotational invariance tells us immediately that the matrix $\underline{\Gamma}$ must be diagonal in $(F,F_3)$. We shall neglect P violation in the decay. Then the non-diagonal matrix elements between S states $(L=0)$ and P states $(L=1)$ must be zero. The only non-diagonal matrix elements (\ref{A0e}) which could be non-zero are, therefore, those for $L=L'=1$, $F=1$, and $(J',J)=(1/2,3/2)$ or $(J',J)=(3/2,1/2)$, respectively. But calculating these matrix elements inserting the usual formulae for E1 transitions on the r.h.s. of (\ref{A0e}) we get zero. Thus, neglecting higher order corrections, the matrix $\underline{\Gamma}$ (\ref{A0e}) is diagonal. The numerical values for the diagonal elements of $\underline{\Gamma}$ are taken from \cite{Sap04,LaShSo05}.

To calculate the matrices $\umat{M}_{\PV}^{(1)}$, $\umat{M}_{\PV}^{(2)}$, $\uvec D$, and $\uvec\mu$ we use the standard Coulomb wave functions for hydrogen. As in \cite{BoBrNa95} (see Appendix B there) we use for these states the phase conventions of \cite{CoSh63} except for an overall sign change in all radial wave functions. The matrices $\umat{\tilde M}_0$, $\underline D_j$ and $\underline \mu_j$ in this basis are collected in Tables \ref{t:H2.M0} to \ref{t:H2.Mu} of Appendix \ref{s:AppendixE} (online material only).

Now we discuss the properties, the ordering and the numbering of the eigenstates of $\umat M(\vmc E,\vmc B)$ as given in (\ref{2.2}).

We are interested only in moderate magnetic fields, that is, we want to stay below the first level crossing in the Breit-Rabi diagram, which implies
\begin{align}\label{A4a}
|\vmc B|< 53.8\,\mathrm{mT}\ .
\end{align}
Note that these crossings are only in the real part of the eigenenergies; see (\ref{2.12}). In the region (\ref{A4a}) degeneracies of the \emph{complex} energies (\ref{2.12}) occur for $\vmc B=\vec 0$ at arbitrary $\vmc E$. This is a consequence of time-reversal (T) invariance. For $\vmc B=\vec 0$ and $\vmc E\not= \vec 0$ we can choose the vector $\vec e_3'=\vmc E/|\vmc E|$ as quantisation axis of angular momentum. Then $F_3'$ is a good quantum number and time reversal invariance implies that there are corresponding eigenstates of $\umat M(\vmc E,\vec 0)$ with quantum numbers $F_3'$ and $-F_3'$ and having the same complex eigenenergies. See Sections 3.3 and 3.4 of \cite{BoBrNa95} for a proof of this result using resolvent methods. Thus, we have degeneracies of the complex eigenenergies of $\umat M(\vmc E,\vmc B)$ (\ref{2.2}) in the parameter subspace
\begin{align}\label{A4b}
\{(\vmc E,\vmc B)\,;\;\vmc E\mbox{ arbitrary, } \vmc B=\vec 0)\}\ .
\end{align}
By numerical methods we checked that, at least for moderate $\vmc B$ fields (\ref{A4a}), there are no further degeneracy points or regions.

{

\begin{table}[htbp]
\begin{center}
\begin{tabular}{cl}
\cline{1-2}\noalign{\smallskip}
\multicolumn{2}{c}{hydrogen}\\ 
\noalign{\smallskip}\cline{1-2}\noalign{\smallskip}
$\alpha$ & $\rket{2\hat L_J,F,F_3,\vmc E,\vmc B}$\\ 
\noalign{\smallskip}\cline{1-2}\noalign{\smallskip}
1 & $\rket{2\hat P_{3/2},2,\phm2,\vmc E,\vmc B}$\\
2 & $\rket{2\hat P_{3/2},2,\phm1,\vmc E,\vmc B}$\\
3 & $\rket{2\hat P_{3/2},2,\phm0,\vmc E,\vmc B}$\\
4 & $\rket{2\hat P_{3/2},2,-1,\vmc E,\vmc B}$   \\
5 & $\rket{2\hat P_{3/2},2,-2,\vmc E,\vmc B}$   \\
6 & $\rket{2\hat P_{3/2},1,\phm1,\vmc E,\vmc B}$\\
7 & $\rket{2\hat P_{3/2},1,\phm0,\vmc E,\vmc B}$\\
8 & $\rket{2\hat P_{3/2},1,-1,\vmc E,\vmc B}$   \\ 
\noalign{\smallskip}\cline{1-2}\noalign{\smallskip}
9 & $\rket{2\hat S_{1/2},1,\phm1,\vmc E,\vmc B}$\\
10 & $\rket{2\hat S_{1/2},1,\phm0,\vmc E,\vmc B}$\\
11 & $\rket{2\hat S_{1/2},1,-1,\vmc E,\vmc B}$   \\
12 & $\rket{2\hat S_{1/2},0,\phm0,\vmc E,\vmc B}$\\
\noalign{\smallskip}\cline{1-2}\noalign{\smallskip}
13 & $\rket{2\hat P_{1/2},1,\phm1,\vmc E,\vmc B}$\\ 
14 & $\rket{2\hat P_{1/2},1,\phm0,\vmc E,\vmc B}$\\ 
15 & $\rket{2\hat P_{1/2},1,-1,\vmc E,\vmc B}$   \\ 
16 & $\rket{2\hat P_{1/2},0,\phm0,\vmc E,\vmc B}$\\ 
\noalign{\smallskip}\cline{1-2} 
\end{tabular}
\caption{The numbering scheme for the atomic $n=2$ states of hydrogen.}
\label{t:state.labelsFD}\vskip10pt
\end{center}
\end{table}

The eigenstates of $\umat M$ (\ref{2.2}) for electric field $\vmc E$ and magnetic field $\vmc B$ equal to zero are the free 2S and 2P states. We write $\hat L$, $\hat P$, $\hat S$ since these states include the parity mixing due to $H_{\PV}$, see (1) of \cite{BeGaNa07_II}. Thus, the eigenstates of the mass matrix (\ref{2.2}), including the PV part but with electric and magnetic fields equal to zero, will be denoted by $\rket{2\hat L_J, F,F_3,\vmc E=\vec 0,\vmc B=\vec 0}$. The corresponding states for the mass matrix without the PV term, that is, with $\umat{\tilde M}_0$ replaced by $\umat M_0$ in (\ref{2.2}), will be denoted by $\rket{2 L_J, F,F_3,\vmc E=\vec 0,\vmc B=\vec 0}$. But it is not convenient to start a numbering scheme at the degeneracy point $(\vmc E,\vmc B)=(\vec 0,\vec 0)$. Therefore, we consider first atoms in a constant $\vmc B$-field pointing in positive 3-direction,
\begin{align}\label{eA:B3FD}
\vmc B = \mcal B\vec e_3\ ,\quad \mcal B> 0\ .
\end{align}
The corresponding eigenstates, denoted by $|2\hat L_J, F,F_3,0,$ $\mcal B\vec e_3)$, and the corresponding quasi projectors (\ref{2.14}) of $\umat M$ in (\ref{2.2}) are obtained from those at $\mcal B=0$ by continuously turning on $\vmc B$ in the form (\ref{eA:B3FD}). Of course, for $\mathcal B>0$, $F_3$ still is a good quantum number but this is no longer true for $F$. The latter is merely a label for the states.
We now choose a reference field $\vmc B_\mathrm{ref}=\mcal B_\mathrm{ref}\vec e_3$, $\mcal B_\mathrm{ref}>0$, below the first crossings in the Breit-Rabi diagram, for instance $\mcal B_{\mathrm{ref}} = 0.05\,\mathrm{mT}$. We are then at a no-degeneracy point and number the $n=2$ states and quasi projectors (\ref{2.14}) with $\alpha=1,\dots,16$ as shown in Table \ref{t:state.labelsFD} setting there $(\vmc E,\vmc B)=(\vec 0,\vmc B_\mathrm{ref})$. In the next step we consider external fields of the form
\begin{align}\label{A5a}
\vmc E'=\left(
\begin{array}{c}
\mathcal E_1\\
0\\
\mathcal E_3
\end{array}\right)\quad ,\quad
\vmc B'=\left(
\begin{array}{c}
0\\
0\\
\mathcal B'
\end{array}\right)\quad ,\quad
\mathcal B'>0\ ,
\end{align}
and a continuous path to these fields from the reference point $(\vmc E,\vmc B)=(\vec 0,\vmc B_\mathrm{ref})$:
\begin{align}\label{A1}
\vmc E'(\lambda)&=\lambda\vmc E'\ ,\nn\\
\vmc B'(\lambda)&=\vmc B_\mathrm{ref}+\lambda(\vmc B'-\vmc B_\mathrm{ref})\ ,\nn\\
\lambda&\in[0,1]\ .
\end{align}
Since we encounter no degeneracies for $\lambda\in[0,1]$ the energy eigenvalues as well as the quasi projectors are continuous functions of $\lambda$ there. This allows us to carry over the numbering of the quasi projectors from $(\vec 0,\vmc B_\mathrm{ref})$ to all fields of the form (\ref{A5a}).

Finally, we consider arbitrary fields $(\vmc E,\vmc B)$ with $\vmc B\not=\vec 0$. We can always find a proper rotation $R$ such that
\begin{align}\label{A6a}
 R\vmc E=\vmc E'\quad , \quad  R\vmc B=\vmc B'
\end{align}
with $(\vmc E',\vmc B')$ of the form (\ref{A5a}). From the considerations of the resolvent in Section \ref{ss:4.1} we conclude that the eigenvalues of $\umat M(\vmc E,\vmc B)$ and $\umat M(\vmc E',\vmc B')$ are equal. There are also no degeneracies here and, therefore, we can unambiguously carry over the numbering of eigenvalues and quasi projectors from the case $(\vmc E',\vmc B')$ to the case $(\vmc E,\vmc B)$. The labels $\alpha=1,\dots,16$ in Table \ref{t:state.labelsFD} for arbitrary $(\vmc E,\vmc B)$ with $\vmc B\not=\vec 0$ correspond to this identification procedure of eigenenergies and quasi projectors. The corresponding eigenstates $\rket{\alpha,\vmc E,\vmc B}$ of $\umat M(\vmc E,\vmc B)$ are defined as the eigenstates of the quasi projectors
\begin{align}
\mathbbm P_\alpha(\vmc E,\vmc B)\rket{\alpha,\vmc E,\vmc B}=\rket{\alpha,\vmc E,\vmc B}
\end{align}
where we also require (\ref{2.11}) to hold. This fixes for given $\alpha$, $\vmc E$, $\vmc B$, the state vector up to a phase factor. In all considerations of flux densities only the quasi projectors enter and thus, such a phase factor in the states is irrelevant. The choice of phase factor \emph{is} relevant for the calculation of the geometric phases via the line integrals (\ref{2.25}) and (\ref{3.2}). Then, we always make sure to choose a phase factor being differentiable along the path considered.

Finally we note that for the case of no P violation, that is for $\delta=0$, the numbering of the quasi projectors and the states $\rket{2 L_J, F,F_3,\vmc E,\vmc B}$ is done in a completely analogous way.

\section{Relations for the geometric flux densities}\label{s:AppendixB}
\setcounter{equation}{0}

In this appendix we derive the representations (\ref{3.14}), (\ref{3.15}) and (\ref{3.16}) for $Y_{\alpha,ab}(K)$ and its derivatives, respectively. In the following we will omit the $K$-dependence of all quantities for abbreviation. We now consider the expression
\begin{align}\label{B1}
X_{\alpha,ab}&:=\frac{\I}{2}\frac{1}{2\pi\I}\sum_{\beta,\gamma}\oint_{S_\alpha}\mathrm d\zeta\,
\frac{\mathrm{Tr}\Big[\mathbbm P_\beta\frac{\partial\umat M}{\partial K_a}\mathbbm P_\gamma\frac{\partial\umat M}{\partial K_b}\Big]}{(\zeta-E_\beta)(\zeta-E_\gamma)^2}\ .
\end{align}
According to the residue theorem the integral vanishes for $\beta=\gamma=\alpha$ since a pole of third order at $\zeta=E_\alpha$ gives a residual of zero
\begin{align}\label{B2}
\oint_{S_\alpha}\mathrm d\zeta\,\frac{1}{(\zeta-E_\alpha)^3}=2\pi\I\,\mathrm{Res}(\frac{1}{(\zeta-E_\alpha)^3};\zeta=E_\alpha)=0\ .
\end{align}
Let $D\subset\mathbbm C$ be a simply connected set with $S_\alpha$ entirely inside $D$ and $E_\sigma\notin D$ for all $\sigma\not=\alpha$; see Figure \ref{ComplexZetaPlane}. Therefore, for $\beta\not=\alpha$ and $\gamma\not=\alpha$ the integrand in (\ref{B1}) is analytic on $D$, and the integral in (\ref{B1}) vanishes due to Cauchy's integral theorem. The only two remaining cases $\beta=\alpha$ and $\gamma\not=\alpha$ as well as $\beta\not=\alpha$ and $\gamma=\alpha$ can be treated using again the residue theorem. We find easily
\begin{align}\label{B3}
X_{\alpha,ab}&=\frac{\I}{2}\sum_{\gamma\not=\alpha}\frac{1}{(E_\alpha-E_\gamma)^2}\mathrm{Tr}\Big[\mathbbm P_\alpha\frac{\partial\umat M}{\partial K_a}\mathbbm P_\gamma\frac{\partial\umat M}{\partial K_b}\Big]\nn\\
&\quad+\frac{\I}{2}\sum_{\beta\not=\alpha}\frac{-1}{(E_\alpha-E_\beta)^2}\mathrm{Tr}\Big[\mathbbm P_\beta\frac{\partial\umat M}{\partial K_a}\mathbbm P_\alpha\frac{\partial\umat M}{\partial K_b}\Big]\nn\\
&=\frac{\I}{2}\sum_{\beta\not=\alpha}\frac{1}{(E_\alpha-E_\beta)^2}\mathrm{Tr}\Big[\mathbbm P_\alpha\frac{\partial\umat M}{\partial K_a}\mathbbm P_\beta\frac{\partial\umat M}{\partial K_b}\Big]\nn\\
&\quad+\frac{\I}{2}\sum_{\beta\not=\alpha}\frac{-1}{(E_\alpha-E_\beta)^2}\mathrm{Tr}\Big[\mathbbm P_\alpha\frac{\partial\umat M}{\partial K_b}\mathbbm P_\beta\frac{\partial\umat M}{\partial K_a}\Big]\nn\\
&=\frac{\I}{2}\sum_{\beta\not=\alpha}\frac{1}{(E_\alpha-E_\beta)^2}\mathrm{Tr}\Big[\mathbbm P_\alpha\frac{\partial\umat M}{\partial K_a}\mathbbm P_\beta\frac{\partial\umat M}{\partial K_b}\Big]-(a\leftrightarrow b)
\end{align}
which is exactly $Y_{\alpha,ab}$, see (\ref{3.13}). Thus, we obtain the integral representation (\ref{3.14}) for $Y_{\alpha,ab}$
\begin{align}\label{B4}
Y_{\alpha,ab}&=X_{\alpha,ab}\nn\\
&=\frac{\I}{2}\frac{1}{2\pi\I}\oint_{S_\alpha}\mathrm d\zeta\,
\mathrm{Tr}\Big[\left(\sum_{\beta}\frac{\mathbbm P_\beta}{\zeta-E_\beta}\right)\frac{\partial\umat M}{\partial K_a}\nn\\
&\qquad\qquad\qquad\qquad\times\left(\sum_{\gamma}\frac{\mathbbm P_\gamma}{(\zeta-E_\gamma)^2}\right)\frac{\partial\umat M}{\partial K_b}\Big]\nn\\
&=\frac{\I}{2}\frac{1}{2\pi\I}\oint_{S_\alpha}\mathrm d\zeta\,
\mathrm{Tr}\Big[\frac{1}{\zeta-\umat M}\frac{\partial\umat M}{\partial K_a}\frac{1}{(\zeta-\umat M)^2}\frac{\partial\umat M}{\partial K_b}\Big]
\end{align}
where we use the relation (\ref{2.17}) for the quasi projectors in the last step.

In order to calculate the derivatives of $Y_{\alpha,ab}$ we first derive some useful relations:
\begin{align}
&0=\frac{\partial}{\partial K_a}\mathbbm 1=\frac{\partial}{\partial K_a}\left[(\zeta-\umat M)^{-1}(\zeta-\umat M)\right]\nn\\
&=\frac{\partial}{\partial K_a}\left[(\zeta-\umat M)^{-1}\right](\zeta-\umat M)\nn\\
&\qquad+(\zeta-\umat M)^{-1}\frac{\partial}{\partial K_a}\left[\zeta-\umat M\right]\nn\\
&\Leftrightarrow\;\;\frac{\partial}{\partial K_a}\frac{1}{\zeta-\umat M}=\frac{1}{\zeta-\umat M}\frac{\partial\umat M}{\partial K_a}\frac{1}{\zeta-\umat M}\ ,\label{B5}\\
&0=\frac{\partial}{\partial K_a}\mathbbm 1=\frac{\partial}{\partial K_a}\left[(\zeta-\umat M)^{-2}(\zeta-\umat M)^2\right]\nn\\
&=\frac{\partial}{\partial K_a}\left[(\zeta-\umat M)^{-2}\right](\zeta-\umat M)^2\nn\\
&\quad+(\zeta-\umat M)^{-2} \frac{\partial}{\partial K_a}\left[(\zeta-\umat M)^2\right]\nn\\
\Leftrightarrow&\;\;\frac{\partial}{\partial K_a}\left[(\zeta-\umat M)^{-2}\right]=(\zeta-\umat M)^{-2}\nn\\
&\quad\times\left(\frac{\partial\umat M}{\partial K_a}(\zeta-\umat M)+(\zeta-\umat M)\frac{\partial\umat M}{\partial K_a}\right)(\zeta-\umat M)^{-2}\nn\\
\Leftrightarrow&\;\;\frac{\partial}{\partial K_a}\frac{1}{(\zeta-\umat M)^2}=\frac{1}{(\zeta-\umat M)^2}\frac{\partial\umat M}{\partial K_a}\frac{1}{\zeta-\umat M}\nn\\
&\quad+\frac{1}{\zeta-\umat M}\frac{\partial\umat M}{\partial K_a}\frac{1}{(\zeta-\umat M)^2}\ .\label{B6}
\end{align}

With (\ref{B5}) and (\ref{B6}) we obtain from (\ref{B4})
\begin{align}\label{B7}
&\frac{\partial}{\partial K_a}Y_{\alpha,bc}=\frac{\I}{2}\frac{1}{2\pi\I}\oint_{S_\alpha}\mathrm d\zeta\Big\{\,\nn\\
&\mathrm{Tr}\Big[\frac{1}{\zeta-\umat M}\frac{\partial^2\umat M}{\partial K_a\partial K_b}\frac{1}{(\zeta-\umat M)^2}\frac{\partial\umat M}{\partial K_c}\Big]\nn\\
&+\mathrm{Tr}\Big[\frac{1}{\zeta-\umat M}\frac{\partial\umat M}{\partial K_b}\frac{1}{(\zeta-\umat M)^2}\frac{\partial^2\umat M}{\partial K_a\partial K_c}\Big]\nn\\
&+\mathrm{Tr}\Big[\frac{1}{\zeta-\umat M}\frac{\partial\umat M}{\partial K_a}\frac{1}{\zeta-\umat M}\frac{\partial\umat M}{\partial K_b}\frac{1}{(\zeta-\umat M)^2}\frac{\partial\umat M}{\partial K_c}\Big]\nn\\
&+\mathrm{Tr}\Big[\frac{1}{\zeta-\umat M}\frac{\partial\umat M}{\partial K_b}\frac{1}{(\zeta-\umat M)^2}\frac{\partial\umat M}{\partial K_a}\frac{1}{\zeta-\umat M}\frac{\partial\umat M}{\partial K_c}\Big]\nn\\
&+\mathrm{Tr}\Big[\frac{1}{\zeta-\umat M}\frac{\partial\umat M}{\partial K_b}\frac{1}{\zeta-\umat M}\frac{\partial\umat M}{\partial K_a}\frac{1}{(\zeta-\umat M)^2}\frac{\partial\umat M}{\partial K_c}\Big]\Big\}\ .
\end{align}
Using the cyclicity of the trace and performing partial integrations of the second and fourth summand in (\ref{B7}) we get
\begin{align}\label{B7b}
&\frac{\partial}{\partial K_a}Y_{\alpha,bc}=\frac{\I}{2}\frac{1}{2\pi\I}\oint_{S_\alpha}\mathrm d\zeta\Big\{\,\nn\\
&\mathrm{Tr}\Big[\frac{1}{\zeta-\umat M}\frac{\partial^2\umat M}{\partial K_a\partial K_b}\frac{1}{(\zeta-\umat M)^2}\frac{\partial\umat M}{\partial K_c}\Big]\nn\\
&-\mathrm{Tr}\Big[\frac{1}{(\zeta-\umat M)^2}\frac{\partial\umat M}{\partial K_b}\frac{1}{\zeta-\umat M}\frac{\partial^2\umat M}{\partial K_a\partial K_c}\Big]\nn\\
&+\mathrm{Tr}\Big[\frac{1}{\zeta-\umat M}\frac{\partial\umat M}{\partial K_a}\frac{1}{\zeta-\umat M}\frac{\partial\umat M}{\partial K_b}\frac{1}{(\zeta-\umat M)^2}\frac{\partial\umat M}{\partial K_c}\Big]\nn\\
&-\mathrm{Tr}\Big[\frac{1}{(\zeta-\umat M)^2}\frac{\partial\umat M}{\partial K_c}\frac{1}{\zeta-\umat M}\frac{\partial\umat M}{\partial K_b}\frac{1}{\zeta-\umat M}\frac{\partial\umat M}{\partial K_a}\Big]\nn\\
&-\mathrm{Tr}\Big[\frac{1}{\zeta-\umat M}\frac{\partial\umat M}{\partial K_c}\frac{1}{(\zeta-\umat M)^2}\frac{\partial\umat M}{\partial K_b}\frac{1}{\zeta-\umat M}\frac{\partial\umat M}{\partial K_a}\Big]\nn\\
&+\mathrm{Tr}\Big[\frac{1}{\zeta-\umat M}\frac{\partial\umat M}{\partial K_b}\frac{1}{\zeta-\umat M}\frac{\partial\umat M}{\partial K_a}\frac{1}{(\zeta-\umat M)^2}\frac{\partial\umat M}{\partial K_c}\Big]\Big\}\ .\nn\\
\end{align}
Using the cyclicity of the trace this can be simplified to
\begin{align}\label{B7c}
&\frac{\partial}{\partial K_a}Y_{\alpha,bc}=\frac{\I}{2}\frac{1}{2\pi\I}\oint_{S_\alpha}\mathrm d\zeta\,\Big\{\nn\\
&\qquad\mathrm{Tr}\Big[\frac{1}{\zeta-\umat M}\frac{\partial^2\umat M}{\partial K_a\partial K_b}\frac{1}{(\zeta-\umat M)^2}\frac{\partial\umat M}{\partial K_c}\Big]\nn\\
&\quad+\mathrm{Tr}\Big[\frac{1}{\zeta-\umat M}\frac{\partial\umat M}{\partial K_a}\frac{1}{\zeta-\umat M}\frac{\partial\umat M}{\partial K_b}\frac{1}{(\zeta-\umat M)^2}\frac{\partial\umat M}{\partial K_c}\Big]\Big\}\nn\\
&\;-(b\leftrightarrow c)\ ,
\end{align}
which proves (\ref{3.15}). With (\ref{2.17}) we find
\begin{align}\label{B7d}
&\frac{\partial}{\partial K_a}Y_{\alpha,bc}=\frac{\I}{2}\frac{1}{2\pi\I}\Big\{\sum_{\beta,\gamma}\oint_{S_\alpha}\mathrm d\zeta\nn\\
&\quad(\zeta-E_\beta)^{-1}(\zeta-E_\gamma)^{-2}\,\mathrm{Tr}\Big[\mathbbm P_\beta\frac{\partial^2\umat M}{\partial K_a\partial K_b}\mathbbm P_\gamma\frac{\partial\umat M}{\partial K_c}\Big]\nn\\
&+\sum_{\beta,\gamma,\sigma}\oint_{S_\alpha}\mathrm d\zeta\,(\zeta-E_\beta)^{-1}(\zeta-E_\gamma)^{-1}(\zeta-E_\sigma)^{-2}\nn\\
&\quad\times\mathrm{Tr}\Big[\mathbbm P_\beta\frac{\partial\umat M}{\partial K_a}\mathbbm P_\gamma\frac{\partial\umat M}{\partial K_b}\mathbbm P_\sigma\frac{\partial\umat M}{\partial K_c}\Big]\Big\}-(b\leftrightarrow c)\ .
\end{align}

The integrals in (\ref{B7d}) are easily evaluated using \linebreak[4] Cauchy's theorems.

With the short hand notations
\begin{align}\label{B9}
\beta:=\mathbbm P_\beta\, ,\quad a:=\frac{\partial\umat M}{\partial K_a}\, ,\quad(ab):=\frac{\partial^2\umat M}{\partial K_a\partial K_b}
\end{align}
we obtain
\begin{align}\label{B10}
&\frac{\partial}{\partial K_a}Y_{\alpha,bc}=\frac{\I}{2}\Big\{\sum_{\gamma\not=\alpha}(E_\alpha-E_\gamma)^{-2}\,\mathrm{Tr}[\alpha(ab)\gamma c]\nn\\
&+\sum_{\beta\not=\alpha}-(E_\alpha-E_\beta)^{-2}\,\mathrm{Tr}[\beta(ab)\alpha c]\nn\\
&-2\sum_{\sigma\not=\alpha}(E_\alpha-E_\sigma)^{-3}\,\mathrm{Tr}[\alpha a\alpha b\sigma c]\nn\\
&+\sum_{\gamma\not=\alpha}(E_\alpha-E_\gamma)^{-3}\,\mathrm{Tr}[\alpha a\gamma b\alpha c]\nn\\
&+\sum_{\beta\not=\alpha}(E_\alpha-E_\beta)^{-3}\,\mathrm{Tr}[\beta a\alpha b\alpha c]\nn\\
&+\sum_{\gamma,\sigma\not=\alpha}(E_\alpha-E_\gamma)^{-1}(E_\alpha-E_\sigma)^{-2}\,\mathrm{Tr}[\alpha a\gamma b\sigma c]\nn\\
&+\sum_{\beta,\sigma\not=\alpha}(E_\alpha-E_\beta)^{-1}(E_\alpha-E_\sigma)^{-2}\,\mathrm{Tr}[\beta a\alpha b\sigma c]\nn\\
&-\sum_{\beta,\gamma\not=\alpha}(E_\alpha-E_\beta)^{-1}(E_\alpha-E_\gamma)^{-2}\,\mathrm{Tr}[\beta a\gamma b\alpha c]\nn\\
&-\sum_{\beta,\gamma\not=\alpha}(E_\alpha-E_\gamma)^{-1}(E_\alpha-E_\beta)^{-2}\,\mathrm{Tr}[\beta a\gamma b\alpha c]\Big\}\nn\\
&-(b\leftrightarrow c)\ .
\end{align}
This can be simplified to
\begin{align}\label{B11}
&\frac{\partial}{\partial K_a}Y_{\alpha,bc}=\frac{\I}{2}\Big\{\sum_{\beta\not=\alpha}(E_\alpha-E_\beta)^{-2}\,\mathrm{Tr}[\alpha(ab)\beta c-\alpha c\beta(ab)]\nn\\
&+\sum_{\beta\not=\alpha}(E_\alpha-E_\beta)^{-3}\,\mathrm{Tr}[-2\alpha a\alpha b\beta c+\alpha a\beta b\alpha c+\beta a\alpha b\alpha c]\nn\\
&+\sum_{\beta,\gamma\not=\alpha}(E_\alpha-E_\beta)^{-1}(E_\alpha-E_\gamma)^{-2}\nn\\
&\qquad\times\mathrm{Tr}[\alpha a\beta b\gamma c+\beta a\alpha b\gamma c-\gamma a\beta b\alpha c-\beta a\gamma b\alpha c]\Big\}\nn\\
&-(b\leftrightarrow c)
\end{align}
which proves (\ref{3.16}).

\section{Useful expressions for PV fluxes}\label{s:AppendixC}
\setcounter{equation}{0}

In this appendix we derive $\mathcal I_{\alpha,jk}^{(\vmc E)\PV}$ (\ref{4.33}) and give the analogous expressions for $\mathcal I_{\alpha,jk}^{(\vmc B)\PV}$ and $\mathcal I_{\alpha,jk}^{(\vmc E,\vmc B)\PV}$. Then, we discuss (\ref{4.22}) and (\ref{4.23}).

Taking into account (\ref{2.2}), (\ref{A0c}) and (\ref{A0d}) we first derive the expansion of $(\zeta-\umat M(K))^{-1}$ around $\delta=0$. Analogously to (\ref{B5}) and (\ref{B6}) we find
\begin{align}\label{C1}
\frac{\partial}{\partial \delta}\frac{1}{\zeta-\umat M}&=\frac{1}{\zeta-\umat M}\frac{\partial\umat M}{\partial \delta}\frac{1}{\zeta-\umat M}
\end{align}
and
\begin{align}\label{C1b}
\frac{\partial}{\partial \delta}\frac{1}{(\zeta-\umat M)^2}&=\frac{1}{(\zeta-\umat M)^2}\frac{\partial\umat M}{\partial \delta}\frac{1}{\zeta-\umat M}\nn\\
&+\frac{1}{\zeta-\umat M}\frac{\partial\umat M}{\partial \delta}\frac{1}{(\zeta-\umat M)^2}\ .
\end{align}
With the short hand notation
\begin{align}\label{C2}
z:=\left.\frac{1}{\zeta-\umat M(\vmc E,\vmc B)}\right|_{\delta=0}=\frac{1}{\zeta-\umat M^{(0)}(K)}
\end{align}
the expansion of the trace in (\ref{4.8}) up to linear order in the PV parameter $\delta$ reads
\begin{align}\label{C3}
&\quad\mathrm{Tr}\Bigg[\frac{1}{\zeta-\umat{M}(\vmc E,\vmc B)}\underline D_j\frac{1}{(\zeta-\umat{M}(\vmc E,\vmc B))^2}\underline D_k\Bigg]\nn\\
&=\mathrm{Tr}\big[(z+z\delta\umat{M}_{\PV} z+\mathcal O(\delta^2))\underline D_j\nn\\
&\quad\times(z^2+z^2\delta\umat{M}_{\PV} z+z\delta\umat{M}_{\PV} z^2+\mathcal O(\delta^2))\underline D_k\big]\nn\\
&=\mathrm{Tr}\big[z\underline D_jz^2\underline D_k\big]+\delta\mathrm{Tr}\big[z\underline D_j z^2\umat{M}_{\PV}z\underline D_k\nn\\
&\quad+z\underline D_j z\umat{M}_{\PV}z^2\underline D_k+z^2\underline D_k z\umat{M}_{\PV}z\underline D_j\big]+\mathcal O(\delta^2)\ .
\end{align}
Inserting (\ref{C3}) in (\ref{4.8}) and performing a partial integration we obtain
\begin{align}\label{C4}
\mathcal I_{\alpha,jk}^{(\vmc E)}(\vmc E,\vmc B)&=\frac{\I}{2}\frac{1}{2\pi\I}\oint_{S_\alpha}\mathrm d\zeta\,\mathrm{Tr}\big[z\underline D_jz^2\underline D_k\big]\nn\\
&\qquad+\delta\mathrm{Tr}\big[z\umat{M}_{\PV}z\underline D_jz^2\underline D_k-(j\leftrightarrow k)\big]
\end{align}
which proves (\ref{4.33}). This derivation also holds for $\mathcal I_{\alpha,jk}^{(\vmc B)\PV}$ and $\mathcal I_{\alpha,jk}^{(\vmc E,\vmc B)\PV}$ where $\underline D_j,\underline D_k$ are replaced by $\underline \mu_j,\underline \mu_k$ and $\underline D_j,\underline \mu_k$, respectively. In this way we obtain from (\ref{4.9}) and (\ref{4.10})
\begin{align}\label{C5}
\mathcal I_{\alpha,jk}^{(\vmc B)\PV}(\vmc E,\vmc B)&=\delta\,\frac{\I}{2}\frac{1}{2\pi\I}\oint_{S_\alpha}\mathrm d\zeta\,\mathrm{Tr}\Bigg[\frac{1}{\zeta-\umat M^{(0)}(\vmc E,\vmc B)}\umat M_{\mathrm \PV}\nn\\
&\hspace{-6em}\times\frac{1}{\zeta-\umat M^{(0)}(\vmc E,\vmc B)}\underline \mu_j\frac{1}{\big(\zeta-\umat M^{(0)}(\vmc E,\vmc B)\big)^2}\underline \mu_k-(j\leftrightarrow k)\Bigg]
\end{align}
and
\begin{align}\label{C6}
&\mathcal I_{\alpha,jk}^{(\vmc E,\vmc B)\PV}(\vmc E,\vmc B)=\delta\,\I\frac{1}{2\pi\I}\oint_{S_\alpha}\mathrm d\zeta\,\mathrm{Tr}\Bigg[\frac{1}{\zeta-\umat M^{(0)}(\vmc E,\vmc B)}\umat M_{\mathrm \PV}\nn\\
&\times\frac{1}{\zeta-\umat M^{(0)}(\vmc E,\vmc B)}\underline D_j\frac{1}{\big(\zeta-\umat M^{(0)}(\vmc E,\vmc B)\big)^2}\underline \mu_k\nn\\
&\hspace{13em}-(\underline D_j\leftrightarrow \underline \mu_k)\Bigg]\ .
\end{align}

Now we discuss (\ref{4.22}) and (\ref{4.23}). From (\ref{4.21}) we see that $\umat M^{(0)}(\vmc E,\vmc B)$ and $\umat M^{(0)}(-\vmc E,\vmc B)$ have the same set of eigenvalues. We still have to check that our numbering scheme leads indeed to (\ref{4.22}) and (\ref{4.23}) for every $\alpha$. We consider again only $\vmc B\not= \vec 0$ and vary $\vmc E$ starting from $\vmc E=\vec 0$:
\begin{align}\label{C29}
\vmc E(\lambda)&=\lambda\vmc E\ ,\nn\\
\lambda&\in[0,1]\ .
\end{align}
For $\lambda=0$ (\ref{4.22}) and (\ref{4.23}) are trivial. Increasing now $\lambda$ continuously we encounter, due to $\vmc B\not=\vec 0$, no level crossings. Therefore, the identification of the eigenvalues and the quasi projectors corresponding to the same index $\alpha$ for $(\vmc E(\lambda),\vmc B)$ and $(-\vmc E(\lambda),\vmc B)$ is always possible. This proves (\ref{4.22}) and (\ref{4.23}).

\section{Detailed calculations of specific flux-density vector fields}\label{s:AppendixD}
\setcounter{equation}{0}

In this appendix we calculate the constants $a^\alpha$ of (\ref{5.6}). From (\ref{4.29}) we find the P-violating part $\vec J_{\alpha}^{(\vmc B)\PV}(\vec 0,\vmc B)$ of $\vec J_{\alpha}^{(\vmc B)}(\vec 0,\vmc B)$ to vanish. Therefore, neglecting terms of second order in the small PV parameter $\delta$, we can calculate $\vec J_{\alpha}^{(\vmc B)}(\vec 0,\vmc B)$ setting $\delta=0$. Then, the 2S states decouple from the 2P states, and we may restrict ourselves to the submatrix $\umat{M}^{\mathrm{2S},(0)}(\vec 0,\vmc B)$ of $\umat{M}^{(0)}(\vec 0,\vmc B)$ (\ref{2.2}) with respect to the 2S states, see Tables \ref{t:H2.M0} and \ref{t:H2.Mu} in Appendix \ref{s:AppendixE}. The derivatives $\partial_{B_i}\umat{M}^{\mathrm{2S},(0)}(\vec 0,\vmc B)$, $i\in\{1,2,3\}$, of this submatrix read in the basis $\rket{\alpha,\vmc E=\vec 0,\vmc B=\vec 0}$ with $\alpha=9,\dots,12$ (see Table \ref{t:state.labelsFD} in Appendix \ref{s:AppendixA})
\begin{align}\label{D1}
\frac{\partial\umat{M}^{\mathrm{2S},(0)}}{\partial B_1}&=\frac{g\mu_B}{2\sqrt 2}\left(
\begin{array}{m{0.9em}m{0.9em}m{0.9em}m{0.9em}@{\hspace{-0.1em}}}
0&1&0&-1\\
1&0&1&0\\
0&1&0&1\\
-1&0&1&0
\end{array}
\right)\ ,\nn\\
\frac{\partial\umat{M}^{\mathrm{2S},(0)}}{\partial B_2}&=\frac{g\mu_B}{2\sqrt 2}\left(
\begin{array}{m{0.9em}m{0.9em}m{0.9em}m{0.9em}@{\hspace{-0.1em}}}
0&-$\I$&0&$\I$\\
$\I$&0&-$\I$&0\\
0&$\I$&0&$\I$\\
-$\I$&0&-$\I$&0
\end{array}
\right)\ ,\nn\\
\frac{\partial\umat{M}^{\mathrm{2S},(0)}}{\partial B_3}&=\frac{g\mu_B}{2}\left(
\begin{array}{m{0.9em}m{0.9em}m{0.9em}m{0.9em}@{\hspace{-0.1em}}}
1&0&0&0\\
0&0&0&1\\
0&0&-1&0\\
0&1&0&0
\end{array}
\right)\ .
\end{align}
Due to rotational invariance of $\vec J_\alpha^{\vmc B}(\vec 0,\vmc B)$, see (\ref{5.6}), we may specify $\vmc B=B_3\,\vec e_3$ for the evaluation of $a^\alpha$ in (\ref{5.7}). This simplifies the calculation of the eigenvalues and of the right and left eigenvectors of $\umat{M}^{\mathrm{2S},(0)}(\vec 0,\vmc B)$. In this case we find
\begin{align}\label{D1b}
&\umat{M}^{\mathrm{2S},(0)}(\vec 0,B_3\,\vec e_3)\nn\\
&\qquad=\left(
\begin{array}{@{\hspace{0.15em}}c@{\hspace{0em}}c@{\hspace{0em}}c@{\hspace{0em}}c@{\hspace{-0.15em}}}
\chi_1+\frac{g\mu_B}{2}\mathcal B_3&0&0&0\\
0&\chi_1&0&\frac{g\mu_B}{2}\mathcal B_3\\
0&0&\chi_1-\frac{g\mu_B}{2}\mathcal B_3&0\\
0&\frac{g\mu_B}{2}\mathcal B_3&0&\chi_2
\end{array}
\right)
\end{align}
where $\chi_1=L+\mathcal A/32-\I\Gamma_S/2$ and $\chi_2=L-3\mathcal A/32-\I\Gamma_S/2$. The eigenvalues of the matrix in (\ref{D1b}) are
\begin{align}
E_9&=\chi_1+\frac{g\mu_B}{2}\mathcal B_3\ ,\nn\\
E_{10}&=L-\frac{\mathcal A}{32}+\frac{\chi_3}{16}-\I\Gamma_S/2\ ,\nn\\
E_{11}&=\chi_1-\frac{g\mu_B}{2}\mathcal B_3\ ,\nn\\
E_{12}&=L-\frac{\mathcal A}{32}-\frac{\chi_3}{16}-\I\Gamma_S/2
\end{align}
where $\chi_3=\sqrt{\HyperFineSplitting^2+(8\mathcal B_3\,g\mu_B)^2}$. From the eigenvectors we calculate explicit representations of the projection operators $\mathbbm P^{\mathrm{2S}}_\alpha$ for the 2S states and obtain
\begin{align}\label{D2}
\mathbbm P^{\mathrm{2S},(0)}_9&=\left(
\begin{array}{m{0.9em}m{0.9em}m{0.9em}m{0.9em}@{\hspace{-0.1em}}}
1&0&0&0\\
0&0&0&0\\
0&0&0&0\\
0&0&0&0
\end{array}
\right)\ ,\nn\\
\mathbbm P^{\mathrm{2S},(0)}_{10}&=\frac{1}{2\chi_3}\left(
\begin{tabular}{@{\hspace{0.2em}}c@{\hspace{0.7em}}c@{\hspace{0.7em}}c@{\hspace{0.7em}}c@{\hspace{-0.1em}}}
0&0&0&0\\
0&$\chi_3+\HyperFineSplitting$&0&$8\mathcal B_3\,g\mu_B$\\
0&0&0&0\\
0&$8\mathcal B_3\,g\mu_B$&0&$\chi_3-\HyperFineSplitting$
\end{tabular}
\right)\ ,\nn\\
\mathbbm P^{\mathrm{2S},(0)}_{11}&=\left(
\begin{array}{m{0.9em}m{0.9em}m{0.9em}m{0.9em}@{\hspace{-0.1em}}}
0&0&0&0\\
0&0&0&0\\
0&0&1&0\\
0&0&0&0
\end{array}
\right)\ ,\nn\\
\mathbbm P^{\mathrm{2S},(0)}_{12}&=\frac{1}{2\chi_3}\left(
\begin{tabular}{@{\hspace{0.2em}}c@{\hspace{0.7em}}c@{\hspace{0.7em}}c@{\hspace{0.7em}}c@{\hspace{-0.1em}}}
0&0&0&0\\
0&$\chi_3-\HyperFineSplitting$&0&$-8\mathcal B_3\,g\mu_B$\\
0&0&0&0\\
0&$-8\mathcal B_3\,g\mu_B$&0&$\chi_3+\HyperFineSplitting$
\end{tabular}
\right)\ .
\end{align}

Now, all ingredients for (\ref{3.13}) are available, and a straightforward calculation yields, with (\ref{3.18}), (\ref{4.9}), and (\ref{4.10b}), the P-conserving flux-density vector field
\begin{align}\label{D3}
\vec J_\alpha^{\vmc B}(\vec 0,\vmc B=B_3\,\vec e_3)=\left\{
\begin{array}{cl}
-\frac{B_3\,\vec e_3}{\phantom{\big(}\hspace{-0.2em}|\mathcal B_3|^3} & ,\quad\mbox{for }\alpha=9\ ,\\[0.3em]
\frac{B_3\,\vec e_3}{\phantom{\big(}\hspace{-0.2em}|\mathcal B_3|^3}  & ,\quad\mbox{for }\alpha=11\ ,\\[0.3em]
0 & ,\quad\mbox{for }\alpha=10,12\ .
\end{array}
\right.
\end{align}
Rotational invariance of $\vec J_\alpha^{\vmc B}(\vec 0,\vmc B)$ then leads to (\ref{5.7}).

We now give the relations between the functions $g_r^\alpha,\;h_r^\alpha$, $r=1,\dots, 15$, introduced in Section \ref{ss:4.3}, and the functions $\tilde g_{i}^\alpha,\;\tilde h_{i}^\alpha$, $i=1,\dots,6$, introduced for (\ref{5.18}) and (\ref{5.19}) in Section \ref{ss:5.3}:
\begin{align}
&\hspace{-0.3em}\tilde g_{1}^\alpha= \mathcal E_0\mathcal B_0\,\Big[g_7^\alpha+\frac13 (2\mathcal E_3^2-\mathcal E_1^2)\,g_8^\alpha+\frac13 (2\mathcal B_3^2-\mathcal B_2^2)\,g_9^\alpha\nn\\
&\hspace{-0.3em}\,\qquad+\frac43\,g_{10}^\alpha\Big] \ ,\\
&\hspace{-0.3em}\tilde g_{2}^\alpha= \mathcal E_0\mathcal B_0\,\big[2\mathcal B_2\mathcal E_3^2\,g_{11}^\alpha+2\mathcal B_2\,g_{12}^\alpha\big] \ ,\\
&\hspace{-0.3em}\tilde g_{3}^\alpha= \mathcal E_0\mathcal B_0\,\Big[\frac12\mathcal B_2\,g_5^\alpha+\mathcal B_2(\mathcal E_3^2-\mathcal E_1^2)\,g_{11}^\alpha+\mathcal B_2\,g_{12}^\alpha\Big] \ ,\\
&\hspace{-0.3em}\tilde g_{4}^\alpha= \mathcal E_0\mathcal B_0\,\Big[-\frac12 g_6^\alpha-\mathcal E_3^2\,g_8^\alpha-\,g_{10}^\alpha\Big] \ ,\\
&\hspace{-0.3em}\tilde g_{5}^\alpha= -\mathcal E_0^2\,\mathcal B_2\,g_2^\alpha \ ,\\
&\hspace{-0.3em}\tilde g_{6}^\alpha= \mathcal E_0^2\,\mathcal B_3^2\,g_3^\alpha \ ,\\
&\hspace{-0.3em}\tilde h_{1}^\alpha= \mathcal E_0\mathcal B_0\,\Big[h_7^\alpha+\frac13 (2\mathcal E_3^2-\mathcal E_1^2)\,h_8^\alpha+\frac13 (2\mathcal B_3^2-\mathcal B_2^2)\,h_9^\alpha\nn\\
&\hspace{-0.3em}\,\qquad+\frac43\mathcal E_3^2\mathcal B_3^2 \,h_{10}^\alpha\Big]\ ,\\
&\hspace{-0.3em}\tilde h_{2}^\alpha= \mathcal E_0\mathcal B_0\,\big[2\mathcal B_2\,h_{11}^\alpha+2\mathcal B_2\mathcal B_3^2\,h_{12}^\alpha\big] \ ,\\
&\hspace{-0.3em}\tilde h_{3}^\alpha= \mathcal E_0\mathcal B_0\,\Big[\frac12\mathcal B_2\,h_5^\alpha+\mathcal B_2(\mathcal E_3^2-\mathcal E_1^2)\,h_{11}^\alpha+\mathcal B_2\mathcal E_3^2\mathcal B_3^2\,h_{12}^\alpha\Big]\ ,\\
&\hspace{-0.3em}\tilde h_{4}^\alpha= \mathcal E_0\mathcal B_0\,\Big[-\frac12\mathcal B_3^2\, h_6^\alpha-\,h_8^\alpha-\mathcal B_3^2\,h_{10}^\alpha\Big] \ ,\\
&\hspace{-0.3em}\tilde h_{5}^\alpha= -\mathcal E_0^2\,\mathcal B_2\,h_2^\alpha \ ,\\
&\hspace{-0.3em}\tilde h_{6}^\alpha=  \mathcal E_0^2\,h_3^\alpha\ .
\end{align}

\onecolumn
\section{The matrix representations of $\umat{\tilde M}_0$, $\uvec{D}$ and $\uvec{\mu}$}\label{s:AppendixE}
\setcounter{equation}{0}

Tables \ref{t:H2.M0}, \ref{t:H2.D} and \ref{t:H2.Mu} show the mass matrix for zero external fields, $\umat{\tilde M}_0$, the electric dipole operator $\uvec{D}$ and the magnetic dipole operator $\uvec{\mu}$ for the $n=2$ states of hydrogen. We give all these matrices in the basis of the pure 2S and 2P states, that is, the states for zero external fields and without the P-violating mixing.

In Tables \ref{t:H2.D} and \ref{t:H2.Mu} we use the spherical unit vectors, which are defined as
\begin{align}
\vec e_0 = \vec e_3\ ,\qquad\vec e_\pm = \mp\frac1{\sqrt2}\klr{\vec e_1 \pm \I\vec e_2}\ ,
\end{align}
where $\vec e_i$ $(i=1,2,3)$ are the Cartesian unit vectors. For $\vec e_\pm$, the following relation holds:
\begin{align}
\vec e_\pm^* = -\vec e_\mp\ .
\end{align}

\begin{table}[htbp]
\begin{center}
\caption{The mass matrix $\umat{\tilde M}_0$ (\ref{2.3}) for the $n=2$ states of hydrogen. For the quantities $\Delta$, $L$, $\mcal A$ and $\delta_{1,2}$ see Table \ref{t:valuesFD}. The lifetimes $\tau_{P,S}$ are $\tau_S=\Gamma_S^{-1}=0.1216\,$s and $\tau_P=\Gamma_P^{-1}=1.596\times 10^{-9}\,$s; see \cite{Sap04,LaShSo05}.
}\label{t:H2.M0}
\vskip10pt
\begin{tabular}{l||c|cccc|}
& \parbox{18mm}{\vskip5pt$\ 2P_{3/2},2,2\ $\vskip4pt} & $\ 2P_{3/2},2,1\ $ & $\ 2P_{3/2},1,1\ $ & $\ 2P_{1/2},1,1\ $ & $\ 2S_{1/2},1,1\ $\\  \hline\hline
$\ 2P_{3/2},2,2\ $ & \mycell{18mm}{$\FineStructure+\frac{\HyperFineSplitting}{160}$\\[-0.0ex]$-\tfrac{\I }{2}\Gamma_P$}
& 0 & 0 & 0 & 0\\ \hline
$\ 2P_{3/2},2,1\ $ & 0 & 
\mycell{18mm}{${\FineStructure}+\frac{\HyperFineSplitting}{160}$\\[-0.0ex]$-\frac{\I}{2}\,{\Gamma_P}$}
& 0 & 0 & 0\\  
$\ 2P_{3/2},1,1\ $ & 0 & 0 & 
\mycell{18mm}{${\FineStructure}-\frac{\HyperFineSplitting}{96}$\\[-0.0ex]$-\frac{\I}{2}\,{\Gamma_P}$} & 
$-\frac{\HyperFineSplitting}{192\,{\sqrt{2}}}$ & 0\\  
$\ 2P_{1/2},1,1\ $ & 0 & 0 & 
$-\frac{\HyperFineSplitting}{192\,{\sqrt{2}}}$ & 
$\frac{\HyperFineSplitting}{96}-\frac\I2{\Gamma_P}$ & 
\mycell{18mm}{$\I{\delta_1}{\LambShift}$\\[-0.0ex]$+\frac\I2\delta_2 \LambShift$} \\ 
$\ 2S_{1/2},1,1\ $ & 0 & 0 & 0 & 
\mycell{18mm}{$-\I{\delta_1}{\LambShift}$\\[-0.0ex]$-\frac\I2 {\delta_2}{\LambShift}$} &
\mycell{18mm}{${\LambShift}+\frac{\HyperFineSplitting}{32}$\\[-0.0ex]$-\frac{\I}{2}\,{\Gamma_S}$}\\  \hline
\end{tabular}\\[15pt]
\centerline{\bf (Table \ref{t:H2.M0}a)}
\end{center}
\end{table}
\begin{table}
\begin{center}
\begin{tabular}{l||cccccc|}
& \parbox{18mm}{\vskip5pt$\ 2P_{3/2},2,0\ $\vskip4pt} & $\ 2P_{3/2},1,0\ $ & $\ 2P_{1/2},1,0\ $ & $\ 2S_{1/2},1,0\ $ & $\ 2P_{1/2},0,0\ $ & $\ 2S_{1/2},0,0\ $\\ \hline\hline
$\ 2P_{3/2},2,0\ $ & 
\mycell{18mm}{${\FineStructure}+\frac{\HyperFineSplitting}{160}$\\[-0.0ex]$-\frac{\I}{2}{\Gamma_P}$} & 0 & 0 & 0
& 0 & 0\\  
$\ 2P_{3/2},1,0\ $ & 0 & 
\mycell{18mm}{${\FineStructure}-\frac{\HyperFineSplitting}{96}$\\[-0.0ex]$-\frac{\I}{2}{\Gamma_P}$} & $-\frac{\HyperFineSplitting}{192{\sqrt{2}}}$ & 0 & 0 & 0 \\  
$\ 2P_{1/2},1,0\ $ & 0 & 
$-\frac{\HyperFineSplitting}{192{\sqrt{2}}}$ & $\frac{\HyperFineSplitting}{96} - \frac\I2{\Gamma_P}$ & 
\mycell{18mm}{$\I\delta_1{\LambShift}$\\[-0.0ex]$+\frac\I2\delta_2{\LambShift}$} & 0 & 0 \\ 
$\ 2S_{1/2},1,0\ $ & 0 & 0 & 
\mycell{18mm}{$-\I{\delta_1}{\LambShift}$\\[-0.0ex]$-\frac\I2{\delta_2}{\LambShift}$} & 
\mycell{18mm}{${\LambShift}+\frac{\HyperFineSplitting}{32}$\\[-0.0ex]$-\frac{\I}{2}{\Gamma_S}$} & 0 & 0 \\ 
$\ 2P_{1/2},0,0\ $ & 0 & 0 & 0 & 0 & 
$-\frac{\HyperFineSplitting}{32} - \frac\I2{\Gamma_P}$ & 
\mycell{18mm}{$\I{\delta_1}{\LambShift}$\\[-0.0ex]$+\frac{3}{2}\I{\delta_2}{\LambShift}$} \\
$\ 2S_{1/2},0,0\ $ & 0 & 0 & 0 & 0 & 
\mycell{18mm}{$-\I{\delta_1}{\LambShift}$\\[-0.0ex]$-\frac{3}{2}\I{\delta_2}{\LambShift}$} & 
\mycell{18mm}{${\LambShift}-\frac{3\HyperFineSplitting}{32}$\\[-0.0ex]$-\frac{\I}{2}{\Gamma_S}$} \\ \hline
\end{tabular}\\[15pt]
\centerline{\bf (Table \ref{t:H2.M0}b)}
\end{center}
\end{table}
\begin{table}
\begin{center}
\begin{tabular}{l||cccc|c|}
& \parbox{20mm}{\vskip5pt$\ 2P_{3/2},2,-1\ $\vskip4pt} & $\ 2P_{3/2},1,-1\ $ & $\ 2P_{1/2},1,-1\ $ & $\ 2S_{1/2},1,-1\ $ & $\ 2P_{3/2},2,-2\ $\\ \hline\hline
$\ 2P_{3/2},2,-1\ $ & 
\mycell{18mm}{${\FineStructure}+\frac{\HyperFineSplitting}{160}$\\[-0.0ex]$-\frac{\I}{2}{\Gamma_P}$} 
& 0 & 0 & 0 & 0 \\  
$\ 2P_{3/2},1,-1\ $ & 0 & 
\mycell{18mm}{${\FineStructure}-\frac{\HyperFineSplitting}{96}$\\[-0.0ex]$-\frac{\I}{2}{\Gamma_P}$} & 
$-\frac{\HyperFineSplitting}{192{\sqrt{2}}}$ & 0 & 0 \\ 
$\ 2P_{1/2},1,-1\ $ & 0 & 
$-\frac{\HyperFineSplitting}{192{\sqrt{2}}}$ & 
$\frac{\HyperFineSplitting}{96} - \frac\I2{\Gamma_P}$ & 
\mycell{18mm}{$\I{\delta_1}{\LambShift}$\\[-0.0ex]$+\frac\I2{\delta_2}{\LambShift}$} & 0 \\ 
$\ 2S_{1/2},1,-1\ $ & 0 & 0 & 
\mycell{18mm}{$-\I{\delta_1}{\LambShift}$\\[-0.0ex]$-\frac\I2{\delta_2}{\LambShift}$} & 
\mycell{18mm}{${\LambShift}+\frac{\HyperFineSplitting}{32}$\\[-0.0ex]$-\frac{\I}{2}{\Gamma_S}$} & 0 \\  \hline
$\ 2P_{3/2},2,-2\ $ & 0 & 0 & 0 & 0 & 
\mycell{18mm}{${\FineStructure}+\frac{\HyperFineSplitting}{160}$\\[-0.0ex]$-\frac{\I}{2}{\Gamma_P}$} \\ \hline
\end{tabular}\\[15pt]
\centerline{\bf (Table \ref{t:H2.M0}c)}
\end{center}
\end{table}

{
\begin{table}
\begin{center}
\caption{The suitably normalised electric dipole operator $\uvec{D}/(e\,r_B)$ for the $n=2$ states of hydrogen where $r_B$ is the Bohr radius for hydrogen.}\label{t:H2.D}
\vskip10pt
\begin{tabular}{l||c|cccc|}
& \mycell{18mm}{$\ 2P_{3/2},2,2\ $} & $\ 2P_{3/2},2,1\ $ & $\ 2P_{3/2},1,1\ $ & $\ 2P_{1/2},1,1\ $ & $\ 2S_{1/2},1,1\ $\\ 
\hline\hline
\mycell{18mm}{$\ 2P_{3/2},2,2\ $} & 0 & 0 & 0 & 0 & $-3\sem$\\ \hline
\mycell{18mm}{$\ 2P_{3/2},2,1\ $} & 0 & 0 & 0 & 0 & 
$\frac{3}{\sqrt{2}}\sez$\\
\mycell{18mm}{$\ 2P_{3/2},1,1\ $} & 0 & 0 & 0 & 0 & 
$-{\sqrt{\frac{3}{2}}}\sez$\\
\mycell{18mm}{$\ 2P_{1/2},1,1\ $} & 0 & 0 & 0 & 0 & 
$-{\sqrt{3}}\sez$\\
\mycell{18mm}{$\ 2S_{1/2},1,1\ $} & $3\sep$ & 
$\frac{3}{{\sqrt{2}}}\sez$ & 
$-{\sqrt{\frac{3}{2}}}\sez$ & 
$-{\sqrt{3}}\sez$ & 0\\ \hline
\end{tabular}\\[15pt]
\centerline{\bf (Table \ref{t:H2.D}a)}
\vskip15pt
\begin{tabular}{l||cccccc|}
& \mycell{18mm}{$\ 2P_{3/2},2,0\ $} & $\ 2P_{3/2},1,0\ $ & $\ 2P_{1/2},1,0\ $ & $\ 2S_{1/2},1,0\ $ & $\ 2P_{1/2},0,0\ $ & $\ 2S_{1/2},0,0\ $\\ \hline\hline
\mycell{18mm}{$\ 2P_{3/2},2,1\ $} & 0 & 0 & 0 & 
$-\frac{3}{\sqrt2}\sem$ & 0 & 0\\ 
\mycell{18mm}{$\ 2P_{3/2},1,1\ $} & 0 & 0 & 0 & 
$-\sqrt{\frac{3}{2}}\sem$ & 0 & 
$-\sqrt{6}\sem$\\ 
\mycell{18mm}{$\ 2P_{1/2},1,1\ $} & 0 & 0 & 0 & 
$-\sqrt{3}\sem$ & 0 & 
$\sqrt{3}\sem$\\ 
\mycell{18mm}{$\ 2S_{1/2},1,1\ $} & 
$\sqrt{\frac{3}{2}}\sem$ & 
$-\sqrt{\frac{3}{2}}\sem$ & 
$-\sqrt{3}\sem$ & 0 & 
$\sqrt{3}\sem$ & 0\\ \hline
\end{tabular}\\[15pt]
\centerline{\bf (Table \ref{t:H2.D}b)}
\end{center}
\end{table}
\begin{table}
\begin{center}
\begin{tabular}{l||cccc}
& \mycell{18mm}{$\ 2P_{3/2},2,1\ $} & $\ 2P_{3/2},1,1\ $ & $\ 2P_{1/2},1,1\ $ & $\ 2S_{1/2},1,1\ $\\ \hline\hline
\mycell{18mm}{$\ 2P_{3/2},2,0\ $} & 0 & 0 & 0 & $-\sqrt{\frac32}\sep$ \\ 
\mycell{18mm}{$\ 2P_{3/2},1,0\ $} & 0 & 0 & 0 & $\sqrt{\frac32}\sep$ \\ 
\mycell{18mm}{$\ 2P_{1/2},1,0\ $} & 0 & 0 & 0 & $\sqrt3\sep$ \\ 
\mycell{18mm}{$\ 2S_{1/2},1,0\ $} & $\frac3{\sqrt2}\sep$ & $\sqrt{\frac32}\sep$ & $\sqrt3\sep$ & 0 \\ 
\mycell{18mm}{$\ 2P_{1/2},0,0\ $} & 0 & 0 & 0 & $-\sqrt3\sep$ \\ 
\mycell{18mm}{$\ 2S_{1/2},0,0\ $} & 0 & $\sqrt6\sep$ & $-\sqrt3\sep$ & 0 \\ \hline
\end{tabular}\\[15pt]
\centerline{\bf (Table \ref{t:H2.D}c)}
\vskip15pt
\begin{tabular}{l||cccccc|}
& \mycell{18mm}{$\ 2P_{3/2},2,0\ $} & $\ 2P_{3/2},1,0\ $ & $\ 2P_{1/2},1,0\ $ & $\ 2S_{1/2},1,0\ $ & $\ 2P_{1/2},0,0\ $ & $\ 2S_{1/2},0,0\ $\\ \hline\hline
\mycell{18mm}{$\ 2P_{3/2},2,0\ $} & 0 & 0 & 0 & 
${\sqrt{6}}\sez$ & 0 & 0\\
\mycell{18mm}{$\ 2P_{3/2},1,0\ $} & 0 & 0 & 0 & 0 & 0 & 
${\sqrt{6}}\sez$\\
\mycell{18mm}{$\ 2P_{1/2},1,0\ $} & 0 & 0 & 0 & 0 & 0 & 
$-{\sqrt{3}}\sez$\\
\mycell{18mm}{$\ 2S_{1/2},1,0\ $} & 
${\sqrt{6}}\sez$ & 0 & 0 & 0 & 
$-{\sqrt{3}}\sez$ & 0\\
\mycell{18mm}{$\ 2P_{1/2},0,0\ $} & 0 & 0 & 0 & 
$-{\sqrt{3}}\sez$ & 0 & 0\\
\mycell{18mm}{$\ 2S_{1/2},0,0\ $} & 0 & 
${\sqrt{6}}\sez$ & 
$-{\sqrt{3}}\sez$ & 0 & 0 & 0\\ \hline
\end{tabular}\\[15pt]
\centerline{\bf (Table \ref{t:H2.D}d)}
\end{center}
\end{table}
\begin{table}
\begin{center}
\begin{tabular}{l||cccc|}
& \mycell{18mm}{$\ 2P_{3/2},2,-1\ $} & $\ 2P_{3/2},1,-1\ $ & $\ 2P_{1/2},1,-1\ $ & $\ 2S_{1/2},1,-1\ $\\ \hline\hline
\mycell{18mm}{$\ 2P_{3/2},2,0\ $} & 0 & 0 & 0 & 
$-\sqrt{\frac{3}{2}}\sem$\\ 
\mycell{18mm}{$\ 2P_{3/2},1,0\ $} & 0 & 0 & 0 & 
$-\sqrt{\frac{3}{2}}\sem$\\
\mycell{18mm}{$\ 2P_{1/2},1,0\ $} & 0 & 0 & 0 & 
$-\sqrt{3}\sem$\\
\mycell{18mm}{$\ 2S_{1/2},1,0\ $} & 
$\frac{3}{\sqrt2}\sem$ & 
$-\sqrt{\frac{3}{2}}\sem$ & 
$-\sqrt{3}\sem$ & 0\\
\mycell{18mm}{$\ 2P_{1/2},0,0\ $} & 0 & 0 & 0 & 
$-\sqrt{3}\sem$\\
\mycell{18mm}{$\ 2S_{1/2},0,0\ $} & 0 & 
$\sqrt{6}\sem$ & 
$-\sqrt{3}\sem$ & 0\\ \hline
\end{tabular}\\[15pt]
\centerline{\bf (Table \ref{t:H2.D}e)}\vskip15pt
\begin{tabular}{l||cccccc|}
& \mycell{18mm}{$\ 2P_{3/2},2,0\ $} & $\ 2P_{3/2},1,0\ $ & $\ 2P_{1/2},1,0\ $ & $\ 2S_{1/2},1,0\ $ & $\ 2P_{1/2},0,0\ $ & $\ 2S_{1/2},0,0\ $\\ \hline\hline
\mycell{18mm}{$\ 2P_{3/2},2,-1\ $} & 0 & 0 & 0 & $-\frac3{\sqrt2}\sep$ & 0 & 0 \\ 
\mycell{18mm}{$\ 2P_{3/2},1,-1\ $} & 0 & 0 & 0 & $\sqrt{\frac32}\sep$ & 0 & $-\sqrt6\sep$ \\ 
\mycell{18mm}{$\ 2P_{1/2},1,-1\ $} & 0 & 0 & 0 & $\sqrt3\sep$ & 0 & $\sqrt3\sep$ \\ 
\mycell{18mm}{$\ 2S_{1/2},1,-1\ $} & $\sqrt{\frac32}\sep$ & $\sqrt{\frac32}\sep$ & $\sqrt3\sep$ & 0 & $\sqrt3\sep$ & 0 \\ \hline
\end{tabular}\\[15pt]
\centerline{\bf (Table \ref{t:H2.D}f)}
\end{center}
\end{table}
\begin{table}
\begin{center}
\begin{tabular}{l||cccc|c|}
& \mycell{18mm}{$\ 2P_{3/2},2,-1\ $} & $\ 2P_{3/2},1,-1\ $ & $\ 2P_{1/2},1,-1\ $ & $\ 2S_{1/2},1,-1\ $ & $\ 2P_{3/2},2,-2\ $\\ \hline\hline
\mycell{18mm}{$\ 2P_{3/2},2,-1\ $} & 0 & 0 & 0 & 
$\frac{3}{{\sqrt{2}}}\sez$ & 0\\  
\mycell{18mm}{$\ 2P_{3/2},1,-1\ $} & 0 & 0 & 0 & 
${\sqrt{\frac{3}{2}}}\sez$ & 0\\  
\mycell{18mm}{$\ 2P_{1/2},1,-1\ $} & 0 & 0 & 0 & 
${\sqrt{3}}\sez$ & 0\\  
\mycell{18mm}{$\ 2S_{1/2},1,-1\ $} & 
$\frac{3}{{\sqrt{2}}}\sez$ & 
${\sqrt{\frac{3}{2}}}\sez$ & 
${\sqrt{3}}\sez$ & 0 & $3\sem$\\ \hline
\mycell{18mm}{$\ 2P_{3/2},2,-2\ $} & 0 & 0 & 0 & $-3\sep$ & 0\\ \hline
\end{tabular}\\[15pt]
\centerline{\bf (Table \ref{t:H2.D}g)}
\end{center}
\end{table}}

{
\begin{table}
\begin{center}
\caption{The suitably normalised magnetic dipole operator $\uvec{\mu}/\mu_B$ for the $n=2$ states of hydrogen, where $\mu_B = e\hbar/(2m_e)$ is the Bohr magneton and $g=2.002319304(76)$ is the Land\'{e} factor of the electron \cite{Moh05}.}\label{t:H2.Mu}
\vskip10pt
\begin{tabular}{l||c|cccc|}
& \mycell{18mm}{$\ 2P_{3/2},2,2\ $} & $\ 2P_{3/2},2,1\ $ & $\ 2P_{3/2},1,1\ $ & $\ 2P_{1/2},1,1\ $ & $\ 2S_{1/2},1,1\ $\\ 
\hline\hline
\mycell{18mm}{$\ 2P_{3/2},2,2\ $} & $-\frac{g+2}{2}\sez$ & $-\frac{\sqrt2(g+2)}{4}\sem$ & $\frac{\sqrt2(g+2)}{4\sqrt3}\sem$ & $-\frac{g-1}{\sqrt3}\sem$ & 0\\ \hline
\mycell{18mm}{$\ 2P_{3/2},2,1\ $} & $\frac{\sqrt2(g+2)}{4}\sep$ & $-\frac{g+2}{4}\sez$ & $-\frac{g+2}{4\sqrt3}\sez$ & $-\frac{g-1}{\sqrt6}\sez$ & 0\\
\mycell{18mm}{$\ 2P_{3/2},1,1\ $} & $-\frac{\sqrt2(g+2)}{4\sqrt3}\sep$ & $-\frac{g+2}{4\sqrt3}\sez$ & $-\frac{5(g+2)}{12}\sez$ & $\frac{g-1}{3\sqrt2}\sez$ & 0\\
\mycell{18mm}{$\ 2P_{1/2},1,1\ $} & $\frac{g-1}{\sqrt3}\sep$ & $-\frac{g-1}{\sqrt6}\sez$ & $\frac{g-1}{3\sqrt2}\sez$ & $\frac{g-4}6\sez$ & 0\\
\mycell{18mm}{$\ 2S_{1/2},1,1\ $} & 0 & 0 & 0 & 0 & $-\frac{g}2\sez$\\ \hline
\end{tabular}\\[15pt]
\centerline{\bf (Table \ref{t:H2.Mu}a)}\vskip15pt
\begin{tabular}{l||cccccc|}
& \mycell{18mm}{$\ 2P_{3/2},2,0\ $} & $\ 2P_{3/2},1,0\ $ & $\ 2P_{1/2},1,0\ $ & $\ 2S_{1/2},1,0\ $ & $\ 2P_{1/2},0,0\ $ & $\ 2S_{1/2},0,0\ $\\ \hline\hline
\mycell{18mm}{$\ 2P_{3/2},2,1\ $} & $-\frac{\sqrt3(g+2)}4\sem$ & $\frac{g+2}{4\sqrt3}\sem$ & $-\frac{g-1}{\sqrt6}\sem$ & 0 & 0 & 0\\ 
\mycell{18mm}{$\ 2P_{3/2},1,1\ $} & $-\frac{g+2}{12}\sem$ & $-\frac{5(g+2)}{12}\sem$ & $-\frac{\sqrt2(g-1)}{6}\sem$ & 0 & $-\frac{\sqrt2(g-1)}{3}\sem$ & 0\\ 
\mycell{18mm}{$\ 2P_{1/2},1,1\ $} & $\frac{\sqrt2(g-1)}{6}\sem$ & $-\frac{\sqrt2(g-1)}{6}\sem$ & $\frac{g-4}{6}\sem$ & 0 & $-\frac{g-4}{6}\sem$ & 0\\ 
\mycell{18mm}{$\ 2S_{1/2},1,1\ $} & 0 & 0 & 0 & $-\frac{g}2\sem$ & 0 & $\frac{g}2\sem$\\ \hline
\end{tabular}\\[15pt]
\centerline{\bf (Table \ref{t:H2.Mu}b)}
\end{center}
\end{table}
\begin{table}
\begin{center}
\begin{tabular}{l||cccc|}
& \mycell{18mm}{$\ 2P_{3/2},2,1\ $} & $\ 2P_{3/2},1,1\ $ & $\ 2P_{1/2},1,1\ $ & $\ 2S_{1/2},1,1\ $\\ \hline\hline
\mycell{18mm}{$\ 2P_{3/2},2,0\ $} & $\frac{\sqrt3(g+2)}4\sep$ & $\frac{g+2}{12}\sep$ & $-\frac{\sqrt2(g-1)}{6}\sep$ & 0 \\ 
\mycell{18mm}{$\ 2P_{3/2},1,0\ $} & $-\frac{g+2}{4\sqrt3}\sep$ & $\frac{5(g+2)}{12}\sep$ & $\frac{\sqrt2(g-1)}{6}\sep$ & 0 \\ 
\mycell{18mm}{$\ 2P_{1/2},1,0\ $} & $\frac{g-1}{\sqrt6}\sep$ & $\frac{\sqrt2(g-1)}{6}\sep$ & $-\frac{g-4}{6}\sep$ & 0 \\ 
\mycell{18mm}{$\ 2S_{1/2},1,0\ $} & 0 & 0 & 0 & $\frac{g}2\sep$ \\ 
\mycell{18mm}{$\ 2P_{1/2},0,0\ $} & 0 & $\frac{\sqrt2(g-1)}{3}\sep$ & $\frac{g-4}6\sep$ & 0 \\ 
\mycell{18mm}{$\ 2S_{1/2},0,0\ $} & 0 & 0 & 0 & $-\frac{g}2\sep$ \\ \hline
\end{tabular}\\[15pt]
\centerline{\bf (Table \ref{t:H2.Mu}c)}\vskip15pt
\begin{tabular}{l||cccccc|}
& \mycell{18mm}{$\ 2P_{3/2},2,0\ $} & $\ 2P_{3/2},1,0\ $ & $\ 2P_{1/2},1,0\ $ & $\ 2S_{1/2},1,0\ $ & $\ 2P_{1/2},0,0\ $ & $\ 2S_{1/2},0,0\ $\\ \hline\hline
\mycell{18mm}{$\ 2P_{3/2},2,0\ $} & 0 & 
$-\frac{g+2}{6}\sez$ & 
$-\frac{\sqrt2(g-1)}{3}\sez$ & 0 & 0 & 0\\
\mycell{18mm}{$\ 2P_{3/2},1,0\ $} & 
$-\frac{g+2}{6}\sez$ & 0 & 0 & 0 & 
$-\frac{{\sqrt{2}}(g-1)}{3}\sez$ & 0\\ 
\mycell{18mm}{$\ 2P_{1/2},1,0\ $} & 
$-\frac{{\sqrt{2}}(g-1)}{3}\sez$ & 0 & 0 & 0 & 
$\frac{g-4}{6}\sez$ & 0\\ 
\mycell{18mm}{$\ 2S_{1/2},1,0\ $} & 0 & 0 & 0 & 0 & 0 & 
$-\frac{g}{2}\sez$\\ 
\mycell{18mm}{$\ 2P_{1/2},0,0\ $} & 0 & 
$-\frac{{\sqrt{2}}(g-1)}{3}\sez$ & 
$\frac{g-4}{6}\sez$ & 0 & 0 & 0\\
\mycell{18mm}{$\ 2S_{1/2},0,0\ $} & 
0 & 0 & 0 & 
$-\frac{g}{2}\sez$ & 
0 & 0\\ \hline
\end{tabular}\\[15pt]
\centerline{\bf (Table \ref{t:H2.Mu}d)}
\end{center}
\end{table}
\begin{table}
\begin{center}
\begin{tabular}{l||cccc|}
& \mycell{18mm}{$\ 2P_{3/2},2,-1\ $} & $\ 2P_{3/2},1,-1\ $ & $\ 2P_{1/2},1,-1\ $ & $\ 2S_{1/2},1,-1\ $\\ \hline\hline
\mycell{18mm}{$\ 2P_{3/2},2,0\ $} & 
$-\frac{\sqrt3(g+2)}{4}\sem$ & 
$\frac{g+2}{12}\sem$ & 
$-\frac{\sqrt2(g-1)}{6}\sem$ & 0\\
\mycell{18mm}{$\ 2P_{3/2},1,0\ $} & 
$-\frac{g+2}{4 \sqrt{3}}\sem$ & 
$-\frac{5 (g+2)}{12}\sem$ & 
$-\frac{\sqrt2(g-1)}{6}\sem$ & 0\\ 
\mycell{18mm}{$\ 2P_{1/2},1,0\ $} & 
$\frac{g-1}{\sqrt{6}}\sem$ & 
$-\frac{\sqrt2(g-1)}{6}\sem$ & 
$\frac{g-4}{6}\sem$ & 0\\ 
\mycell{18mm}{$\ 2S_{1/2},1,0\ $} & 0 & 0 & 0 & 
$-\frac{g}{2}\sem$\\ 
\mycell{18mm}{$\ 2P_{1/2},0,0\ $} & 0 & 
$\frac{\sqrt2(g-1)}{3}\sem$ & 
$\frac{g-4}{6}\sem$ & 0\\ 
\mycell{18mm}{$\ 2S_{1/2},0,0\ $} & 0 & 0 & 0 & 
$-\frac{g}{2}\sem$\\ \hline
\end{tabular}\\[15pt]
\centerline{\bf (Table \ref{t:H2.Mu}e)}\vskip15pt
\begin{tabular}{l||cccccc|}
& \mycell{18mm}{$\ 2P_{3/2},2,0\ $} & $\ 2P_{3/2},1,0\ $ & $\ 2P_{1/2},1,0\ $ & $\ 2S_{1/2},1,0\ $ & $\ 2P_{1/2},0,0\ $ & $\ 2S_{1/2},0,0\ $\\ \hline\hline
\mycell{18mm}{$\ 2P_{3/2},2,-1\ $} & 
$\frac{\sqrt3(g+2)}{4}\sep$ & 
$\frac{g+2}{4 \sqrt{3}}\sep$ & 
$-\frac{g-1}{\sqrt{6}}\sep$ & 0 & 0 & 0 \\ 
\mycell{18mm}{$\ 2P_{3/2},1,-1\ $} & 
$-\frac{g+2}{12}\sep$ & 
$\frac{5 (g+2)}{12}\sep$ & 
$\frac{\sqrt2(g-1)}{6}\sep$ & 0 & 
$-\frac{\sqrt2(g-1)}{3}\sep$ & 0 \\ 
\mycell{18mm}{$\ 2P_{1/2},1,-1\ $} & 
$\frac{\sqrt2(g-1)}{6}\sep$ & 
$\frac{\sqrt2(g-1)}{6}\sep$ & 
$-\frac{g-4}{6}\sep$ & 0 & 
$-\frac{g-4}{6}\sep$ & 0 \\ 
\mycell{18mm}{$\ 2S_{1/2},1,-1\ $} & 0 & 0 & 0 & 
$\frac{g}{2}\sep$ & 0 & $\frac{g}{2}\sep$ \\ \hline
\end{tabular}\\[15pt]
\centerline{\bf (Table \ref{t:H2.Mu}f)}
\end{center}
\end{table}
\begin{table}
\begin{center}
\begin{tabular}{l||cccc|c|}
& \mycell{18mm}{$\ 2P_{3/2},2,-1\ $} & $\ 2P_{3/2},1,-1\ $ & $\ 2P_{1/2},1,-1\ $ & $\ 2S_{1/2},1,-1\ $ & $\ 2P_{3/2},2,-2\ $\\ \hline\hline
\mycell{18mm}{$\ 2P_{3/2},2,-1\ $} & 
$\frac{g+2}{4}\sez$ & 
$-\frac{g+2}{4{\sqrt{3}}}\sez$ & 
$-\frac{g-1}{{\sqrt{6}}}\sez$ & 
0 & $-\frac{\sqrt2(g+2)}{4}\sem$\\
\mycell{18mm}{$\ 2P_{3/2},1,-1\ $} & 
$-\frac{g+2}{4{\sqrt{3}}}\sez$ & 
$\frac{5(g+2)}{12}\sez$ & 
$-\frac{g-1}{3{\sqrt{2}}}\sez$ & 
0 & $-\frac{\sqrt2(g+2)}{4\sqrt3}\sem$\\ 
\mycell{18mm}{$\ 2P_{1/2},1,-1\ $} & 
$-\frac{g-1}{{\sqrt{6}}}\sez$ & 
$-\frac{g-1}{3{\sqrt{2}}}\sez$ & 
$-\frac{g-4}{6}\sez$ & 
0 & $\frac{g-1}{\sqrt3}\sem$\\ 
\mycell{18mm}{$\ 2S_{1/2},1,-1\ $} & 
0 & 0 & 0 & 
$\frac{g}{2}\sez$ & 0\\ \hline
\mycell{18mm}{$\ 2P_{3/2},2,-2\ $} & $\frac{\sqrt2(g+2)}{4}\sep$ & $\frac{\sqrt2(g+2)}{4\sqrt3}\sep$ & $-\frac{g-1}{\sqrt3}\sep$ & 0 & 
$\frac{2 + g}{2}\sez$\\ \hline
\end{tabular}\\[15pt]
\centerline{\bf (Table \ref{t:H2.Mu}g)}
\end{center}
\end{table}}
\vfill

\twocolumn


\bibliographystyle{epj}
\bibliography{myapvbib_FluxDensities}

\end{document}